\documentclass[twocolumn]{aastex63} 

\usepackage[colorinlistoftodos]{todonotes}
\usepackage[flushleft]{threeparttable}
\usepackage{graphicx}

\let\Oldtodo\todo
\renewcommand{\todo}[1]{\Oldtodo[inline]{#1}}

\hypersetup{linkcolor=cyan,citecolor=cyan,filecolor=cyan,urlcolor=cyan}

\usepackage{amsmath}
\usepackage{float}

\shorttitle{The Neptunian ``Final Parsec" Problem}
\shortauthors{Hallatt, Barker \& Millholland}

\graphicspath{{./}{figures/}}

\begin{document}

\title{The Neptunian Final Parsec Problem:
How Inertial Waves Enable Neptunes and Gas Giants to Survive Tidal Migration}
\correspondingauthor{Tim Hallatt}
\email{thallatt@mit.edu}

\author[0000-0003-4992-8427]{Tim Hallatt}
\affil{MIT Kavli Institute for Astrophysics and Space Research, Massachusetts Institute of Technology, Cambridge, MA 02139, USA}

\author[0000-0003-4397-7332]{Adrian J. Barker}
\affil{School of Mathematics, University of Leeds, Leeds, LS2 9JT, UK}

\author[0000-0003-3130-2282]{Sarah C. Millholland}
\affil{MIT Kavli Institute for Astrophysics and Space Research, Massachusetts Institute of Technology, Cambridge, MA 02139, USA}
\affil{Department of Physics, Massachusetts Institute of Technology, Cambridge, MA 02139, USA}

\begin{abstract}
Observations indicate that some hot Neptunes and Jupiters underwent high eccentricity migration (HEM). A longstanding issue facing HEM theory remains, however: in order to shrink their orbits by ${\sim}1{-}2$ orders of magnitude, planets must dissipate ${\gtrsim}10{\times}$ more orbital energy than is contained in their structural binding energy. Previous work suggests that excitation of planets’ fundamental modes (``$f$-modes'') at very high eccentricities, $e{\sim}$0.99, can shrink semi-major axes by ${\sim}$an order of magnitude (${\sim}5{\rightarrow}0.5$ au) while safely bleeding tidal energy to space. We examine how planets traverse the remaining order of magnitude in orbital shrinkage, $a{\sim}$0.5${\rightarrow}$0.05 au, which still poses a danger for Neptunian planets due to their lower binding energies (their ``final parsec'' problem). To assess whether tidally-migrating Neptunes balance their energy budgets, we perform coupled orbit/structure calculations tracking their frequency-dependent tidal response under inertial waves (IWs). We show that tidal radius inflation at moderate eccentricities sweeps Neptunes’ fundamental modes back into resonance with tidal forcing; only a small amount of tidal heat relative to a Neptune's binding energy is needed to drive this $f$-mode excitation. This excitation of Neptunes’ $f$-modes may offer a route to safe circularization, provided the $f$-mode dissipates without harming the planet. In contrast, eccentric gas giants ($e{\sim}0.9$) circularize without drama over ${\sim}$Gyr. Our theory predicts that warm Neptunes on a HEM track ($a{\gtrsim}0.1 \ \mathrm{au}, \ e{\gtrsim}0.8$) should be rare, in opposition to warm gas giants and lower eccentricity Neptunes. Observational confirmation of our predictions would constrain the interior structures of Neptunian planets via tidal theory.
\end{abstract}

\section{Introduction} \label{sec:intro}

Observations indicate that some ``hot Neptunes" (with masses $M_{\rm p}{\sim}30 \ M_{\oplus}$, and orbital periods ${\sim}$4-10 days) are emplaced in their current orbits via high eccentricity migration (HEM). Under HEM, highly eccentric ($e{\gtrsim}0.99$), distant orbits (semi-major axis $a{\gtrsim}$au) are thought to shrink via dissipation of tides excited in the planet (and possibly the star) at periastron. Distinct from the anomalously dense planets in the sub-Jovian desert \citep[periods ${\lesssim}$3 days;][]{visbeh25,halmil26b}, and warm Neptunes further out \citep[periods ${\gtrsim}10$ days, in the so-called ``savannah"; e.g.][]{bouattmall23}, hot Neptunes at periods ${\sim}4{-}10$ days indeed appear consistent with HEM along a number of axes, including their broadly distributed stellar obliquities \citep{bouattmall23,espwinbra26}, their possible pile-up at orbital periods ${\sim}5$ days in a ``ridge'' \citep[][]{casboulil24}, non-zero eccentricities \citep{corboudel20}, low planetary companion occurrence rate \citep[][]{almcasmar26,thoniecri26}, and their potentially similar stellar companion occurrence rate as their gas giant cousins, hot Jupiters \citep{doyarmacu25}.

While there is strong observational support behind HEM, a longstanding issue at the heart of theory remains: can planets survive the dissipation necessary to shrink orbits by ${\sim}1{-}2$ orders of magnitude? As we will show, even moderately eccentric Neptunes must dissipate ${\gtrsim}10{\times}$ more orbital energy than the gravitational binding energy contained in their structures. This energy budget issue is even more pronounced for Neptunes than it is for gas giants, for which the problem has long been recognized (and only partially solved, as we outline below). The purpose of this paper is to help determine how Neptunes and gas giants survive migration in the face of intense tidal heating.

The rate and deposition location of tidal heating dictate whether the planet survives HEM. Planets may survive unscathed if tidal heat is efficiently radiated to space (deposited at low pressures), instead of increasing the temperature of atmospheric gas and therefore driving mass loss. For planets on highly eccentric orbits ($e{\gtrsim}0.99$), tidal dissipation may be facilitated through excitation of planets' fundamental modes (``$f$-modes"), which oscillate on timescales comparable to the planet's dynamical time ${\sim}1/\omega_{\rm dyn}{=}(GM_{\rm p}/R_{\rm p}^{3})^{-1/2}$ \citep[with $\omega_{\rm dyn}$ the dynamical frequency, $G$ the gravitational constant, and $M_{\rm p}$ and $R_{\rm p}$ planet mass and radius, respectively;][]{unn79}. Provided $f$-modes are not driven so strongly that their dissipation unbinds the atmosphere via tidal streams \citep{guiramlin11} or deep shocks (\cite{zanmacbry26}; i.e.~periastron distances must exceed ${\gtrsim}0.01$ au), tidal energy can be safely bled to space via photon diffusion from the planet's upper layers where $f$-modes break \cite[see][for discussion of analogous processes in stars]{quaferkas16}.

The $f$-mode, however, can only shrink planetary orbits to a certain extent: only highly eccentric orbits permit the mode energy to grow chaotically from orbit to orbit, leading to wave breaking and orbital shrinkage \citep[e.g.][see also \citealt{zanmacbry26} for recent developments]{mar95,ivapap04,wu18,viclai18}. As a result of this stalling in the $f$-mode excitation as the orbit shrinks, planets remain distant and eccentric ($a{\sim}0.5$ au, $e{\sim}0.9$) even after intense dissipation \citep[e.g.][see also \citealt{yuweiarr22}, their Equation 5]{viclai19,teylaivic19,verful19}. How planets traverse the last order of magnitude in orbital distance, ${\sim}0.5{\rightarrow}0.05$ au, remains unclear. This final stage of orbital circularization is the focus of our paper; as we will show, tidal heating during this less extreme stage of migration continues to pose a danger to Neptunian planets. We dub this tension the ``Neptunian final parsec problem" \citep[in analogy with the challenge of shrinking binary black hole orbits by ${\sim}100{\times}$ from a separation ${\sim}1$ pc;][]{milmer03}.

Coupled planet structure/orbit calculations confirm that Neptunes' internal binding energies are readily overwhelmed by tidal heating. \cite{halmil26} showed that, under a simple tidal model with a constant quality factor \citep[e.g.][]{barogi09}, Neptunes are in fact susceptible to runaway tidal inflation, whereby atmospheric tides produce radius puff-up, which in turn enhances dissipation. They suggest that, unless Neptunes are extremely metal-rich (bulk metal mass fractions ${\gtrsim}0.8$) and exhibit finely-tuned tidal quality factors, tidal heating must be localized in the planet's upper layers to avoid runaway inflation. A major question that remains is whether this runaway inflation (and possible planetary demise) persists under more realistic tidal physics, as evolution under equilibrium tides differs qualitatively from more realistic models \cite[such as ``dynamical tides"; e.g.][]{ogi14}. Furthermore, the true strength of equilibrium tides may be far weaker than the free parameter used by \cite{halmil26}, such that equilibrium tides may not represent a major channel of dissipation \citep[e.g.][]{ivapap04b,Nils2023,lazbarvri24}. We will show that runaway inflation persists under more sophisticated tidal physics, and indeed may play a key role in resolving hot Neptunes' final parsec problem.

In the absence of efficient $f$-mode excitation and equilibrium tides, moderately eccentric planets ($a{\lesssim}0.5$ au, $e{\lesssim}0.9$) must dissipate orbital energy through other means. Chief alternatives include viscoelastic heating from a rocky/icy core \cite[e.g.][]{der79}, and dynamical tides in the form of ``gravity waves'' \citep[GWs; restored by buoyancy in radiative zones, e.g.][]{golnic89}, and ``inertial waves" \cite[IWs; restored by the Coriolis force in rotating convective zones, e.g.][]{gre68,ogilin04}. These mechanisms do not dissipate heat equally, however. Core dissipation in the solid or fluid state \citep[e.g.][]{corboulas14,stolai14} may circularize Neptune orbits over ${\sim}$Gyr \citep{petbha26}. Superheated fluid cores may also support their own dynamical tide however \citep[][]{farchi26}, further boosting core heating. GW dissipation on the other hand may only become appreciable in strongly irradiated planets, whose outer radiative zones extend to deep pressures/densities \citep[][]{lubtouliv97,lazbarvri24}\footnote{An additional concern is that the GW wavelength $\lambda_{\rm GW}$ can be comparable to the width of Neptunes' radiative zones $\Delta r_{\rm rad}$, precluding the standard dissipation rate estimation that relies on $\lambda_{\rm GW}{\ll}\Delta r_{\rm rad}$ for validity of the WKB approximation \citep[e.g.][]{goodic98}.}. If they exist, stably stratified regions at deeper pressures may also aid dissipation via gravity/gravito-inertial waves \citep[e.g.][]{jertouogi17,ponbarhol24}.

In contrast to GWs, IWs also appear capable of circularizing Neptunes; frequency-averaged \citep{lazbarvri24} and frequency-dependent calculations \citep[see e.g.][for application to gas giants]{ivapap07} indicate considerable IW dissipation regardless of irradiation level or planetary evolution stage. Whether such dissipation can be realized as a planet's orbit evolves remains unclear. Linear calculations predict that IW heating can differ by orders of magnitude over narrow bands of frequency \citep[while tidal frequency itself evolves with the orbit and planetary rotation; e.g.][]{ogilin04}, potentially shutting off heating unless particular frequencies are tidally excited. In this work we help determine whether IWs may contribute to, or even dominate, circularization. 

The main goal of our paper is to help resolve the Neptunian final parsec problem, i.e., to articulate how Neptunes dissipate ${\sim}10{\times}$ more energy from their orbits than is contained in their structures. To achieve this goal, we first determine whether the runaway inflation instability discovered by \cite{halmil26} persists under frequency-dependent tidal dissipation due to IWs. At face value, the erratic nature of IW dissipation may help Neptunes survive tidal heating; during an ``on" cycle, the planet will puff up, but provided tidal frequencies evolve into an ``off" cycle, it will cool and radiate the extra heat (as opposed to continuously puffing up under ever-growing tidal heating). We will show, however, that runaway inflation persists even under frequency-dependent tidal heating; IW dissipation is inherently unstable to planetary inflation.

We next estimate what happens $\textit{after}$ runaway inflation. Using our coupled structure/orbit code, we find that planets' $f$-mode frequencies sweep into resonance with tidal forcing during planetary puff-up. We highlight that even a small amount of tidal heat relative to planets' binding energies is sufficient to induce strong $f$-mode excitation. We suggest that the solution to the Neptunian final parsec problem is that $f$-mode breaking dissipates orbital energy without harming the planet, analogous to the highly eccentric case explored by previous authors. In other words: runaway tidal inflation under IWs is a feature, not a bug, allowing moderately eccentric planets to drain orbital energy without being destroyed.

A second goal of our paper is to explore the evolution of eccentric gas giants under IWs. We will establish that, unlike their Neptunian counterparts, giant planets circularize over ${\sim}$Gyr, experiencing only modest structural upheaval. Our calculations help resolve the longstanding question of whether giant planets can survive dissipating more energy from their orbits than is contained in their structures.

This paper is organized as follows. In Section \ref{sec:prelim}, we quantify how Neptunes' structural binding energies are dwarfed by the energy lost from their orbits during migration, further motivating our calculations. Section \ref{sec:dynamicaltides} describes our planet structure models and our numerical methodology to compute their IW dissipation. We summarize how IW dissipation depends on planet structure/thermal properties in Section \ref{sec:results1}. Section \ref{sec:tidal_EOM} presents the equations of motion used to follow planets' tidal/structural/thermal evolution, while the results of these integrations are provided in Section \ref{sec:evo}. We itemize the big-picture implications of our results for theory and observations in Section \ref{sec:discussion}, and close with a summary of our paper in Section \ref{sec:conclusion}.

\begin{figure*}[t]
\centering
\includegraphics[width=0.8\textwidth]{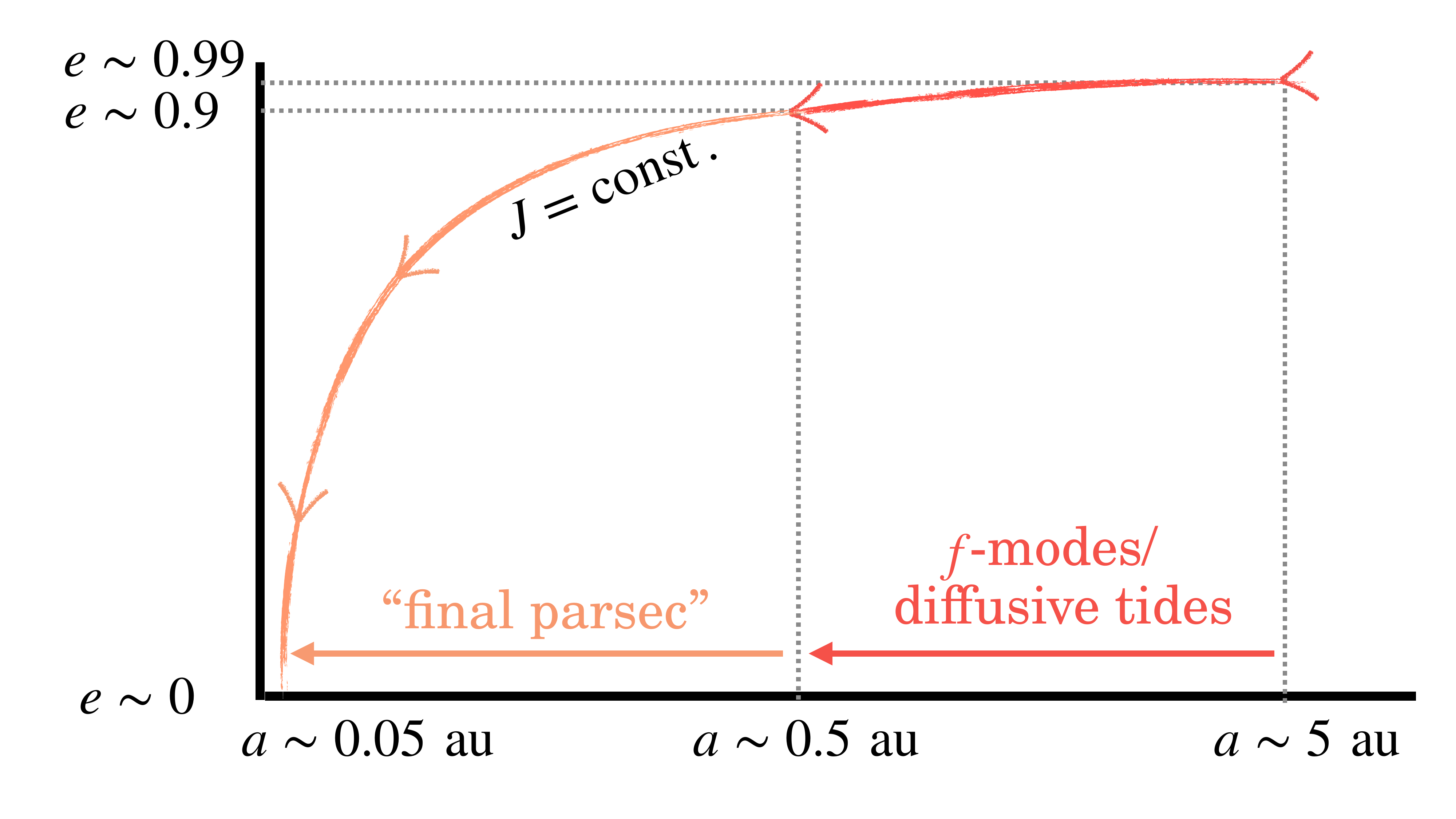}
\caption{Schematic of the problem addressed in this paper. In the high eccentricity migration paradigm, semi-major axes must shrink by ${\sim}$2 orders of magnitude ($a{\sim}5{\rightarrow}0.05$ au) to emplace a hot Neptune/Jupiter. Planets migrate under tidal dissipation following the constant angular momentum track depicted (``$J$=const.''); in so doing, they must dissipate more energy from their orbits than is contained in their structural binding energies. Previous work suggests that excitation of planets' fundamental modes at very high eccentricities, $e{\sim}0.99$, can shrink semi-major axes by ${\sim}$an order of magnitude while safely draining tidal/orbital energy to space \citep[this phase is depicted in red; e.g.][]{wu18,viclai19,yuweiarr22,zanmacbry26}. 
This paper is concerned with how planets traverse the remaining order of magnitude in orbit shrinkage, $a{\sim}0.5{\rightarrow}0.05$ au, which still poses a danger for Neptunian planets -- what we term the ``final parsec'' problem (illustrated in orange). 
\label{figure:schematic}}
\end{figure*}

\section{The Neptunian ``Final Parsec" Problem}\label{sec:prelim}

In this section we first establish the relevant energy scales involved in the circularization of Neptunian planets' orbits. We will show that even moderately eccentric Neptunian planets must dissipate more energy from their orbits than is contained in their structural binding energy.

The orbital energy of a Neptunian planet is,
\begin{align}\label{equation:Eorb}
\begin{split}
    E_{\rm orb}&=-\frac{GM_{\star}M_{\rm p}}{2a}\\
    &\approx  -2\times 10^{42}\bigg(\frac{a}{0.5 \ \rm au}\bigg)^{-1}\bigg(\frac{M_{\rm p}}{0.1 \ M_{\rm J}}\bigg)\bigg(\frac{M_{\star}}{M_{\odot}}\bigg) \ \rm erg,
\end{split}
\end{align}
\noindent where $M_\star$ is the stellar mass and $a$ the semi-major axis. In migrating from initial to final semi-major axes $a_{i}{\rightarrow}a_{f}$, the energy lost from the orbit is:
\begin{align}\label{equation:DeltaEorb}
\begin{split}
    \Delta E_{\rm orb}&=-\frac{GM_{\star}M_{\rm p}}{2}\bigg[\frac{1}{a_{f}}-\frac{1}{a_{i}}\bigg]\\
    &\approx  -10^{43}\bigg(\frac{M_{\rm p}}{0.1 \ M_{\rm J}}\bigg)\bigg(\frac{M_{\star}}{M_{\odot}}\bigg) \ \rm erg,
\end{split}
\end{align}

\noindent 
which we have evaluated using
$a_{i}{=}0.5 \ \rm au$ and $a_{f}{=}0.05 \ \rm au$. Meanwhile, the planet's structural binding energy is
\begin{align}\label{equation:Ebind}
\begin{split}
    E_{\rm b}&\sim-\frac{GM^{2}_{\rm p}}{R_{\rm p}}\\
    &\sim  -6\times 10^{41}\bigg(\frac{R_{\rm p}}{6 \ R_{\oplus}}\bigg)^{-1}\bigg(\frac{M_{\rm p}}{0.1 \ M_{\rm J}}\bigg)^{2} \ \rm erg.
\end{split}
\end{align}

\noindent 
The ratio of energy lost from the orbit to the planet's structural binding is,
\begin{align}\label{equation:Eratio}
\begin{split}
    \frac{\Delta E_{\rm orb}}{E_{\rm b}}&\sim17 \bigg(\frac{M_{\rm p}}{0.1 \ M_{\rm J}}\bigg)^{-1}\bigg(\frac{M_{\star}}{M_{\odot}}\bigg)\bigg(\frac{R_{\rm p}}{6 \ R_{\oplus}}\bigg).
\end{split}
\end{align}

Equation \ref{equation:Eratio} shows that the orbital energy that must be lost during migration exceeds the planet's structural binding energy by ${\gtrsim}$ an order of magnitude. Even for a giant planet, the energy ratio is ${\sim}$unity. In light of this energy budget mismatch, our goal is to quantitatively assess how the energy budget of a tidally migrating Neptune evolves under the relevant heating and cooling mechanisms. We will show that even Neptunes that are moderately eccentric ($e{\sim}0.5$ as opposed to $e{\sim}0.99$) cannot balance their energy budgets through IW dissipation alone. A schematic of the problem facing tidal migration theory is displayed in Figure \ref{figure:schematic}.

\section{Inertial Wave Spectra of hot Neptunes}\label{sec:dynamicaltides}

Our ultimate goal is to study the coupled evolution of a hot Neptune's structure and orbit under the influence of tidal dissipation due to IWs. In this section we first discuss how we compute IW spectra of $\textit{static}$ Neptunian planets (i.e. neglecting evolution). The parameter space probed by our calculations is summarized in Table \ref{tab:fiducial_table}.

\subsection{Planet Structure}

Our tidal calculations employ the planet structure models of \cite{halmil26}. We briefly re-iterate the assumptions underpinning these models here. Readers interested in the planet structure microphysics (e.g.~equation of state, opacities) and benchmarking the model's performance against the $\texttt{MESA}$ code \citep[][]{paxbildot11,paxcanarr13,paxmarsch15,paxschbau18,paxsmosch19} may refer to \cite{halmil26}. 

To follow planets' structural/thermal evolution, we constructed a large grid of planet structures in thermal and mechanical equilibrium (i.e., with spatially constant luminosity $L$). This grid varies the planetary core mass ($m_{\rm c}{\in}\{10,20\} \ M_{\oplus}$), envelope mass (yielding total masses $M_{\rm p}{\in}\{30,300\} \ M_{\oplus}$; envelopes are assumed to have solar composition with metal mass fraction $Z{=}0.02$), equilibrium temperatures ($T_{\rm eq}{\in}[509{-}1250]$ K in steps of ${\Delta}T_{\rm eq}{=}5 \ \rm K$, roughly corresponding to the range of equilibrium temperatures expected 
from 0.3${-}$0.05 au), and entropy in the innermost convective zone ($S{\in}[6,11] \ k_{\rm B}/m_{\rm H}$, with $k_{\rm B}$ Boltzmann's constant and $m_{\rm H}$ hydrogen's mass; the lower entropy is comparable to Jupiter's present day entropy of $S{\sim}6 \ k_{\rm B}/$baryon, e.g. \citealt{surtejsu25}). Core radii $R_{\rm c}$ follow the mass/radius relation for Earth-like composition of \cite{formarbar07}, approximately given by $R_{\rm c}{\propto}m^{0.25}_{\rm c}$. 

\begin{deluxetable*}{CCCCCCc}\label{tab:fiducial_table}
\tablecaption{Parameters used in our planet structure and tidal calculations.} 
\label{tab:parameters}
\tablecolumns{5}
\tablewidth{0pt}
\tablehead{
\colhead{Parameter} &
\colhead{Definition} &
\colhead{Value} & 
\colhead{Units} & 
}
\startdata
$M_{\rm p}$ & \rm planet \ mass & \{3{\times}10^{1}, \ 3{\times}10^{2}\} & M_{\oplus} \\
$m_{\rm c}$ & \rm \ planet \ core \ mass & \{10, \ 20\} & M_{\oplus} \\
$S$ & \rm \ planet \ entropy & [6, \ 11] & k_{\rm B}/m_{\rm H} \\
$T_{\rm eq}$ & \rm \ equilibrium \ temperature & [509, \ 1250] & \rm K \\
$a$ & \rm initial \ semi${-}$major \ axis & 0.3 & \rm au \\
$e$ & \rm initial \ eccentricity & 0.896 & \nodata \\
$Ek$ & \rm Ekman \ number & \{10^{-10}, \ 10^{-5}\} & \rm \nodata \\
$c_{\rm V}$ & \rm core \ heat \ capacity & 10^{7} & \rm erg \ g^{-1} \ K^{-1} \\
\enddata
\end{deluxetable*}

Unlike other recent planet structure codes \citep[e.g.][]{sursutej24}, our models do not account for composition gradients, semi/double-diffusive convection, or convective inhibition at the silicate core/hydrogen atmosphere interface \citep[e.g.][]{marguiste22}. Future work should address to what extent composition gradients and convective stability would alter the coupled tidal/structure evolution we explore here \citep[see e.g.][]{ponbarhol24}; we outline specific issues future work should resolve in Section \ref{subsec:fw}. Unless stated otherwise, planetary radii in this section (``$R_{\rm p}$'') correspond to the radiative-convective boundary (RCB), above which IWs cannot propagate.

\subsection{Tidal Problem Formulation}\label{subsec:tidal_problem}

We now specify our method to compute the linear tidal IW response of a given planet model. Our goal is to describe the basic physics and assumptions adopted by our calculations, and outline deviations in our methodology from previous efforts. Readers interested in a full exposition of the governing equations may consult Appendix \ref{sec:EOM}.

We consider a planet rigidly rotating at frequency $\boldsymbol{\Omega}_{\rm s}{=}\Omega_{\rm s}\hat{\boldsymbol{z}}$ (with rotation period $P_{\rm s}{=}2\pi/\Omega_{\rm s}$), and orbiting at frequency $\Omega_{\rm o}{=}\sqrt{GM_{\star}/a^{3}}$ (with orbital period $P_{\rm o}{=}2\pi/\Omega_{\rm o}$). Following standard procedure \citep[e.g.][]{kau61}, we decompose the tidal potential into spherical and temporal harmonics. For a given spherical harmonic degree $\ell$, order $m$, temporal harmonic $N$, and tidal frequency $\omega_{mN}{=}N\Omega_{\rm o}{-}m\Omega_{\rm s}$, the $\ell,m,N$ tidal potential component reads \citep[in a frame rotating at $\Omega_{\rm s}$; e.g.][]{polsme90},
\begin{equation}\label{equation:psi_tid}
    \Psi_{\ell,m,N}(\boldsymbol{r},t)=\mathrm{Re} [\mathcal{A}_{\ell,m,N} \, r^{\ell}Y^{m}_{\ell}(\theta,\phi)e^{-\mathrm{i}\omega_{mN} t}],
\end{equation}

\noindent where $\mathcal{A}_{\ell,m,N}$ is the tidal potential amplitude, and $Y^{m}_{\ell}(\theta,\phi)$ is an orthonormalized spherical harmonic. We focus on the quadrupolar $\ell{=}m{=}2$ tidal component, as it dominates tidal forcing for the eccentric, coplanar systems we explore in this work. The amplitude of the $\ell{=}m{=}2$ tidal potential, $\mathcal{A}_{2,2,N}$, reads \citep[e.g.][see also \citealt{bar25}]{murder99}: 
\begin{align}\label{equation:amplitude}
\begin{split}
    \mathcal{A}_{2,2,N}&=-C_{\rm t}\omega^{2}_{\rm dyn},
\end{split}
\end{align}

\noindent where we have defined $C_{\rm t}{=}\epsilon_{\rm t}A_{2,2,N}$, with $\epsilon_{\rm t}{=}(M_{\star}/M_{\rm p})(R_{\rm p}/a)^{3}$ a dimensionless measure of the fractional strength of tides to self-gravity \citep[e.g.][]{zah08,tre23}, and $A_{2,2,N}$ a numerical prefactor (e.g. $A_{2,2,2}{=}\sqrt{6\pi/5}$).

Following \cite{ogi13}, we separate the planet response to this potential into ``non-wavelike" and ``wavelike" components. The non-wavelike component describes the instantaneous hydrostatic response of the planet to the tidal potential \citep[analogous to the conventional equilibrium tide, but corrected for use in approximately neutrally stratified convection zones; e.g.][see also the discussion in \citealt{bar20}]{goodic98,terpapnel98}.\footnote{Our planet structure models indicate efficient convection, with (squared) Brunt–Väisälä  frequency $N^{2}{\sim}{-}10^{[-13,-15]} \ \rm s^{-2}$, much smaller in absolute magnitude than the (squared) tidal frequency.} We compute this non-wavelike displacement and its associated velocity field $\boldsymbol{u}_{\rm nw}$ following \cite{bar20} (re-iterated for clarity in the \hyperref[sec:appendix]{Appendix}). The wavelike component, on the other hand, corrects for the excitation of IWs through the Coriolis force acting on the non-wavelike flow. We next describe how we compute the wavelike response from the non-wavelike solution.

Our method to compute the wavelike solution builds on that of \cite{ogi09}, who studied IW excitation in a homogeneous, incompressible model atmosphere. To treat IWs in our more realistic planet models, we instead account for a radially-varying density profile following Vanon et al.~(under review). Adopting a radially-varying density is necessary, as density can vary by ${\sim}5$ orders of magnitude from 
the upper to lower boundaries in our atmosphere models (RCB to core surface, respectively; see Figure \ref{figure:densityprofiles}). To model the wavelike tide in planet structures exhibiting this strongly varying density, we employ the anelastic approximation:
\begin{equation}\label{equation:anelastic}
    \boldsymbol{\nabla}\cdot(\rho \boldsymbol{u}_{\rm w})=0,
\end{equation}

\noindent where $\rho(r)$ is the background density profile supplied from our structure models, and $\boldsymbol{u}_{\rm w}$ is the wavelike velocity field that we solve for. This is a consistent approximation for describing low-frequency inertial waves \citep[e.g.][]{ogi13}.

Armed with the anelastic constraint, we solve the following linearized, viscous momentum equation for the wavelike flow \citep[see][for the inviscid case, and Vanon et al.~(under review) for the viscous case and for discussion of non-linear effects]{ogi13}:
\begin{align}\label{equation:NS}
    & \hspace{-0.05cm}\partial_{t}\boldsymbol{u}_{\rm w}+2\boldsymbol{\Omega}_{\rm s}\times \boldsymbol{u}_{\rm w}=-\boldsymbol{\nabla}W_{\rm w}+\frac{1}{\rho}\boldsymbol{\nabla}\cdot\sigma_{\rm w}+\boldsymbol{f}_{\rm w},\\
    & \sigma_{\rm w}=\nu\rho\bigg[\boldsymbol{\nabla}\boldsymbol{u}_{\rm w}+(\boldsymbol{\nabla}\boldsymbol{u}_{\rm w})^{T}-\frac{2}{3}(\boldsymbol{\nabla}\cdot\boldsymbol{u}_{\rm w})\rm I\bigg],
    \label{equation:NS2}
\end{align}

\noindent with $W_{\rm w}{=}h_{\rm w}{+}\Psi$ composed of the wavelike contribution to specific enthalpy $h_{\rm w}$ and tidal gravitational potential given by Equation \ref{equation:psi_tid}, $\sigma_{\rm w}$ the stress tensor \citep[e.g.][]{bat00}, $\rm{I}$ the identity tensor, and $\nu$ the constant kinematic viscosity (detailed in Section \ref{sec:viscosity}). The last term in Equation \ref{equation:NS}, $\boldsymbol{f}_{\rm w}$, gives rise to IWs by the action of the Coriolis acceleration on the non-wavelike flow: 
\begin{equation}\label{equation:f_IW}
    \boldsymbol{f}_{\rm w}=- 2\boldsymbol{\Omega}_{\rm s}\times\boldsymbol{u}_{\rm nw}.
\end{equation}

We solve Equation \ref{equation:NS} similarly to \cite{ogi09}; to summarize, we decompose the solution into spheroidal/toroidal contributions and project onto spherical harmonics, solving each $r,\theta,\phi$-component (along with continuity, Equation \ref{equation:anelastic}) as a system of \
linear algebraic equations (an overview of our numerical methods is given in Section \ref{sec:numerics}). We adopt units such that length scales are measured in planetary radii, and timescales are measured in $\Omega_{\rm s}^{-1}$. We employ impenetrable and stress-free boundary conditions at the bottom and top of our atmospheres. In principle, our bottom boundary at the core surface may be better represented with a no-slip condition ($\boldsymbol{u}_{\rm w}{=}0$) instead; our choice of bottom boundary condition is unlikely to substantially alter our results, however. We next detail how we extract a tidal dissipation rate from our flow solutions. 

\subsection{Dissipation}\label{sec:viscosity}

Given the velocity field of the wavelike component $\boldsymbol{u}_{\rm w}$, the time-averaged viscous dissipation rate is (for our zero viscous flux boundary conditions):
\begin{equation}\label{equation:dissipation}
    D_{\nu}=2\int_{\rm conv}\rho\nu e_{ij}e_{ij} \ dV,
\end{equation}

\noindent where $dV$ is the volume element and the integral is computed over the convective fluid layer, and the rate-of-strain tensor is given by \citep[e.g.][]{bat00},
\begin{equation}
e_{ij}{=}\frac{1}{2}(\partial_{i}u_{\mathrm{w},j}+\partial_{j}u_{\mathrm{w}, i})-\frac{1}{3}(\boldsymbol{\nabla} \cdot \boldsymbol{u}_{\rm w})\delta_{ij},
\end{equation}

\noindent with $\delta_{ij}$ the Kronecker delta. In steady state, the viscous dissipation rate is balanced by the rate of tidal work on the wavelike flow, $P_{\rm t}$:

\begin{equation}\label{equation:Ptid}
    P_{\rm t}=\int_{\rm conv}\rho \boldsymbol{u}_{\rm w}\cdot \boldsymbol{f}_{\rm w} \ dV,
\end{equation}

\noindent i.e., $D_{\nu}{\approx}P_{\rm t}$ in steady-state. We regard our computations to be converged once $D_{\nu}{\approx}P_{\rm t}$ to sufficient accuracy (outlined in Section \ref{sec:numerics}).

\subsection{Tidal Quality Factor}

Once we have computed the dissipation rate $D_{\nu}$, we express it in terms of the more commonly used modified tidal quality factor, $Q^{\prime}$. The inverse of $Q^{\prime}$ is proportional to the amount of energy dissipated per tidal period as a fraction of the maximum energy stored in the tide \citep[e.g.][]{gol63}. Defining the dimensional dissipation rate $\tilde{D}_{\nu}$ (i.e.~in erg s$^{-1}$), and noting that the maximum energy stored in the equilibrium tide is ${\propto}(\omega_{\rm dyn}C_{\rm t}R_{\rm p})^{2}M_{\rm p}{\propto}R^{5}_{\rm p}|\mathcal{A}|^{2}/G$ \citep[e.g.][]{zah08}, the conversion between $\tilde{D}_{\nu}$ and $Q^{\prime}$ for an $\ell{=}2$ tidal component is \citep[e.g.][]{ogi14},
\begin{align}\label{equation:Q}
\begin{split}
    Q^{\prime}(\omega_{mN})=\frac{15}{16\pi G}\frac{R^{5}_{\rm p}|\mathcal{A}_{mN}|^{2}|\omega_{mN}|}{\tilde{D}_{\nu}(\omega_{mN})}.
\end{split}
\end{align}

To elucidate the scaling behavior of $Q^{\prime}$, we evaluate $Q^{\prime}$ in dimensionless units in the following way. We first write $\tilde{D}_{\nu}{=}\bar{\rho}\Omega^{3}_{\rm s}R^{5}_{\rm p}D_{\nu}$, where the dimensionless dissipation rate $D_{\nu}{\propto}C^{2}_{\rm t}$ is measured in units of the characteristic IW dissipation rate $\bar{\rho}\Omega^{3}_{\rm s}R^{5}_{\rm p}$, and $\bar{\rho}{=}M_{\rm p}/(4\pi R^{3}_{\rm p}/3)$ is the planet's bulk density. Substituting the non-dimensional dissipation rate $D_{\nu}$ into Equation \ref{equation:Q}, and including the tidal potential amplitude from Equation \ref{equation:amplitude}, we may write $Q^{\prime}$ as,
\begin{equation}\label{equation:Q_dimensionless}
\begin{split}
    Q^{\prime}(\omega_{mN})&=\frac{5}{4}\frac{|\omega_{mN}|}{\Omega_{\rm s}}\bigg(\frac{C_{\rm t}}{\epsilon_{\Omega}}\bigg)^{2}\frac{1}{D_{\nu}},
\end{split}
\end{equation}

\noindent where the (squared) ratio of spin to dynamical frequency, 
\begin{equation}\label{equation:eps_omega}
\begin{split}
\epsilon^{2}_{\Omega}=\frac{\Omega^{2}_{\rm s}}{\omega^{2}_{\rm dyn}}
    = \frac{\Omega^{2}_{\rm s}R^{3}_{\rm p}}{GM_{\rm p}}.
\end{split}
\end{equation}

\noindent We set $C_{\rm t}{=}1$ without loss of generality, since $D_{\nu}{\propto}C^{2}_{\rm t}$ in linear theory, so this factor cancels when computing $Q'$.

Equations \ref{equation:Q_dimensionless} and \ref{equation:eps_omega} inform us that $Q^{\prime}$ depends on the relative strengths of rotation and self-gravity. At fixed dimensionless forcing frequency $\omega_{mN}/\Omega_{\rm s}$, faster rotating planets experience relatively greater Coriolis driving of IWs, and hence greater dissipation (smaller $Q'$). Similarly, at fixed mass and rotation rate, a larger planetary radius increases $\epsilon_{\Omega}$ and therefore reduces $Q^{\prime}$. This scaling may also be understood by noting that the characteristic energy in the IWs scales as $(\Omega_{\rm s}C_{\rm t}R_{\rm p})^{2}M_{\rm p}$, whereas the maximum energy stored in the equilibrium tide scales as $(\omega_{\rm dyn}C_{\rm t}R_{\rm p})^{2}M_{\rm p}$. 

We compute $Q^{\prime}$ at fixed rotation rate $\Omega_{\rm ref}{=}2\pi/(1\rm \ day)$, and construct a grid of $Q^{\prime}(\omega_{mN}{,}\Omega_{\rm s}{=}1 \ \rm day)$ as a function of planet entropy and dimensionless tidal frequency. As the planetary spin and orbit evolve, we evaluate $D_{\nu}$ at the instantaneous value of $\hat{\omega}_{mN}$ and rescale the corresponding $Q^{\prime}$ by $(\Omega_{\rm ref}/\Omega_{\rm s})^{2}$.

We also compute the tidal quality factor corresponding to the frequency-averaged dissipation, partly to benchmark our results against other (simpler) calculations of IW dissipation. We compute the resulting $Q^{\prime}$ quantifying the frequency-averaged dissipation in two different ways. Firstly, we compute the quality factor corresponding to the frequency-averaged dissipation using the approach devised by
\cite{ogi13}, which gives a prediction that we denote by $\bar{Q^{\prime}}$. This approach assumes that IWs are excited impulsively and are then fully damped (through a mechanism that we do not need to specify), and requires solving ordinary differential equations that model the planetary structure \citep[see also][]{bar20}. Secondly, we average our frequency-dependent dissipation spectra (computed from solving the inertial wave response directly) over frequency, which gives a result that we denote by $\langle Q^{\prime}\rangle$. This weighted frequency average is given by,
\begin{align}\label{equation:Q_av}
\begin{split}
    \langle Q^{\prime}\rangle&=\frac{3}{2}\frac{1}{\langle \mathrm{Im}\{k^{2}_{2}\} \rangle},\\
    \langle \mathrm{Im}\{k^{m}_{2}\}\rangle&=\int^{2\Omega_{\rm s}}_{-2\Omega_{\rm s}}\frac{\mathrm{Im}\{k^{}_{2}(\omega_{mN})\}}{\omega_{mN}}d\omega_{mN},\\
\end{split}
\end{align}

\noindent where $\mathrm{Im}\{k_{2}(\omega_{mN})\}{=}\mathrm{sgn}(\omega_{mN})3/(2Q^{\prime}(\omega_{mN}))$ is the imaginary component of the Love number, i.e. the ratio of the induced gravitational potential to that imposed for a given component. The integral spans the range over which IWs propagate ($\omega_{mN}{\in}[-2,2]\Omega_{\rm s}$). We compute the integral using $\texttt{scipy}$'s implicit $\texttt{BDF}$ integrator with relative and absolute tolerances of $10^{-12}$ \citep{virgomoli20}. We use the level of agreement between the independent calculations leading to $\bar{Q^{\prime}}$ and $\langle Q^{\prime}\rangle$ as an important validation of our codes.

\subsection{Viscosity}\label{sec:viscosity}

The viscosity associated with damping of tidal IWs is poorly understood. Owing to this uncertainty, we explore a wide range of values (specified below), and for simplicity, assume the viscosity is spatially constant. 

\begin{figure*}[t]
\centering
\includegraphics[width=1\textwidth]{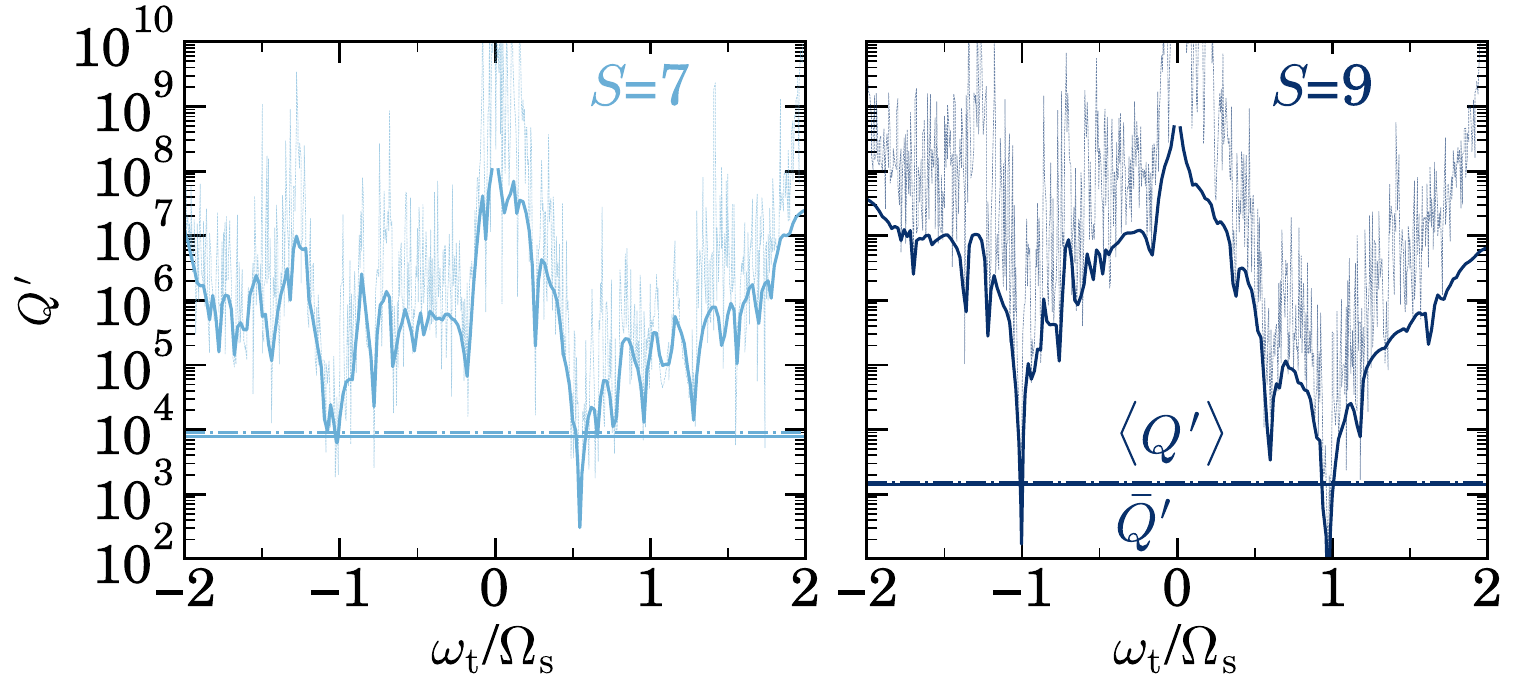}
\caption{Tidal quality factor $Q^{\prime}$ of a Neptunian planet due to inertial wave dissipation, as a function of tidal to spin frequency ratio $\omega_{\rm t}/\Omega_{\rm s}$. This example planet has mass $M_{\rm p}{=}30 \ M_{\oplus}$, core mass $M_{\rm c}{=}20 \ M_{\oplus}$, equilibrium temperature $T_{\rm eq}{=}509$ K (roughly the equilibrium temperature at 0.3 au around a Sun-like star), and is assumed to spin with a period of 1 day. The left and right panels employ planet entropies of $S{=}7$ and $S{=}9 \ k_{\rm B}/m_{\rm H}$, respectively. Solid and dashed lines span a range of astrophysically plausible viscosities (parameterized via the Ekman number $Ek$), with solid employing $Ek{=}10^{-5}$ and dashed using $Ek{=}10^{-10}$. Solid and dot-dashed horizontal lines are quality factors corresponding to frequency-averaged dissipation; solid are obtained by averaging our $Q^{\prime}(\omega_{\rm t}/\Omega_{\rm s})$ over frequency (denoted ``$\langle Q^{\prime}\rangle$"; we integrate over the $Ek{=}10^{-5}$ curve), whereas dot-dashed are independently computed following \cite{ogi13} (labeled ``$\bar{Q}^{\prime}$"). The two independent frequency-averaged dissipation rates are in reasonable agreement, validating our computations. With weaker viscosities to smooth out features in the flow, low viscosity planets exhibit fine structure in their dissipation rate. Resonant peaks in dissipation (troughs in $Q^{\prime}$; e.g., near $\omega_{\rm t}{\sim}{\pm}\Omega_{\rm s}$ in right panel) are produced regardless of viscosity, however, aiding planet migration.
\label{figure:compareEk}}
\end{figure*}   

To illustrate the spread in possible values of $\nu$, we adopt as an example mechanism inertial wave damping via turbulent convection. Under this hypothesis, interaction with convective eddies damps IW energy, akin to turbulent dissipation of the equilibrium tide \citep[e.g.][]{zah66,zah08}. Turbulent dissipation may exhibit strong rotational corrections \citep[][Duguid et al. in review]{BDL2014,mataucgue16,Nils2023} and tidal frequency dependence \citep[e.g.][see also \citealt{vidbar20,dugbarjon20,zholai26} for recent developments]{zah89,golkee77,golnic77}. We provide three estimates of this viscosity to flesh out the range in possible values. 

First, the (non-rotating) convective viscosity given by mixing length theory (MLT) yields a typical value $\nu_{\rm MLT}{\sim}\ell_{\rm MLT}v_{\rm c}{\sim}10^{-5}R^{2}_{\rm p}\Omega_{\rm s}$, with $\ell_{\rm MLT}$ the mixing length (taken to be the scale height) and $v_{\rm c}$ the convective velocity \cite[computed as in Appendix A of][]{halmil26}. Second, we estimate the suppression of $\nu_{\rm MLT}$ due to rapid rotation following \cite{Nils2023}. Since our planet models exhibit convective turnover times much longer than their rotation periods, the shorter convective length scales and velocities under rotating mixing length theory (RMLT) further decrease viscosity via $\nu_{\rm RMLT}{\sim}Ro^{4/5}\nu_{\rm MLT}{\sim}10^{-8}R^{2}_{\rm p}\Omega_{\rm s}$, with $Ro{=}v_{\rm c}/(2 \ell_{\rm MLT}\Omega_{\rm s})$ the convective Rossby number \citep[see Equation 50 of][]{Nils2023}. Third, we estimate the suppression of $\nu_{\rm MLT}$ in the ``fast tides" regime, wherein the dominant convective eddies may only traverse a fraction of their mixing length over a tidal period \citep[i.e., the tidal frequency exceeds the turnover frequency, $|\omega_{mN}|{\gg}\omega_{\rm c}{=}v_{\rm c}/\ell_{\rm c}$;][]{zah66,golnic77}. Adopting a scaling $\nu_{\rm fast}{\propto}(\omega_{\rm c}/\omega_{mN})^{2}$ \citep[originally proposed by][though this is supported by the latest direct numerical simulations of e.g.~\citealt{dugbarjon20,vidbar20}, even if the physical mechanism responsible is likely to be different]{golnic77} and assuming rotation, orbit, and convective turnover periods of 1, 60, and 365 days, respectively, such a suppressed viscosity reaches ${\nu}_{\rm fast}{\sim}10^{-6}\nu_{\rm MLT}$, yielding a ${\sim}6$ order of magnitude spread in possible viscosity values. These choices translate to Ekman numbers $Ek{\equiv}\nu/(R^{2}\Omega){\sim}10^{-5}{-}10^{-11}$. As a result, we use two values, $Ek{\in}\{10^{-5},10^{-10}\}$ to parameterize the viscosity in our calculations (converged calculations at $Ek{=}10^{-11}$ would be more computationally demanding and are not expected to produce qualitatively different results).

To further contextualize our choices of viscosity, we compare against the molecular viscosity obtained in our structure models. The molecular viscosity yields a range in Ekman number $Ek{\sim}10^{-12}{-}10^{-17}$ across the atmosphere, significantly below that obtained from turbulent convection in the ``fast-tides'' regime. Our adopted lower bound of $Ek{=}10^{-10}$ therefore may still be an overestimate, if convection does not efficiently damp IWs. We find, however, that our results are not sensitive to the choice of $Ek$ (when $Ek$ is sufficiently small; only $Ek{\lesssim}10^{-5}$ is explored in this paper), as different choices of $Ek$ produce the same resonant peaks in dissipation that dominate evolution. Future work should confirm whether this conclusion holds if only molecular viscosity is used ($Ek{\sim}10^{-17}$). 

\subsection{Numerics}\label{sec:numerics}

We solve Equations \ref{equation:anelastic}--\ref{equation:NS2} for the linear tidal response, and then calculate a dissipation rate using Equation \ref{equation:dissipation}. We do this for a chosen tidal frequency ($\omega_{mN}$), Ekman number ($Ek$), planet entropy ($S$), and core/envelope mass. Our tidal dissipation spectra fix the planets' equilibrium temperature to $T_{\rm eq}{=}$509 K, corresponding to their initial condition in our evolution calculations ($a{=}0.3$ au around a Sun-like star). This choice of evaluating dissipation spectra at a fixed $T_{\rm eq}$ is motivated by the fact that $T_{\rm eq}$ has a subdominant effect on $Q^{\prime}$, since $T_{\rm eq}$ only weakly alters planets' convective zone structures compared to other parameters we vary (core/envelope mass, entropy, $Ek$).\footnote{As an explicit example, $Q^{\prime}$ for a $300 \ M_{\oplus}$ gas giant with a 10 $M_{\oplus}$ core and entropy $S{=}7 \ k_{\rm B}/m_{\rm H}$ at $\omega_{\rm t}/\Omega_{\rm s}{=}0.549$ changes by ${\sim}6\%$ across equilibrium temperatures 509-160 K.} We will indeed show that these other parameters change $Q^{\prime}$ by orders of magnitude. Moreover, $T_{\rm eq}$ does not evolve strongly during most of our evolution calculations, since runaway inflation ensues before planets' orbits shrink. Our structure models supply $\rho(r)$ and $g(r)$ as well as the core and RCB radii, respectively. 

Solutions are discretized in radius using a Chebyshev collocation method on $N_{\rm ch}$ Chebyshev-Gauss-Lobatto nodes \citep{preflateu86}. The latitudinal dependence is expanded in spherical harmonics and treated using a Galerkin method, with fixed azimuthal wavenumber $m_{\rm lat}{=}2$ and spherical harmonic degrees $n_{\rm lat}{\in}[2,N_{\rm lat}]$. This spectral method converts the problem into a large sparse system of algebraic equations, which we solve using the \texttt{MUMPS} parallel sparse direct solver through the \texttt{python-mumps} interface\footnote{\href{https://gitlab.kwant-project.org/kwant/python-mumps}{Available here.}}. We find convergence within ${\lesssim}0.1\%$ using a default resolution of $N_{\rm lat}{=}N_{\rm ch}{=}120$ (i.e. $|D_{\nu}{-}P_{\rm t}|/P_{\rm t}{\lesssim}0.1\%$), except for planets with low viscosities ($Ek{\lesssim}10^{-9}$) that require higher resolution. In these higher resolution cases we employ $N_{\rm lat}{=}N_{\rm ch}{=}520$ to produce convergence\footnote{Some results were obtained using an independent code employing a weak Chebyshev Galerkin method in radius, which also adopted poloidal and toroidal scalars to reduce the number of variables. After benchmarking both codes, this was used to compute some of the most demanding cases with $S{\in}[7,8]$ with $Ek{=}10^{-10}$. This code used a typical resolution of $N_{\rm ch}{=}600$ and $N_{\rm lat}{=}800$ for these cases (found to be sufficient by comparing to calculations at even higher resolution at selected frequencies near dissipative peaks), with a frequency interval of $0.001$.} within ${\lesssim}0.1\%$. Larger grid resolutions are memory-prohibitive, requiring ${\gtrsim}400$ GB of RAM during linear algebra solves.

To compute a spectrum of $Q^{\prime}$ in a given model, we step through frequency intervals $\Delta\omega_{mN}{=}0.02\Omega_{\rm s}$ for $Ek{=}10^{-5}$, and $\Delta\omega_{mN}{\in}\{0.001,0.005\}\Omega_{\rm s}$ for $Ek{=}10^{-10}$. We add finer frequency resolution (steps of $\Delta\omega_{mN}{=}0.001\Omega_{\rm s}$) in the interval ${\pm}0.01\Omega_{\rm s}$ surrounding resonant troughs to capture the minimum $Q^{\prime}$, which we find to be necessary to match frequency-averaged predictions ($\bar{Q^{\prime}}$). These sampling intervals are fine enough to accurately resolve peaks and troughs in the dissipation rate, while minimizing computing time. Computing $Q^{\prime}(\omega_{mN})$ with $\omega_{mN}{\in}[{-}2,2]\Omega_{\rm s}$ at high resolution ($N_{\rm lat}{=}N_{\rm ch}{=}520$) takes ${\sim}24$ hours after parallelizing the linear algebra across 4 CPU cores with $\texttt{python-mumps}$.

\subsection{Eigenmode Computation}\label{subsec:eigenmode}

To help interpret the frequency dependence of $Q^{\prime}(\omega_{mN})$, we also compute the spectrum of free (unforced) inertial eigenmodes in each of our planets. Following a similar approach to \cite{ponbarhol24}, we omit tidal forcing and recast our equations as an eigenvalue problem for the (complex) frequency $\omega_{\rm mode}$. The solution provides the eigenvectors (e.g., velocity field $\mathbf{u}_{\rm mode}$), mode frequency ($\rm Re[\omega_{\rm mode}]$), and mode damping rate ($\rm Im[\omega_{\rm mode}]$). We focus on eigenmodes that share the $\ell{=}m{=}2$ equatorial symmetry of the tidal forcing, as they are the only modes that can be linearly excited.

\section{Inertial Wave Dissipation as a Function of Planet Structure}\label{sec:results1}

Our goal in this section is to overview how IW dissipation varies with frequency, planet entropy, atmospheric pressure level, viscosity, core mass, and total mass. We first characterize the frequency dependence.

\subsection{Frequency}\label{subsec:freq}

Figure \ref{figure:compareEk} provides an illustration of how frequency, entropy, and viscosity impact the dissipative IW response. As expected, IW dissipation exhibits erratic frequency dependence \citep[e.g.][]{ogilin04,ogi09}. The nature of this frequency dependence is not fully understood, although one aspect that is thought to be relevant is that IWs' local group propagation direction (and therefore global structure) relative to the spin axis depends on frequency via $\Theta{=}\arcsin|\omega_{mN}/2\Omega_{\rm s}|$ \citep[e.g.][]{riegeoval01}; as the frequency evolves, so too does the overlap between the IW and tidal forcing (as elaborated further below).

Our frequency averaged value accurately agrees with the value predicted by the impulsive formalism of 
\cite{ogi13}, but only if the minima in $Q^{\prime}$ are finely resolved in frequency; the frequency-averaged response is sensitive to resonant troughs (dissipative peaks). Troughs in $Q^{\prime}$ at $\omega_{\rm t}/\Omega_{\rm s}{\sim}{\pm}1$ are evident in both low and high entropy planets, but shift deeper and toward larger frequencies as entropy grows. The remainder of this subsection elucidates the frequency dependence of $Q^{\prime}$ by determining what sets the \textit{frequency} of resonant troughs, before exploring what sets the \textit{strength} of tidal dissipation at such troughs (i.e., why are some peaks in dissipation larger than others?).

\setcounter{figure}{2}
\begin{figure*}[htbp]
        \centering
        \includegraphics[width=0.495\linewidth]{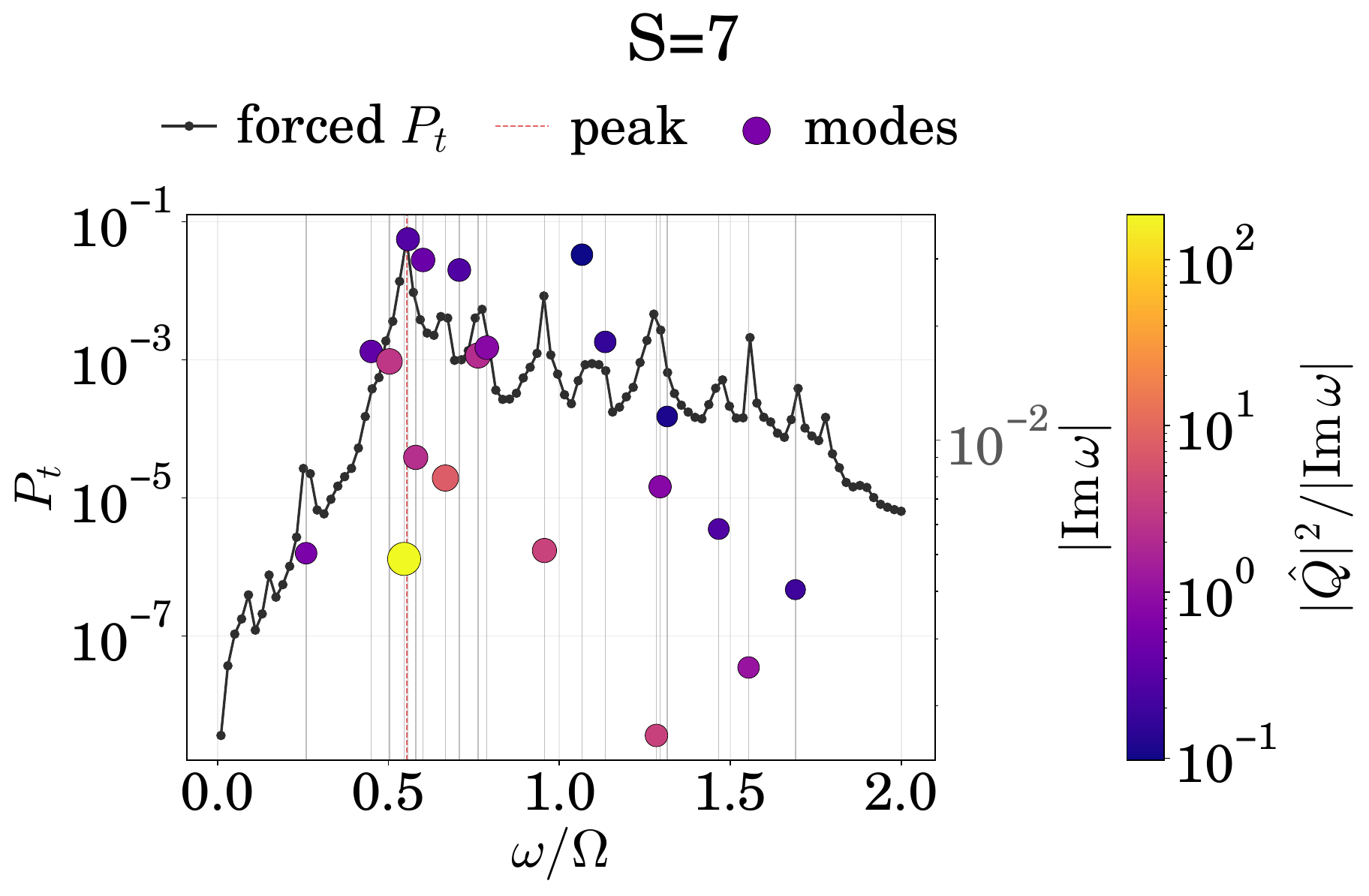}
        \centering
         \includegraphics[width=0.495\linewidth]{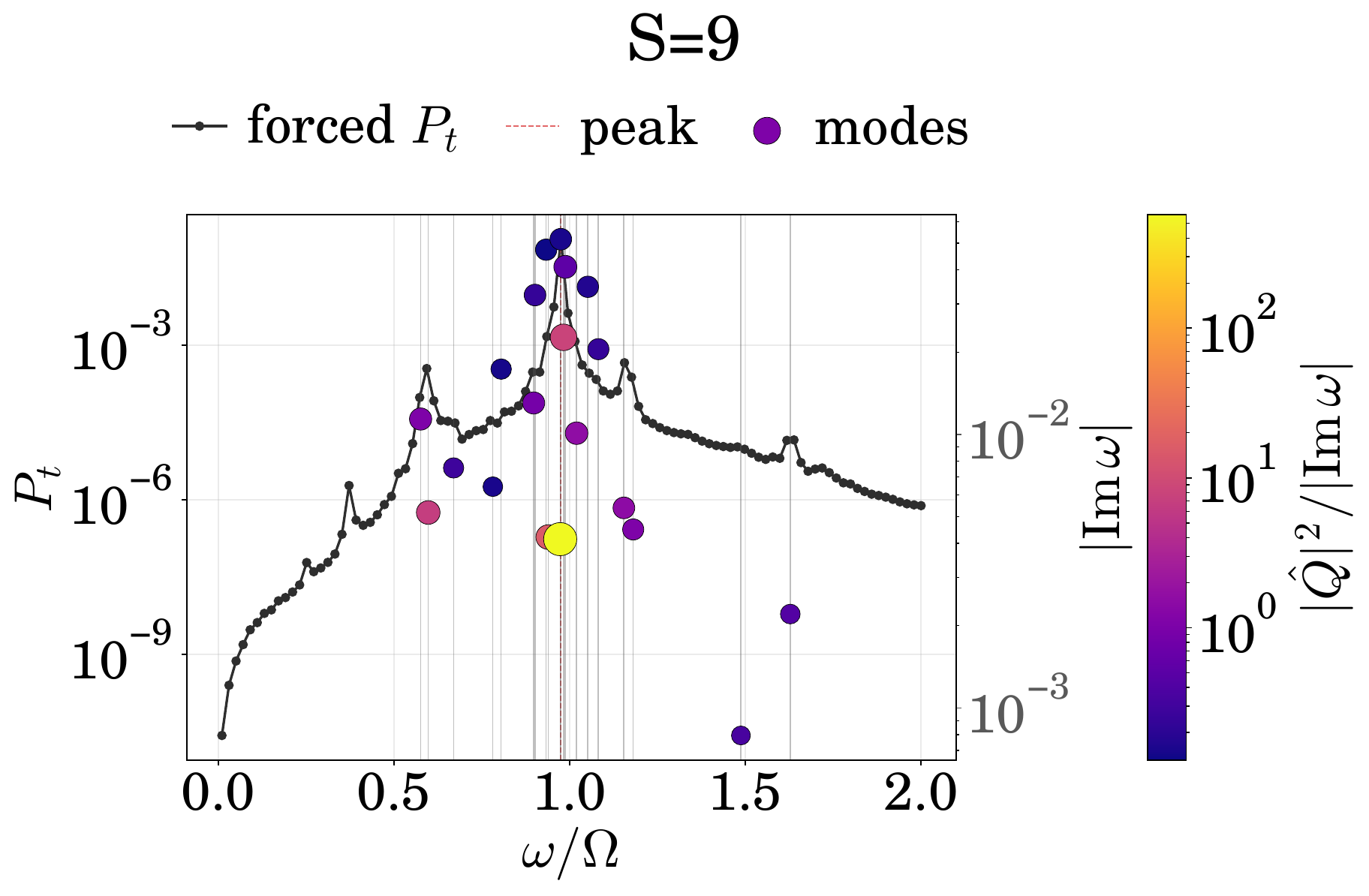}
        \caption{Exploring the frequency and magnitude of inertial wave dissipation rates. Both panels display dimensionless tidal power $P_{\rm t}$ of forced inertial waves (measured in units of the characteristic inertial wave dissipation rate $\bar{\rho}\Omega^{3}_{\rm s}R^{5}_{\rm p}$, in black curves/points; left $y$-axes), damping rates of free inertial eigenmodes (circled data points, right $y$-axes, $|\mathrm{Im}[\omega]|$; we only plot 20 least-damped modes for visual clarity), and a measure of the spatial overlap between these free eigenmodes and the tidal forcing (color bar/color of circles; normalized by the damping rate) as a function of frequency. The spatial overlap is computed as $\int \mathbf{f}_{\rm tid}{\cdot}\mathbf{u}_{\rm mode} dV$, quantifying the degree to which $\ell{=}m{=}2$ tidal forcing aligns with the structure of the mode velocity. The left and right panels use the low and high entropy planets from Figure \ref{figure:compareEk}, respectively. Peaks in dissipation (minima in $Q^{\prime}$) occur at tidal frequencies resonant with the least-damped eigenmodes. The relative magnitude of the dissipation partly reflects the degree of spatial overlap between eigenmodes and the tidal forcing. \label{figure:eigenmodes}}
\end{figure*}

\setcounter{figure}{3}
\begin{figure*}[htbp]
    \centering
    \includegraphics[width=\linewidth]{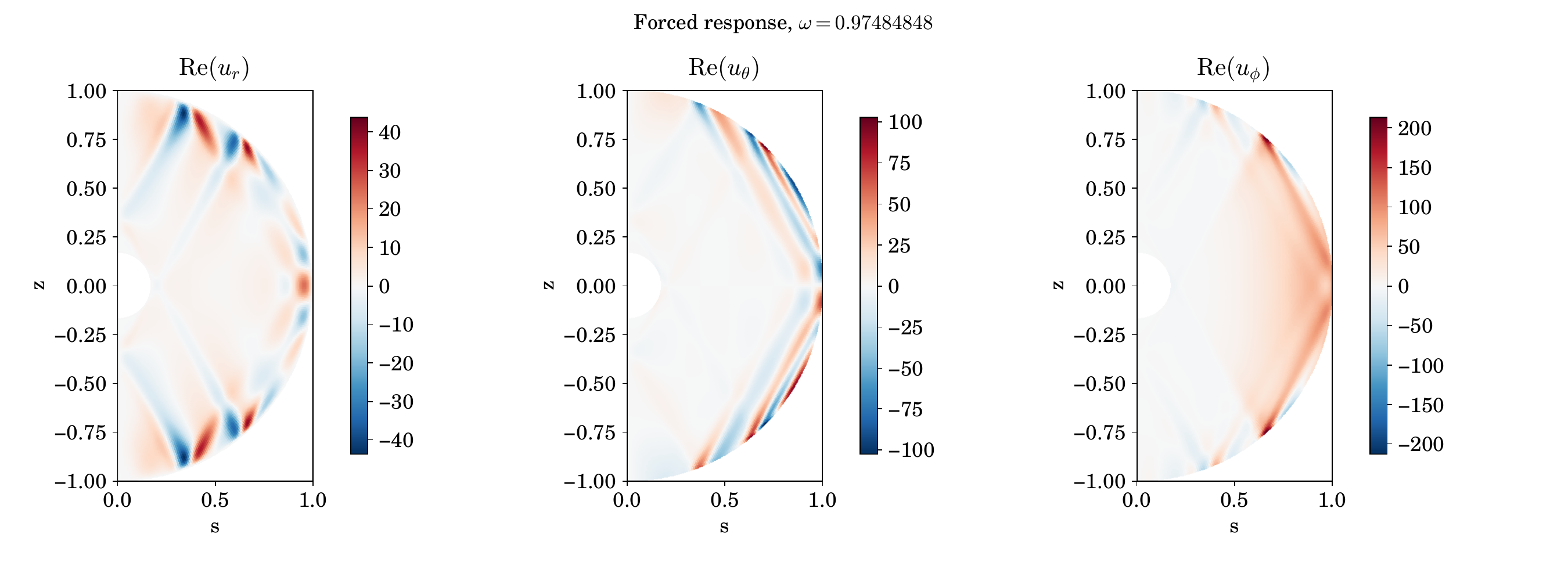}
    
    \vspace{-1em} 
    
    \includegraphics[width=\linewidth]{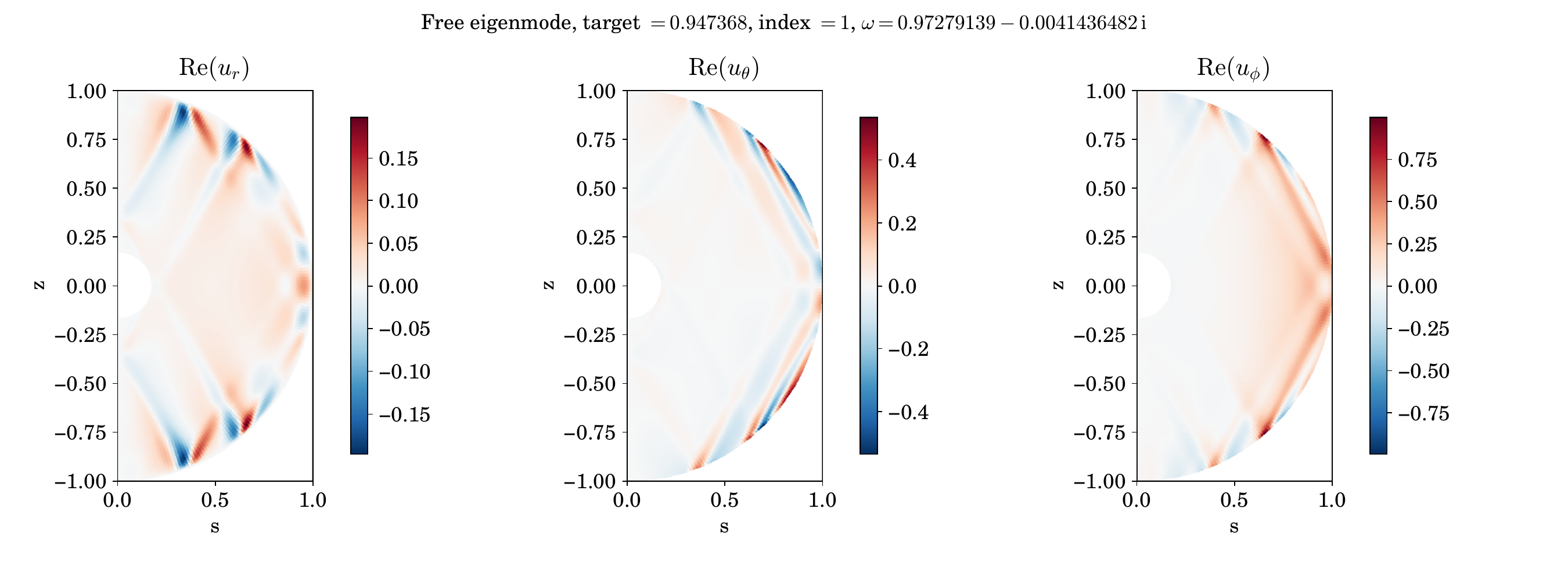}
    \caption{Visual comparison of the velocity fields of tidally-excited inertial waves (top panels; left to right are radial, latitudinal, and azimuthal velocities) and the closest free inertial eigenmode (bottom panels), for a frequency $\omega_{\rm t}/\Omega_{\rm s}{\sim}1$ for the same high-entropy planet as Figure \ref{figure:compareEk} ($S{=}9$). The resonant trough in $Q^{\prime}$ (dissipative peak) at this frequency, $\omega_{\rm t}/\Omega_{\rm s}{\sim}1$ (see Figure \ref{figure:compareEk}), thus reflects efficient tidal excitation of this free inertial eigenmode. \label{figure:velocity_fields}}
\end{figure*}

To help understand the frequency dependence of $Q^{\prime}$, we have separately computed the spectrum of free inertial eigenmodes for the planets in Figure \ref{figure:compareEk}, following the procedure described in Section \ref{subsec:eigenmode}. As displayed in Figure \ref{figure:eigenmodes}, tidal power $P_{\rm t}$ is maximized ($Q^{\prime}$ minimized) at the same frequencies that produce least-damped inertial eigenmodes (we only plot the 20 least-damped modes, for readability). In other words: maxima in dissipation occur at tidal frequencies that are resonant with weakly-damped eigenmodes. Figure \ref{figure:velocity_fields} visually confirms that the velocity field of a tidally-excited IW at the most dissipative frequency peak clearly reflects that of a free inertial eigenmode at a very similar frequency.

Having determined that the troughs in $Q^{\prime}$ arise due to resonances with (nearby least-damped) free inertial eigenmodes, we now consider the relative magnitude of the dissipation at these frequencies.
We find that the amount of tidal power dissipated at a given resonance reflects the spatial overlap between the tidal forcing and the corresponding mode -- i.e., the degree to which the $\ell{=}m{=}2$ tidal forcing aligns with the mode's velocity field. The color bar in Figure \ref{figure:eigenmodes} displays a measure of this spatial overlap normalized by the mode damping rates\footnote{This measure is intentionally crude, and should not be considered as a reliable prediction of the modal response amplitude. The forced IW problem is non-self-adjoint, so the forcing projects onto the corresponding adjoint (left) eigenvector -- which we have not explicitly computed -- rather than directly onto the right eigenvector (as assumed here). Moreover, choosing  $|\mathrm{Im}[\omega_{\rm mode}]|$ as the denominator is appropriate only sufficiently close to an isolated resonance. This approximate measure was chosen for simplicity, but it demonstrates its usefulness in Figure \ref{figure:eigenmodes}.} between the effective tidal forcing (given in Equation \ref{equation:f_IW}) and a given eigenmode, computed as $\int\mathbf{u}_{\rm mode}{\cdot}\mathbf{f}_{\rm w}dV/\mathrm{Im}[\omega_{\rm mode}]$. We observe that the height of each local maximum in $P_{\rm t}$ indeed correlates strongly with the spatial overlap of the nearest eigenmode. As an example, the lower entropy planet's global maximum in dissipation at $\omega_{\rm t}/\Omega_{\rm s}{\sim}0.546$ (see Figure \ref{figure:compareEk}) lies very close in frequency to the mode with the highest overlap in Figure \ref{figure:eigenmodes}. Similarly, the resonant feature at $\omega_{\rm t}/\Omega_{\rm s}{\sim}1.14$ in the high entropy planet lies near multiple high-overlap eigenmodes, with one mode in particular exhibiting the highest degree of overlap. The denser spectrum of modes that overlap with tidal forcing may be responsible for producing a broader trough in $Q^{\prime}$ than for the lower entropy planet.

To summarize: troughs in $Q^{\prime}(\omega_{\rm t})$ arise due to resonance between tidal forcing and free inertial eigenmodes. The amount of tidal power injected at each resonance is correlated with the degree of spatial overlap between the tidal forcing and the mode, normalized by the mode damping rate.

\subsection{Entropy}

\begin{figure}
\epsscale{1.3}
\plotone{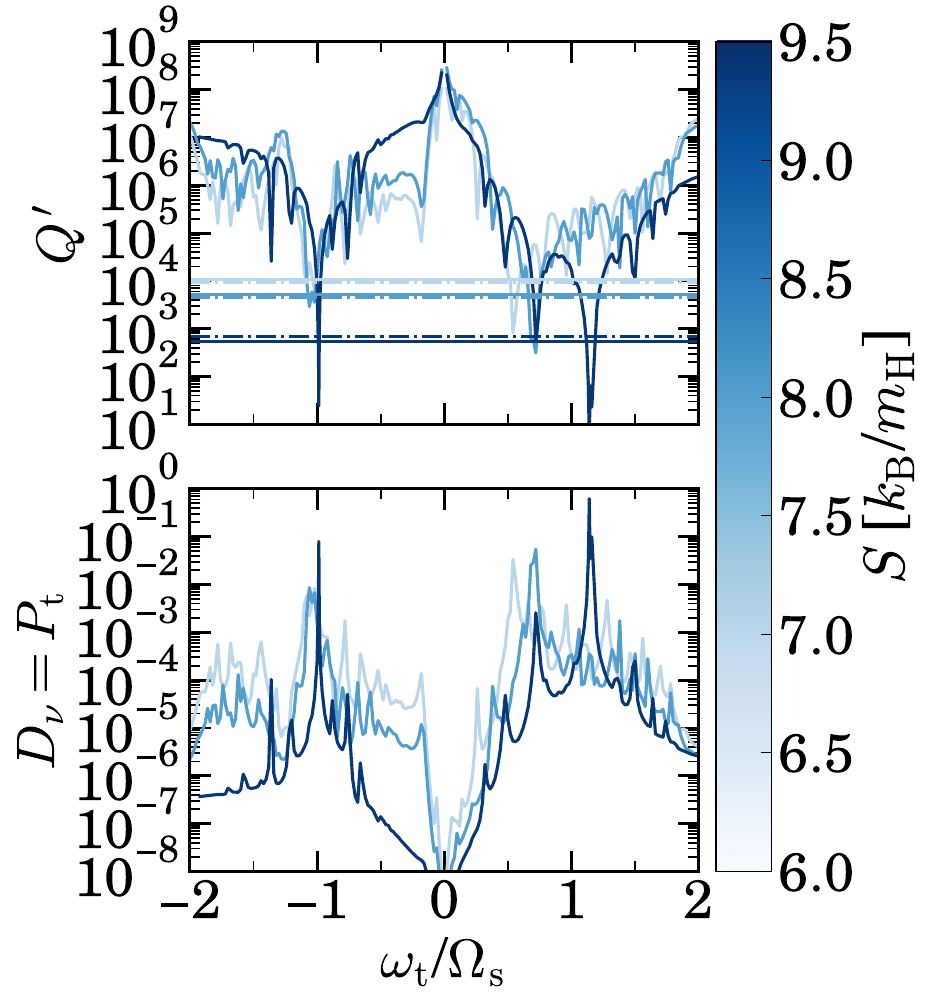}
\caption{Top panel: similar to Figure \ref{figure:compareEk}, except we vary the planet entropy in $S{\in}\{7,8,9.4\} \ k_{\rm B}/m_{\rm H}$ following the color bar on the right, but keep the Ekman number fixed ($Ek{=}10^{-5}$). Frequency-averaged quality factors decline with increasing $S$, producing more dissipation. Bottom panel: dimensionless tidal power (i.e. the energy injection rate into the tidal wave) for the same planet models as in the top panel. On average, dimensionless tidal power is $\textit{lower}$ in higher entropy planets (with the exception of a strong resonance at ${\omega_{\rm t}/\Omega_{\rm s}}{\sim}1$). The decrease in quality factor toward higher entropies is driven by planets' longer dynamical times; high-entropy planets' weaker self-gravity seeds a stronger IW response. \label{figure:compareS}}
\end{figure}

The upper panel of Figure \ref{figure:compareS} provides a closer look at the effect of entropy on $Q^{\prime}$ (at fixed rotation rate). We find that higher entropy planets exhibit smaller frequency-averaged $\langle Q^{\prime}\rangle$, but larger $Q^{\prime}(\omega_{\rm t})$ across much of the spectrum. Our goal here is to isolate why $\langle Q^{\prime}\rangle$ drops so strongly with entropy. We will show that the dependence of $\langle Q^{\prime} \rangle$ on planet entropy stems from both increased tidal power at resonant frequencies, and the weaker self-gravity of puffier planets. 

\begin{figure}
\epsscale{1.3}
\plotone{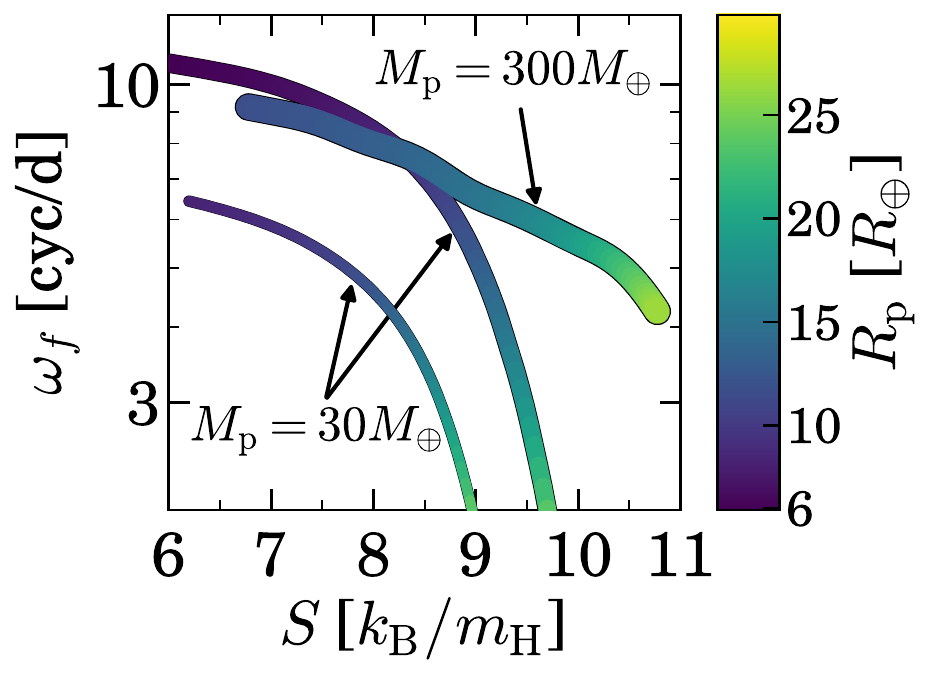}
\caption{Frequency of the $\ell{=}2$ fundamental mode (``$f$-mode", $\omega_{f}$), as a function of planet entropy ($x$ axis).
We plot $\omega_{f}(S)$ for two Neptunes (total mass $M_{\rm p}{=}30 \ M_{\oplus}$ with 10 and 20 $M_{\oplus}$ cores in thin and thicker lines, respectively), and one giant planet (300 $M_{\oplus}$ with a 10 $M_{\oplus}$ core). Each curve is color-coded following planetary radii at each entropy. The mode frequency is computed using $\texttt{GYRE}$ \citep[assuming adiabatic modes with radial node number $n_{\rm pg}{=}0$, and neglecting rotation;][]{towtei13}. Our $\texttt{GYRE}$ calculations set the radial displacement to zero at the bottom boundary/core surface (\texttt{inner}$\_$\texttt{bound=}\texttt{ZERO}$\_$\texttt{R}), while the top boundary is set by matching to the solution in an isothermal atmosphere extending above the photosphere \citep[\texttt{outer}$\_$\texttt{bound=}\texttt{JCD}; following][]{chr08}. As planetary radii (entropies) grow, the frequency of the planets' fundamental mode declines. \label{figure:om_dyn}}
\end{figure}

To determine the dominant contribution to variations in $Q^{\prime}$ with entropy, we first note that $Q^{\prime}(\omega_{\rm t}{,}S){\propto}\omega^{2}_{\rm dyn}(S)/D_{\nu}(\omega_{\rm t}{,}S)$. Motivated by this scaling, we plot in Figure \ref{figure:compareS} (lower panel) dimensionless tidal power $D_{\nu}$ (${=}P_{\rm t}$) as a function of frequency, along with the planets' radii/dynamical frequencies as a function of $S$ in Figure \ref{figure:om_dyn}. Figure \ref{figure:compareS} highlights that, for most frequencies, tidal power is $\textit{smaller}$ in the high entropy planet across much of the spectrum. The exception occurs at $\omega_{\rm t}/\Omega_{\rm s}{\sim}{\pm}1$, where strong resonance with an inertial eigenmodes boosts $D_{\nu}$. This behavior in $D_{\nu}$ is qualitatively reflected in $Q^{\prime}$; only resonances at $\omega_{\rm t}/\Omega_{\rm s}{\sim}{\pm}1$ drive $Q^{\prime}$ below that of lower entropies. The drop in $Q^{\prime}$ at these resonances is more pronounced than the linear relationship produced by $D_{\nu}$ alone, $Q^{\prime}{\propto}1/D_{\nu}$, however.\footnote{As an example, $Q^{\prime}$ at $\omega_{\rm t}/\Omega_{\rm s}{\sim}{-}1$ for $S{=}9.4$ is ${\sim}0.01{\times}$ that of $S{=}8$, whereas $D_{\nu}$ is only a factor ${\sim}8{\times}$ larger.} This exacerbated decline in $Q^{\prime}$ points to planets' shifting dynamical frequencies also contributing to $Q^{\prime}$; spatial overlap of inertial eigenmodes with tidal forcing (which sets $D_{\nu}{=}P_{\rm t}$; Section \ref{subsec:freq}) does not fully explain the increased frequency-averaged dissipation in high entropy planets. 

Figure \ref{figure:om_dyn}, on the other hand, confirms that high entropy planets' shorter dynamical frequencies do indeed contribute to their smaller $\langle Q^{\prime}\rangle$; given $Q^{\prime}{\propto}\omega^{2}_{\rm dyn}{\propto}R^{-3}_{\rm p}$, the $S{=}9.4$ planet's shorter dynamical frequency produces ${\sim}(24 \ R_{\oplus}/9 \ R_{\oplus})^{-3}{\sim}19{\times}$ smaller $Q^{\prime}$ than for the $S{=}7$ planet, in approximate agreement with the frequency-averaged result in Figure \ref{figure:compareS}. Identical reasoning explains the differences in $Q^{\prime}$ between models of different entropies in Figure \ref{figure:compareS_mc=10}, for planets with 10 $M_{\oplus}$ cores. 

In summary, $\langle Q^{\prime}\rangle$ declines in puffier (higher entropy) planets due to two combined effects. First, more tidal power $P_{\rm t}$ is injected into the planets' resonant eigenmodes (i.e.~their spatial overlap with tidal forcing is larger). Second, puffy planets' weaker self-gravity increases the energy in IWs relative to the maximum energy stored in the equilibrium tide.

\subsection{Dependence of Heating on Atmospheric Depth}\label{subsec:deposition}

We next probe how viscous dissipation of IWs depends on depth in the atmosphere. Figure \ref{figure:deposition} reveals that IWs deposit heat across a large dynamic range in pressure. Tidal dissipation is more centrally concentrated for the lower entropy planet, with the lion's share of heating (${\gtrsim}50\%$) generated across the uppermost decade in pressure ${\gtrsim}10^{11}$ dyn/cm$^{2}$ (containing ${\sim}80\%$ of the atmosphere by mass). Such deep heating at low entropy may persist in the presence of a stratified layer at the core interface \citep[e.g.][see also Section \ref{subsec:thermal_evo}]{marguiste22}, which still permits a large, low-entropy convective zone. Future work should address whether deep heating also occurs in planets with large-scale compositional stratification (see Section \ref{subsec:thermal_evo} and \ref{subsec:fw}). The higher entropy planet on the other hand has a larger fraction of tidal heat generated in upper layers approaching the RCB, likely owing to the large velocity shear at those heights produced in its response (see Figure \ref{figure:velocity_fields}). We conclude that IWs distribute tidal heat throughout the convective zone, rather than heating occurring in localized pressure bands.\footnote{Experimenting with a non-resonant frequency $\omega_{\rm t}/\Omega_{\rm s}{=}1.6$ for our low-entropy model in Figure \ref{figure:deposition} yields a quantitatively similar heating profile as the resonant frequency used there. The pressure at which $50\%$ of the tidal heat is deposited is lower than at the resonant frequency used in Figure \ref{figure:deposition} by an order-unity factor.} However, a significant portion of the tidal heat is deposited in the deepest atmospheric layers, which can power strong planetary radius inflation (see Section \ref{sec:evo}).

\subsection{Viscosity}

As shown in Figure \ref{figure:compareEk}, low viscosity planets ($Ek=10^{-10}$) exhibit extremely fine structure in frequency, but nonetheless produce the same dominant resonant troughs as those with higher viscosity ($Ek=10^{-5}$). As we will explore in Section \ref{sec:evo}, planets' coupled structural/orbital evolution is dominated by these resonant features, so we do not expect orbital evolution to differ qualitatively for low viscosity planets. The exception is likely to occur for planets on nearly circular orbits, who sample fewer tidal harmonics that may or may not include the resonant troughs in $Q^{\prime}$.

\subsection{Core Mass}

\begin{figure}
\epsscale{1.3}
\plotone{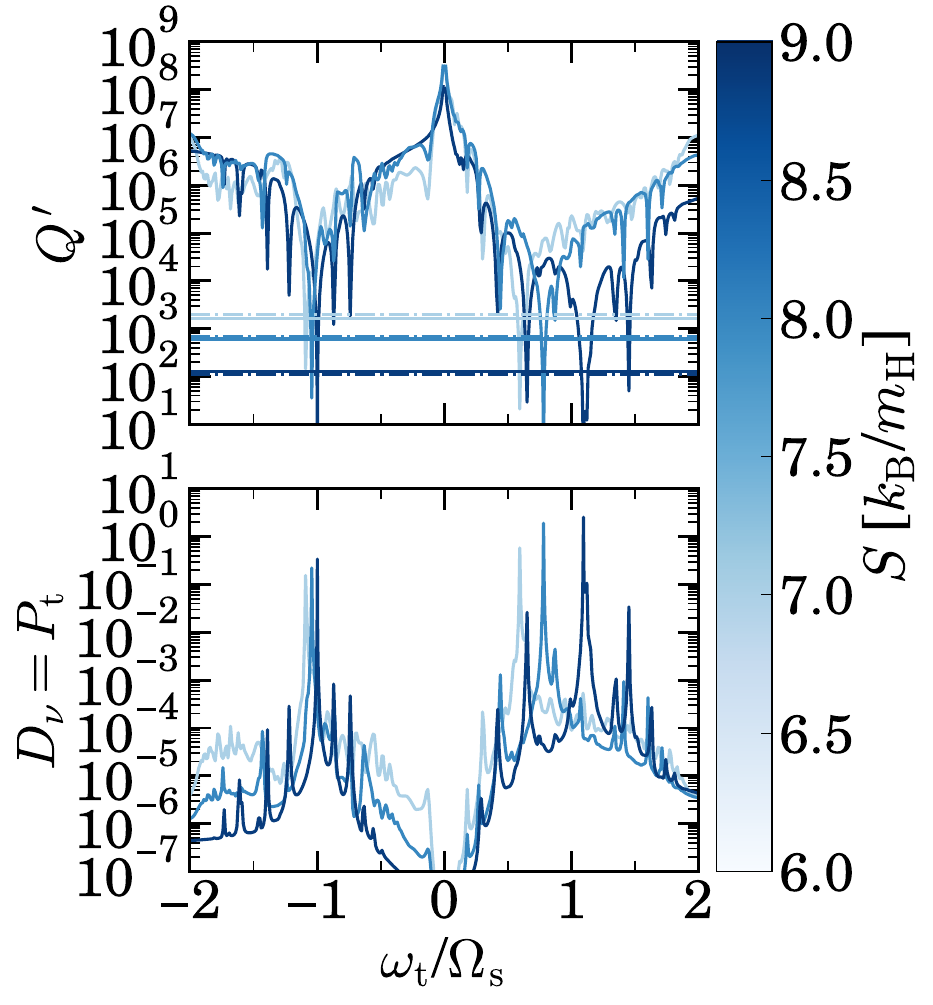}
\caption{Similar to Figure \ref{figure:compareS}, except using a core mass $m_{\rm c}{=}10 \ M_{\oplus}$ and entropies $S{\in}\{7,8,8.85\} \ k_{\rm B}/m_{\rm H}$ (light to dark blue, respectively; note the different bottom $y$ axis range from Figure \ref{figure:compareS}). General features remain the same as the $m_{\rm c}{=}20 \ M_{\oplus}$ planet from Figure \ref{figure:compareS}; $Q^{\prime}$ declines with $S$ due to resonant troughs (dissipative peaks) at tidal frequencies commensurate with planet rotation ($\omega_{\rm t}{\sim}{\pm}\Omega_{\rm s}$). Planets with smaller cores exhibit stronger dissipation at a given entropy, owing to larger spatial overlap between eigenmodes and tidal forcing. \label{figure:compareS_mc=10}}
\end{figure}

Comparing Figure \ref{figure:compareS_mc=10} and Figure \ref{figure:compareS}, we find that decreasing the core mass by a factor of two ($m_{\rm c}{=}20{\rightarrow}10 \ M_{\oplus}$) produces ${\sim}10{\times}$ lower $\langle Q^{\prime}\rangle$ at a given entropy. Figure \ref{figure:densityprofiles} displays how the ratio of core/envelope radii declines from models with large to small cores, for varying entropy; we see that the ratio of core/envelope radii at a given entropy is ${\sim}2{\times}$ lower for models with 10 $M_{\oplus}$ cores than for those with 20 $M_{\oplus}$ cores. Similar to increasing entropy, shrinking the size of the core makes $Q^{\prime}$ smaller due to greater overlap between eigenmodes and tidal forcing (which leads to larger $P_{\rm t}$ for a given $\epsilon_\Omega$), and planets' shorter dynamical frequencies. As noted in previous studies using polytropic models \citep[see e.g.~Section 4.10 of][]{ogi13} and confirmed in our Figure \ref{figure:compareS_mc=10}, increasing the radius of the atmosphere relative to the core produces dissipation that is dominated to a greater extent by resonances with global eigenmodes. 

\subsection{Total Mass}\label{subsec:total_mass}

As depicted in Figure \ref{figure:compareS_mp=300}, giant planets exhibit a smaller dynamic range in $\langle Q^{\prime}{\rangle}(S)$ than Neptunes. This lesser change in $Q^{\prime}$ with entropy reflects the fact that gas giants' structures change less significantly with entropy than Neptunes (see Figure \ref{figure:om_dyn}). Qualitative features of the $Q^{\prime}(\omega_{\rm t}/\Omega_{\rm s})$ plot are similar to the lower mass planets', but the locations and depths of troughs differ.

\subsection{Summary of Inertial Wave Dissipation vs. Planet Structure}

Figure \ref{figure:Q_av} summarizes how the \textit{frequency}-\textit{averaged} tidal quality factor varies with entropy, core mass, and total mass for our Neptunian and gas giant case studies. Table \ref{tab:Qtab} provides the tidal quality factors displayed in Figure \ref{figure:Q_av} for future studies that may wish to estimate them. Our evolution calculations described below employ the full frequency-dependent tidal quality factors, however.

\begin{deluxetable}{cccc}
\tabletypesize{\scriptsize}
\tablecaption{Frequency-averaged inertial wave tidal quality factors for our Neptunian and giant planet models with spin periods of 1 day, as a function of entropy (see Figure \ref{figure:Q_av}).
\label{tab:Qtab}}
\tablehead{
\colhead{$S$} &
\colhead{$\langle Q^{\prime}\rangle$} &
\colhead{} &
\colhead{} \\
\colhead{} &
\colhead{$\{m_{\rm c},M_{\rm p}\}{=}\{10,30\}$} &
\colhead{$\{20,30\}$} &
\colhead{$\{10,300\}$}
}
\startdata
6.0  & 4435.3 & 16295.0 & \nodata \\
6.5  & 2963.9 & 12013.1 & 9451.4 \\
7.0  & 1956.9 & 9103.9  & 8466.1 \\
7.5  & 1186.4 & 6566.3  & 7339.3 \\
8.0  & 679.0  & 4534.0  & 6337.1 \\
8.5  & 264.3  & 2721.5  & 5262.8 \\
9.0  & \nodata & 1506.3  & 4215.3 \\
9.5  & \nodata & \nodata & 3136.0 \\
10.0 & \nodata & \nodata & 1982.8 \\
10.5 & \nodata & \nodata & 974.9 \\
11.0 & \nodata & \nodata & 387.2 \\
\enddata
\tablecomments{See Figure \ref{figure:om_dyn} for planetary radii as a function of entropy.
}
\end{deluxetable}

\begin{figure}
\epsscale{1.3}
\plotone{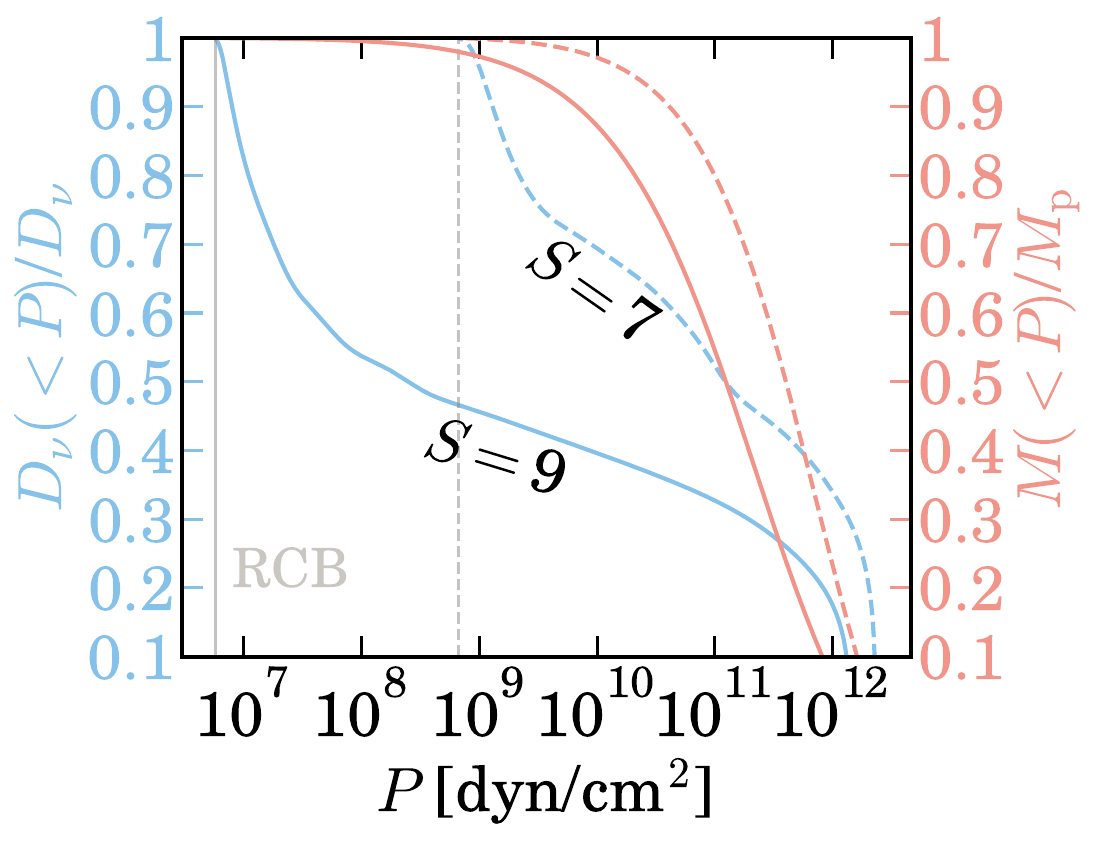}
\caption{Inertial wave dissipation profiles throughout Neptunian planets' atmospheres. Pressure ($x$-axis) runs from the core surface to radiative/convective boundary (``RCB") from right to left. Blue curves (left $y$-axis) depict the cumulative dissipation rate as a function of pressure, as a fraction of the total dissipation rate. Red curves (right $y$-axis) plot the atmosphere's cumulative mass fraction (the mass fraction is zero at the core surface). We display heating profiles for the two planets from Figure \ref{figure:compareEk} at entropies $S{=}\{7{,}9\} \ k_{\rm B}/m_{\rm H}$ in dashed and solid curves (see Figure \ref{figure:velocity_fields} for the $S{=}9$ planet's velocity field), at resonant frequencies $\omega_{\rm t}/\Omega_{\rm s}{=}\{0.546{,}0.973\}$, respectively. Inertial waves produce deep heating, aiding radius inflation. \label{figure:deposition}}
\end{figure}

\begin{figure}
\epsscale{1.3}
\plotone{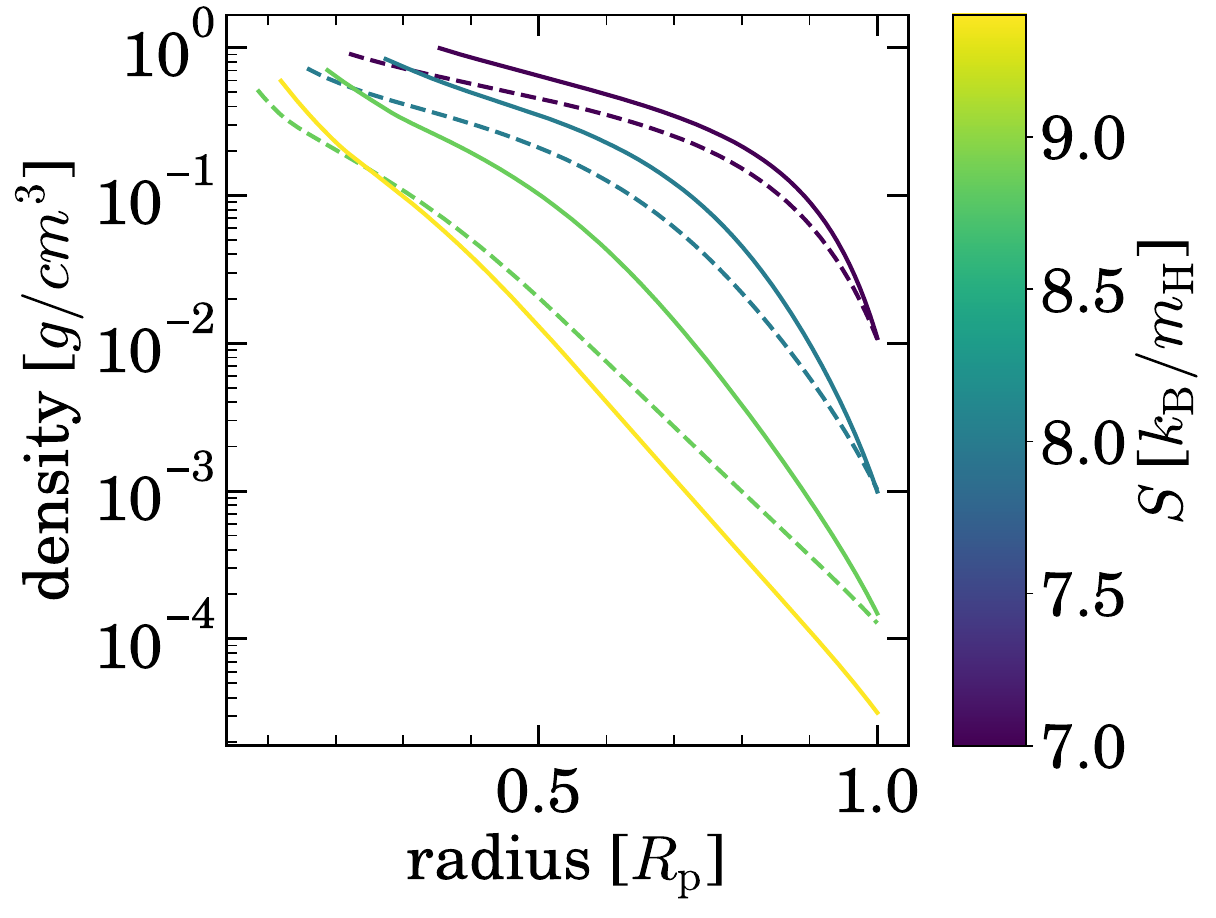}
\caption{Radial density profiles using core masses of 20 and 10 $M_{\oplus}$ in solid and dashed lines, respectively. We plot $\rho(r)$ for entropies of $S{=}\{7,8,8.85\} \ k_{\rm B}/m_{\rm H}$ following the color bar, and include $S{=}9 \ k_{\rm B}/m_{\rm H}$ for the 20 $M_{\oplus}$ core (yellow solid line). Radius along the $x$-axis is normalized to the outer radius of each model (the radiative/convective boundary). Higher entropy planets exhibit a larger dynamic range in density and envelope/core radius ratio. The tidal response of planets with structures dominated by their envelopes (high entropies, small cores) becomes more driven by global inertial eigenmodes, aiding dissipation. \label{figure:densityprofiles}}
\end{figure}

\begin{figure}
\epsscale{1.3}
\plotone{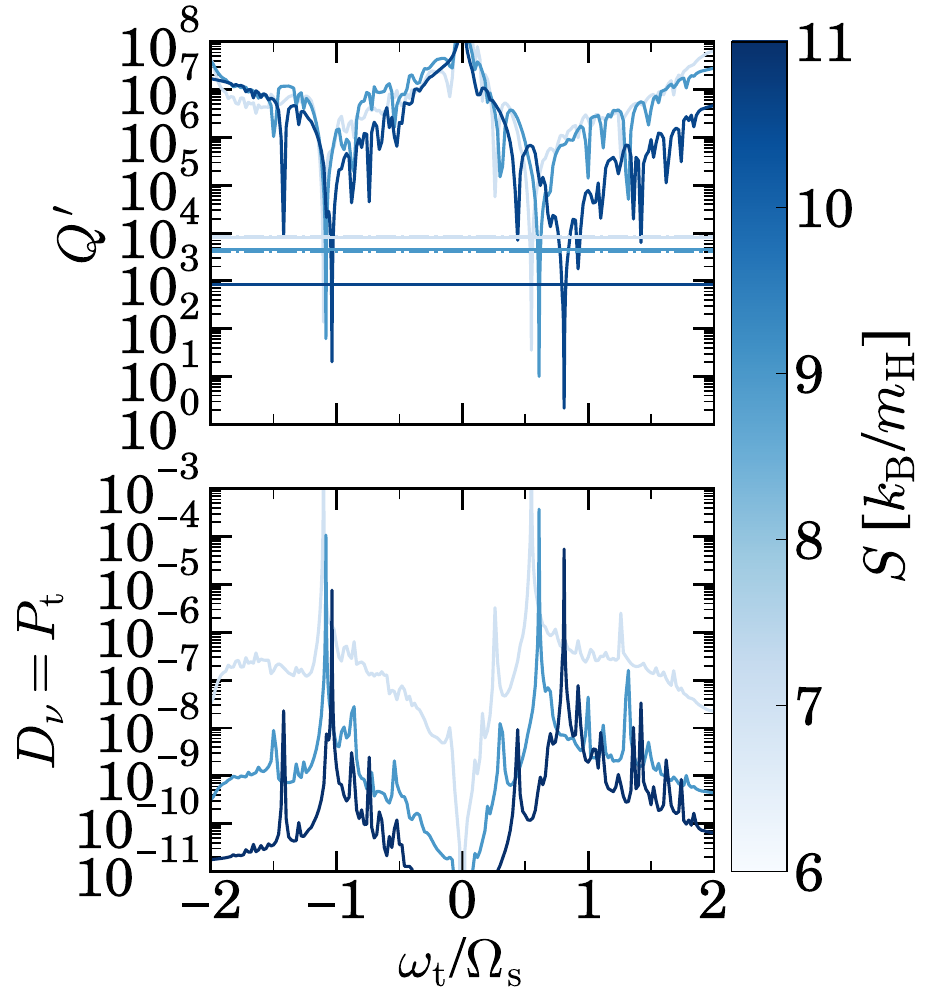}
\caption{Similar to Figure \ref{figure:compareS}, for a giant planet (300 $M_{\oplus}$) and entropies $S{\in}\{7,9,10.6\} \ k_{\rm B}/m_{\rm H}$ (light to dark blue, respectively; note the different bottom $y$-axis range from Figure \ref{figure:compareS}). Giant planets' tidal quality factors grow with entropy less strongly than Neptunes. \label{figure:compareS_mp=300}}
\end{figure}

\begin{figure}
\epsscale{1.15}
\plotone{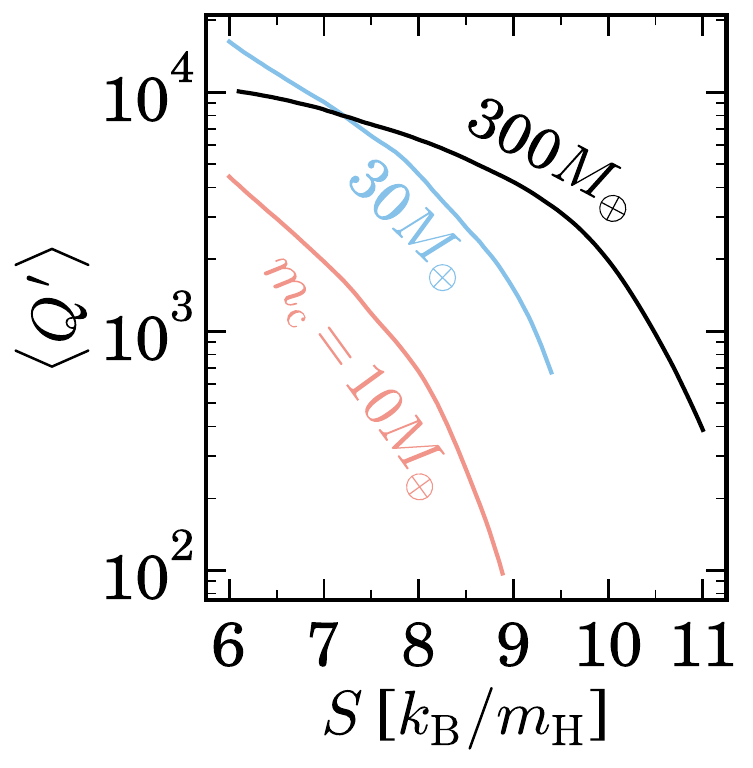}
\caption{Modified tidal quality factor for frequency-averaged dissipation of inertial waves, as a function of planet entropy $S$, for a giant planet (black solid line) and two Neptunian planets with 20 and 10 $M_{\oplus}$  cores (blue and red, respectively). Each curve assumes a planetary rotation period of 1 day. Puffy planets' (i.e.~those with high entropy) weaker self-gravity increases the fraction of tidal energy dissipated per cycle via inertial waves, potentially yielding runaway inflation. \label{figure:Q_av}}
\end{figure}

\section{Coupled Orbit, Structure, $\&$ Thermal Evolution: Methods}\label{sec:tidal_EOM}

 Our goal in this section is to explain how we follow planets' coupled thermal/structural/orbital evolution, under frequency-dependent IW dissipation. We first outline how we follow planetary thermal evolution, before providing the equations of motion under frequency-dependent tidal dissipation. Results of our integrations are overviewed in Section \ref{sec:evo}. In the following, we refer to the planet radius, ``$R_{\rm p}$'', as the photospheric radius (optical depth 2/3 to internal radiation).  This is in contrast to our tidal response calculations, where we referred to the radiative/convective boundary as the planet ``radius''. 

\subsection{Structural  $\&$ Thermal Evolution}\label{subsec:thermal_evo}

Planets' thermal evolution is computed by evolving the entropy of the convective zone as the planet heats or cools \citep[e.g.][]{hub77,arrbil06}. This ``following the adiabats" approach is valid under two conditions \citep[e.g.][]{huacum12}. First, the convective turnover time (${\lesssim}$yr) must be shorter than the time over which the planet's entropy changes ($S/|\dot{S}|{\gtrsim}10^{4}$ yr), so that the convective zone remains isentropic \citep[e.g.][]{pacsie72}. Second, the thermal timescale in the radiative zone ${\sim}Pc_{\rm p}T/(gF){\lesssim}10^{3}\rm \ yr$ \citep[with $P$ the pressure, $c_{\rm p}$ the specific heat capacity, $g$ the gravity, and $F{=}L/(4\pi R^{2}_{\rm p})$ the outgoing radiative flux; see e.g.~Figure 10 of][for estimates]{halmil26} must also be shorter than the timescale over which the entropy changes, such that the planet's luminosity is constant with radius. We confirm a posteriori that both the convective turnover time and thermal timescale in the radiative zone are indeed shorter than the timescale over which entropy evolves, even under intense tidal heating.

Under the assumptions listed above, the entropy of the convective zone evolves as:
\begin{align}\label{equation:adiabats}
\begin{split}
    \frac{dS}{dt}&=\frac{-L + L_{\rm t} + L_{\rm c}}{\int_{\rm conv} Tdm},
\end{split}
\end{align}

\noindent where $\int_{\rm conv} Tdm$ is the integral of the temperature over the mass of the convective zone, $L_{\rm t}$ is the tidal luminosity (detailed in Section \ref{sec:orbital_EOM}), and $L_{\rm c}$ reflects energy exchange with the core. Equation \ref{equation:adiabats} describes the decrease (increase) in the entropy of the convective zone as the planet's interior cools (heats) through radiative (tidal) luminosity. The thermal inertia of a rocky core also brakes changes in $S$ as energy is lost (gained) to increases (decreases) in the temperature of the core material. $L_{\rm c}$ accounts for the core's thermal inertia, i.e. the energy exchanged as its temperature evolves \citep[e.g.][]{lopformil12}:
\begin{align}\label{equation:Lc}
\begin{split}
    L_{\rm c}&=-c_{\rm V}m_{\rm c}\langle\frac{dT_{\rm c}}{dt}\rangle,
\end{split}
\end{align}

\noindent with $c_{\rm V}{=}10^{7}$ erg/K/g the specific heat of supercritical silicate \cite[i.e.~temperatures above the silicate/hydrogen critical temperature ${\sim}4000$ K;][see also \citealt{bolbra13}]{youstirog24,gilsti26}, and $T_{\rm c}$ the core/envelope interface temperature (tabulated in our planet structure grid as a function of entropy). To compute the derivative in Equation \ref{equation:Lc}, we track a running mean ($\langle{\dots}\rangle$) of $dT_{\rm c}/dt{=}(\partial\log T_{\rm c}/\partial \log S){\times}d\log S/dt$ over the previous ten time steps (of typical size $\delta t{\sim}5{\times}10^{3}$ yr), where $\partial\log T_{\rm c}/\partial \log S$ is computed and supplied from our structure grid. We have confirmed that our results are not sensitive to the particular  averaging window adopted. 

In employing Equation \ref{equation:adiabats} we have assumed that tidal heat is distributed at the base of the convective zone, such that the entire convective zone evolves as a single adiabat. Section \ref{subsec:deposition} and Figure \ref{figure:deposition} confirm that IWs $\textit{do}$ deposit a significant portion of heat in the
deepest atmosphere layers, with low-entropy planets suffering particularly deep heating. Depositing heat anywhere in the convective zone does generally produce radius inflation, so we expect our ``following the adiabats'' approach to yield a reasonable approximation to the structural evolution under IW heating.\footnote{We have performed exploratory experiments with $\texttt{MESA}$ to confirm that planetary radius inflation is not sensitive to the depth of tidal heating (assuming an initial entropy $S{\sim}7 \ k_{\rm B}/m_{\rm H}$, an extra heat of $10^{26}$ erg/s, and using a $30 \ M_{\oplus}$ model with a 20 $M_{\oplus}$ core as in Figure \ref{figure:deposition}). We find that the planet continues to inflate (albeit to a lesser degree), and a large convective zone continues to exist, when using deposition layers containing ${\sim}99\%$ of the planet mass (pressures ${\sim}10^{9}$ dyn/cm$^{2}$; the heat in our experiments is deposited across the atmospheric layers with mass fractions $0.99{-}0.995$). This radius growth stems from the increase in entropy in layers above the deposition pressure. We therefore conclude that planetary radii are likely to inflate under IW heating, even if the heating is entirely localized at shallow depths that contain the vast majority of the planet's mass.}

We also note that IW heating at low pressures ${\lesssim}10^{9}$ dyn/cm$^{2}$ could shut down convection locally, depending on the underlying luminosity, the rate of tidal heating, and the deposition pressure. Heating deposited above a given layer reduces the intrinsic luminosity that must be transported through the deeper interior, thereby reducing the radiative temperature gradient and potentially rendering the layer convectively stable. Such convective shutdown becomes progressively more difficult at greater depths, where radiative diffusion is inefficient and the radiative temperature gradient can remain larger than the adiabatic gradient even when the luminosity carried from the deeper interior is substantially reduced \citep[e.g.][]{komyou17}. Radiative window(s) opened at depth would in turn experience heating via gravity/gravito-inertial modes, and confine IWs to surrounding convective cavities \citep[][]{jertouogi17}. Future work should address to what extent the dissipation produced by IWs may alter planets' convective stability, and therefore tidal heating rate from inertial/gravity/gravito-inertial waves \citep[see e.g.][]{Lin2023,Dewberry2023,ponbarhol23,ponbarhol24,2024Dhouib}. Tracking such pressure-dependent heating profiles in addition to the frequency dependence, orbital and structural evolution explored here is a worthwhile avenue for future work (see Section \ref{subsec:fw} for a discussion of other outstanding questions).

\subsection{Orbital Equations of Motion}\label{sec:orbital_EOM}

We follow orbital evolution under eccentricity tides by summing over the planet's linear, $\ell{=}m{=}2$ IW response at each frequency. The tidal torque and energy dissipation rates for the $\ell{=}m{=}2$ component read \citep[e.g.][see also \citealt{bouefr19}]{stolai14}:
\begin{align}\label{equation:tidal_sums}
\begin{split}
    \mathcal{T}_{\parallel}&=\frac{3}{4}\mathcal{T}_{\rm 0}\sum_{N=0}^{\infty}F^{2}_{2N}(e)\bigg[\mathrm{Im}\{k_{2}(\omega_{2N})\}-\mathrm{Im}\{k_{2}(-\omega_{2N})\}\bigg],\\
    \dot{E}_{\rm t}&=\frac{3}{8}\mathcal{T}_{0}\Omega_{\rm o}\sum_{N=0}^{\infty}NF^{2}_{2N}(e)\bigg[\mathrm{Im}\{k_{2}(\omega_{2N})\}\\&\hspace{4cm}-\mathrm{Im}\{k_{2}(-\omega_{2N})\}\bigg],\\
    L_{\rm t}&=\dot{E}_{\rm t}-\Omega_{\rm s}\mathcal{T}_{\parallel},
\end{split}
\end{align}

\noindent with $\mathcal{T}_{0}{=}(\omega_{\rm dyn}\epsilon_{\rm t}R_{\rm p})^{2}M_{\rm p}$, and $F_{2N}$ the Hansen coefficients \citep[e.g.][]{hug81,murder99}. $\dot{E}_{\rm t}$ in Equations \ref{equation:tidal_sums} is the energy transfer rate from the orbit to the planet in the inertial frame, while $L_{\rm t}$ is the energy transfer rate in the planet's corotating frame (i.e.~the tidal heating rate). Equations \ref{equation:tidal_sums} follow from rewriting Equations 16 and 17 of \cite{stolai14} using $m{=}2$ and noting that $F_{2(-N)}{\sim}0$ for $N{\geq}0$ (see below). We do not include the $m{=}0$ component, but we do not expect its inclusion to change our qualitative conclusions. We adopt the approximate fitting formulae for the Hansen coefficients developed by \cite{sulai22}:\footnote{We have confirmed via direct numerical integration of the Hansen coefficients \cite[e.g.][]{hug81} that Equation \ref{equation:FN2} provides a reasonable fit for all semi-major axes/eccentricities explored in this paper.}
\begin{equation}\label{equation:FN2}
\begin{split}
F_{2N}&\approx \begin{cases} 
C_{2}N^{2}\exp(-N/\eta_{2}), & N\geq0 \\ 
0, & N<0 
\end{cases}
\end{split}\quad ,
\end{equation}

\noindent with $C^{2}_{2}{=}4f_{5}/(3(1-e^{2})^{9/2}\eta^{5}_{2})$, $\eta_{2}{=}4f_{2}/(5f_{5}(1-e^{2})^{3/2}))$, and $f_{2,5}$ the standard eccentricity expansions \citep[e.g.][which arise even though we do not adopt such a crude constant time-lag tidal model]{hut81}:
\begin{equation}\label{equation:fe}
\begin{split}
f_{2}(e^{2})&=1+\frac{15}{2}e^{2}+\frac{45}{8}e^{4}+\frac{5}{16}e^{6},\\ 
f_{5}(e^{2})&=1+3e^{2}+\frac{3}{8}e^{4}.
\end{split}
\end{equation}

We truncate Equations \ref{equation:tidal_sums} by solving for the maximum $N$ that yields tidal frequencies in the range $\omega_{\rm t}/\Omega_{\rm s}{=}(N\Omega_{\rm o}/\Omega_{\rm s}{-}m){\in}[-2,2]$ (such that IWs propagate). This maximum $N$ is $N_{+2}{\sim}4P_{\rm o}/P_{\rm s}$, which ranges from $N_{+2}{\sim}1000$ for wide orbits ($a{\sim}$au) to $N_{+2}{\sim}20$ for close-in ($a{\sim}0.1$ au). We find that orbital evolution is dominated by tidal harmonics with $N\Omega_{\rm o}{\sim}\Omega_{\rm peri}$ where $\Omega_{\rm peri}{\sim}2\pi/(1 \ \rm day)$ is the orbital frequency at periapse, i.e.~the tidal evolution is dominated by $N{{\sim}50}$, for the initial semi-major axes and eccentricities we consider.

The orbit and planet rotation rate evolve under tidal dissipation in the planet (we neglect stellar dissipation) according to,
\begin{align}
    \frac{\dot{a}}{a}&=-\frac{2a \dot{E}_{\rm t}}{GM_{\star}M_{\rm p}} \label{equation:adot},\\
    \frac{e\dot{e}}{1-e^{2}}&=-\frac{a \dot{E}_{\rm t}}{GM_{\star}M_{\rm p}}+\frac{\mathcal{T}_{\parallel}}{J} \label{equation:edot},\\
    \dot{\Omega}_{\rm s}\big|_{\rm tid}&=\frac{\mathcal{T}_{\parallel}}{I_{\rm p}} \label{equation:omega_dot},
\end{align}

\noindent where $I_{\rm p}$ is the planet's moment of inertia, and $J{=}M_{\rm p}\sqrt{GM_{\star}a(1{-}e^{2})}$ is the orbital angular momentum (for $M_{\rm p}{\ll}M_{\star}$). Equation \ref{equation:omega_dot} torques the planet spin toward local zeros in $\mathcal{T}_{\parallel}$. The planet's moment of inertia evolves with entropy, and is tabulated in our structure grid. Under conservation of rotational angular momentum, the planet's rotation rate spins down (in the absence of tidal torque) as the moment of inertia grows with entropy. The contribution from this angular momentum conservation to the rotation rate evolution follows,
\begin{equation}
\dot{\Omega}_{\rm s}\big|_{\rm inflate}{=}-\Omega_{\rm s}\frac{\partial \log I_{\rm p}}{\partial S}\dot{S},
\end{equation}

\noindent where $\partial \log I_{\rm p}/\partial S$ is computed and tabulated in our structure grid. In total, the planet rotation rate evolves under tides and radius inflation via $\dot{\Omega}_{\rm s}{=}\dot{\Omega}_{\rm s}|_{\rm tid}{+}\dot{\Omega}_{\rm s}|_{\rm inflate}$.

\subsection{Estimating What Happens After Runaway Inflation}

The last ingredient we include in our evolution calculations is a measure of the degree to which a planet's fundamental mode may be tidally excited during radius inflation. Our goal is to estimate when the approximations used in our evolution tracks are likely to break down, as tidal inflation gives way to non-linear $f$-mode excitation/breaking (which we do not calculate). To estimate when a planet's $f$-mode may be strongly excited, we follow the non-rotating model of \cite{preteu77} in writing the energy imparted to the mode after a single periastron passage (assuming zero initial mode amplitude):
\begin{align}\label{equation:overlap}
    \Delta E_{\rm kick}&=
    2\pi^{2}\omega^{2}_{\rm dyn}\bigg(\frac{GM^{2}_{\star}}{R_{\rm p}}\bigg)\bigg(\frac{R_{\rm p}}{D_{\rm peri}}\bigg)^{2\ell+2}|Q_{n\ell}|^{2}|K_{\ell m}|^{2},\\
    K_{\ell m}&=\frac{W_{\ell m}}{2\pi}\int^{P_{\rm o}/2}_{-P_{\rm o}/2} \bigg(\frac{D_{\rm peri}}{r(t)}\bigg)^{\ell +1}e^{-\mathrm{i}[\omega_{f}t-m\Phi(t)]}\, dt\label{equation:Klm},
\end{align}

\noindent where $D_{\rm peri}{=}a(1{-}e)$ is the periastron distance, $\Phi(t)$ is the true anomaly, and $W_{\ell m}{\sim}1$ for $\ell{=}m{=}{\pm}2$. The last two terms in Equation \ref{equation:overlap} quantify how strongly the tidal potential couples to a given mode. The overlap integral $Q_{n\ell}$ measures the spatial coupling between the tidal potential and the mode eigenfunction \citep[see Equation~38 of][]{preteu77}, whereas $K_{\ell m}$, the ``orbit integral'', measures the temporal coupling between the tidal forcing and the mode during the periastron passage. We neglect the evolution of $Q_{n\ell}$, which typically takes an order-unity value, assuming that its variation with planetary structure is modest compared with that of $K_{\ell m}$; as shown below, $K_{\ell m}$ increases by $\gtrsim4$ orders of magnitude during the IW-driven evolution.

\begin{figure*}[t]
\centering
\includegraphics[width=1\textwidth]{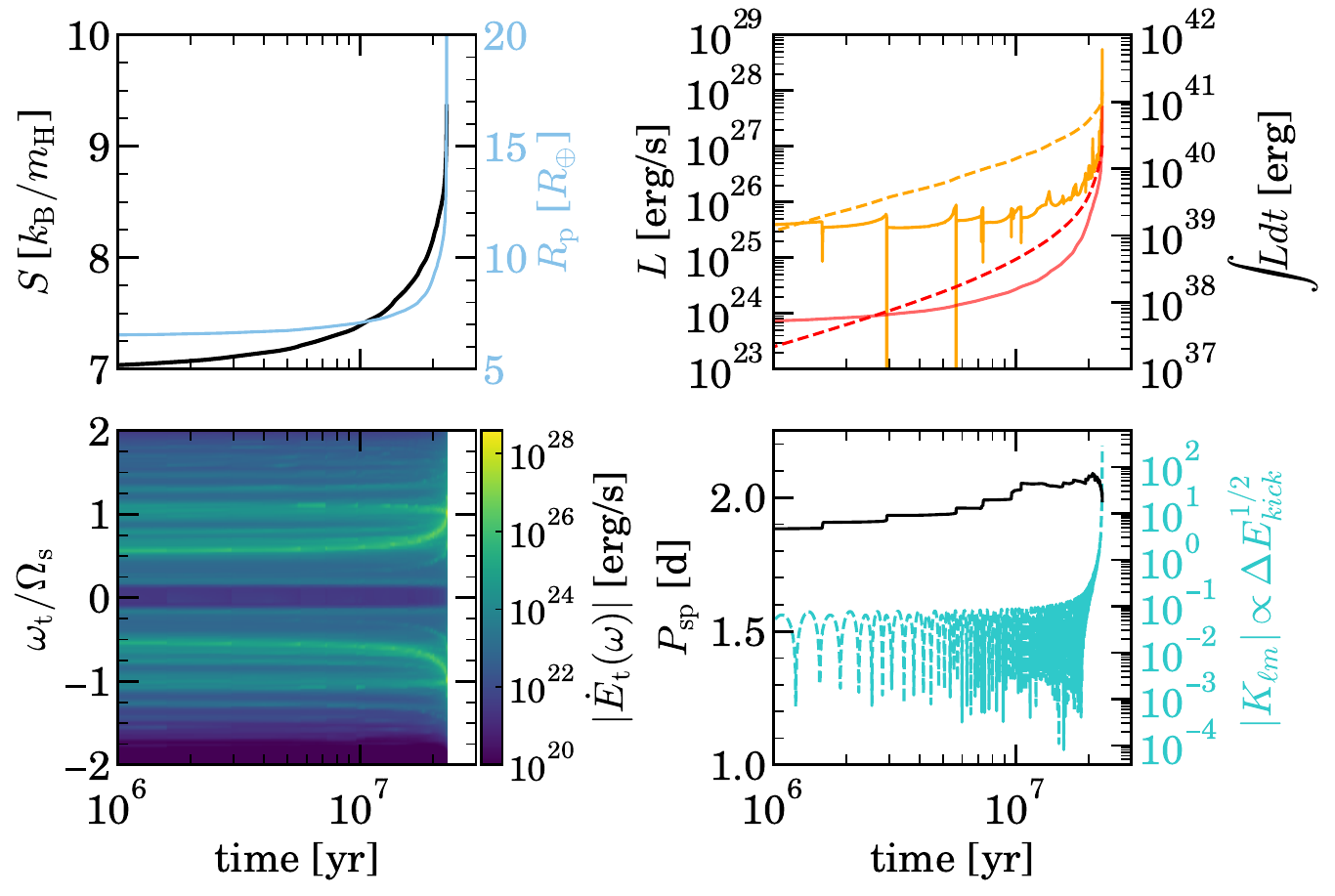}
\caption{Coupled thermal/structure/orbital evolution of a Neptunian planet (30 $M_{\oplus}$, 20 $M_{\oplus}$ core) under inertial wave tidal dissipation. This planet begins with a spin period ${\sim}$1.76 days (corresponding to ${\sim}10\%$ of the break-up frequency), and semi-major axis and eccentricity $a{=}0.3$ au, and $e{=}0.896$ respectively ($a,e$ do not evolve significantly before the end of the calculation at ${t}{\sim}2{\times}10^{7}$ yr). Top row, left panel: entropy (left axis, black curve) and planet's photospheric radius (right axis, blue curve).
Top row, right panel: luminosity (left axis, solid curves) and integrated energy lost (right axis, dashed curves), for tidal heating (orange curves) and radiative cooling (red curves). Bottom panels: tidal luminosity spectrum as a function of frequency and time (left panel; as measured in the inertial frame, see Equation \ref{equation:tidal_sums}), and planet spin period (black solid in left $y$-axis, right panel) and overlap between the planet's fundamental mode and orbit (turquoise dashed line and right $y$-axis). Inertial wave heating grows in lockstep with planet radii, producing runaway inflation. As planet radii puff up, planets' $f$-mode frequencies sweep into resonance with tidal forcing; their $f$-modes are excited.
\label{figure:evo1}}
\end{figure*}

Equation \ref{equation:Klm} expresses the fact that modes with pattern speeds $\omega_{f}/m$ that match the advance of the true anomaly near periastron are most strongly excited; in other words, prograde ($m{=}{+}2$) modes with periods commensurate with the periastron passage time $\tau_{\rm peri}{=}\sqrt{D^{3}_{\rm peri}/G M_{\star}}$. The integral \ref{equation:Klm} is computed by post-processing our evolutionary track at each timestep using $\omega_{f}(S(t))$ following Figure \ref{figure:om_dyn}. The numerical integral is computed via $\texttt{scipy}$'s $\texttt{DOP853}$ explicit Runge-Kutta method with a relative tolerance $10^{-12}$.

\subsection{Summary of Evolution Calculation}

We study the coupled orbit/structure/thermal evolution of planets undergoing frequency-dependent, linear IW dissipation. Planetary structure/thermal evolution is evolved by ``following the adiabats'' (Equation \ref{equation:adiabats}), while the tidal dissipation rate is computed using frequency-dependent IW spectra (Equation \ref{equation:Q_dimensionless}). Orbital shrinkage and circularization follows Equations \ref{equation:adot}, \ref{equation:edot}, and \ref{equation:omega_dot}. The integration is carried out with $\texttt{scipy}$'s implicit $\texttt{BDF}$ method with relative and absolute tolerances of $10^{-12}$ \citep[][]{virgomoli20}.

\begin{figure}
\epsscale{1.2}
\plotone{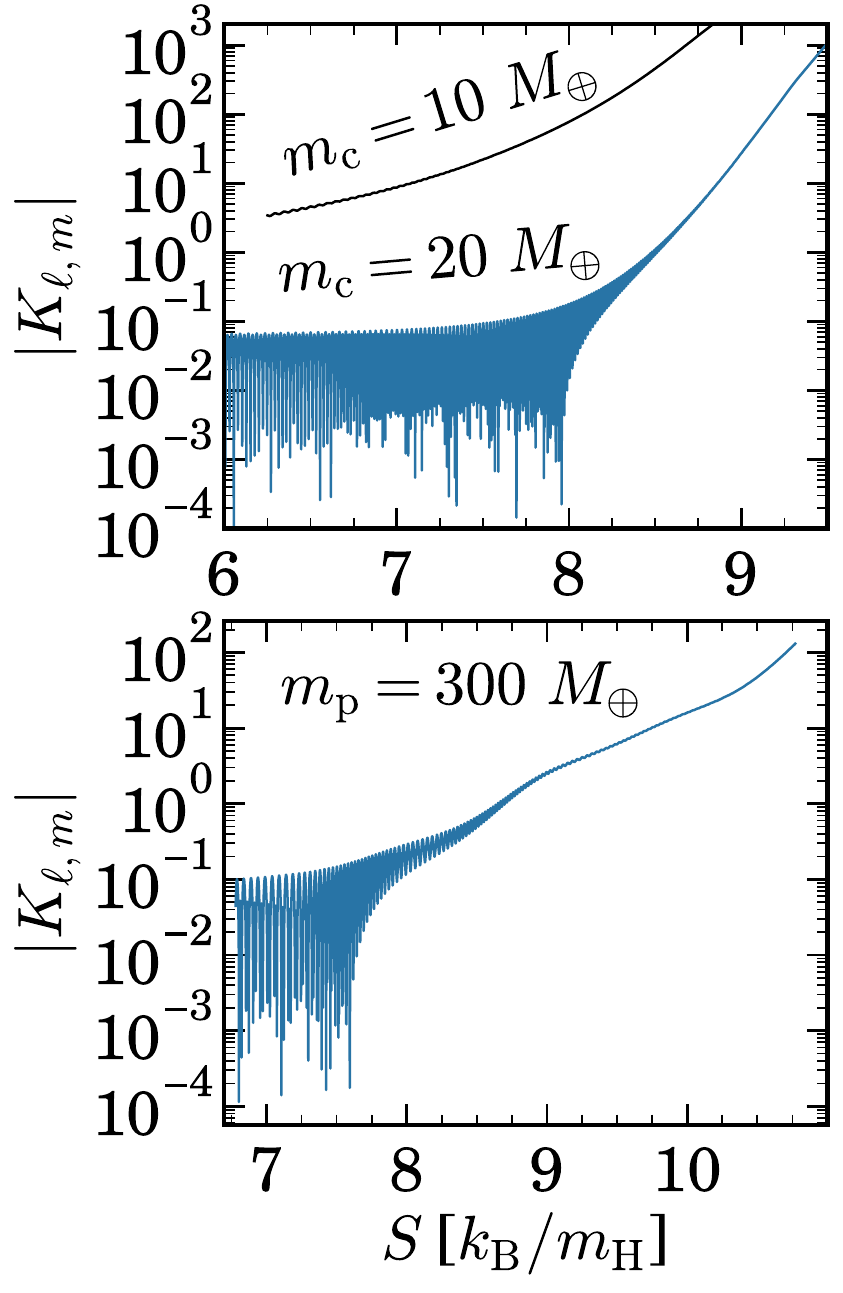}
\caption{An estimate of the strength of coupling between the tidal potential and planets' fundamental modes, as quantified by the orbit integral $|K_{\ell,m}|$ for the $\ell{=}2,m{=}{+}2$ (prograde) $f$-mode. We plot $|K_{\ell,m}|$ as a function of planet entropy for the Neptunian and giant planets in top and bottom panels, respectively. We assume an orbital semi-major axis $a{=}0.3$ au and eccentricity $e{=}0.896$. As planets inflate, their fundamental modes sweep into resonance with tidal forcing. \label{figure:overlap}}
\end{figure}

\begin{figure}
\epsscale{1.3}
\plotone{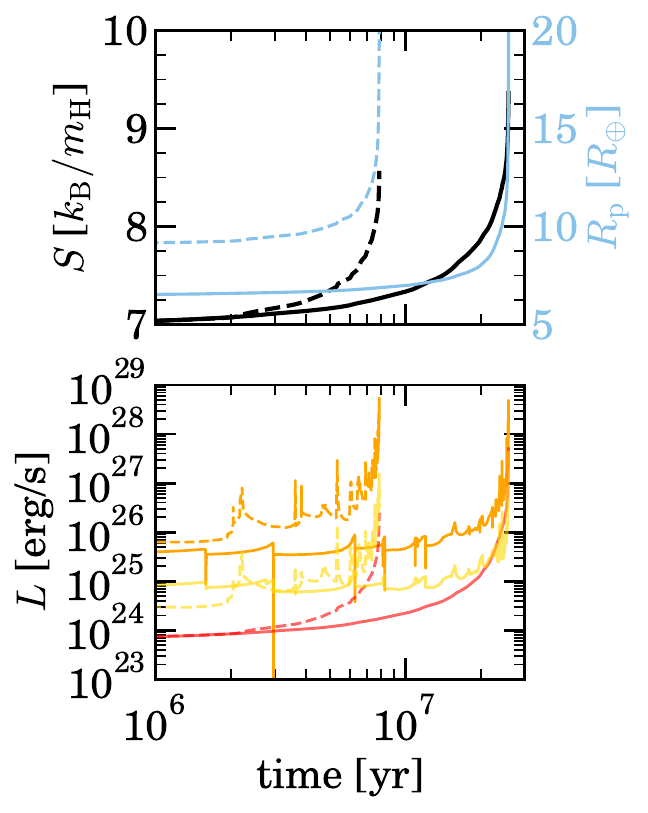}
\caption{Studying the prevalence of runaway for planets incorporating the effects of core thermal inertia (for the same planet as Figure \ref{figure:evo1}; solid lines) and for a planet with a smaller core than Figure \ref{figure:evo1} (10 $M_{\oplus}$, dashed lines; also including core thermal inertia). The color scheme follows Figure \ref{figure:evo1}, with the addition of core luminosity (absolute value) in yellow. Core luminosity is dwarfed by tidal heating, and cannot prevent runaway inflation. Planets with smaller cores are puffier, making them more susceptible to runaway tidal inflation (inflation time ${\sim}8{\times}10^{6}$ vs. ${\sim}20{\times}10^{6}$ yr).
\label{figure:evo2}}
\end{figure}

\begin{figure}
\epsscale{1.3}
\plotone{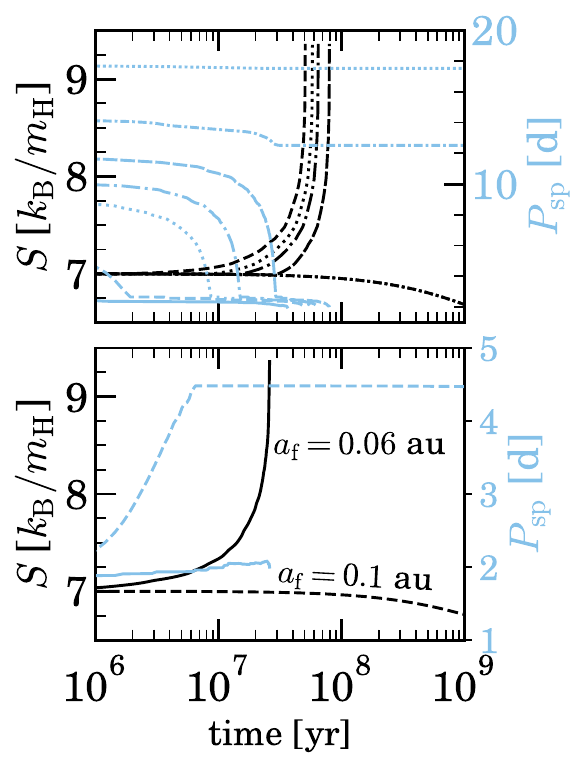}
\caption{Studying when tidal dissipation via inertial waves stalls. Top and bottom panels vary the initial planet spin period and orbital angular momentum, respectively. Both panels depict the planetary evolution in entropy (black; left $y$-axis) and spin period (blue; right $y$-axis). Each $S(t),P_{\rm sp}(t)$ pair of evolutionary tracks share the same linestyle. In the top panel, each planet begins with the same semi-major axis and eccentricity, $a{=}0.3$ au and $e{=}0.896$ respectively. In the bottom panel, we use two different initial orbital angular momenta, parameterized via the post-circularization orbital distance $a_{\rm f}{=}a(1{-}e^{2})$. Both curves use initial semi-major axes $a{=}0.3$ au but vary initial eccentricity $e{=}\{0.896,0.8\}$, respectively. Inertial wave heating leads to an all-or-nothing divergence of outcomes: violent tidal heating scenarios are avoided for planets that begin rotating slower than a threshold (${\sim}14$ days in this example), or those with sufficiently large orbital angular momentum (low $e$), while those with slightly faster initial spins/less angular momentum undergo runaway inflation. Inertial waves only dominate the evolution of Neptunes with low enough orbital angular momentum to circularize inside ${\sim}$0.1 au.
\label{figure:pspin}}
\end{figure}

\begin{figure*}[t]
\centering
\includegraphics[width=1\textwidth]{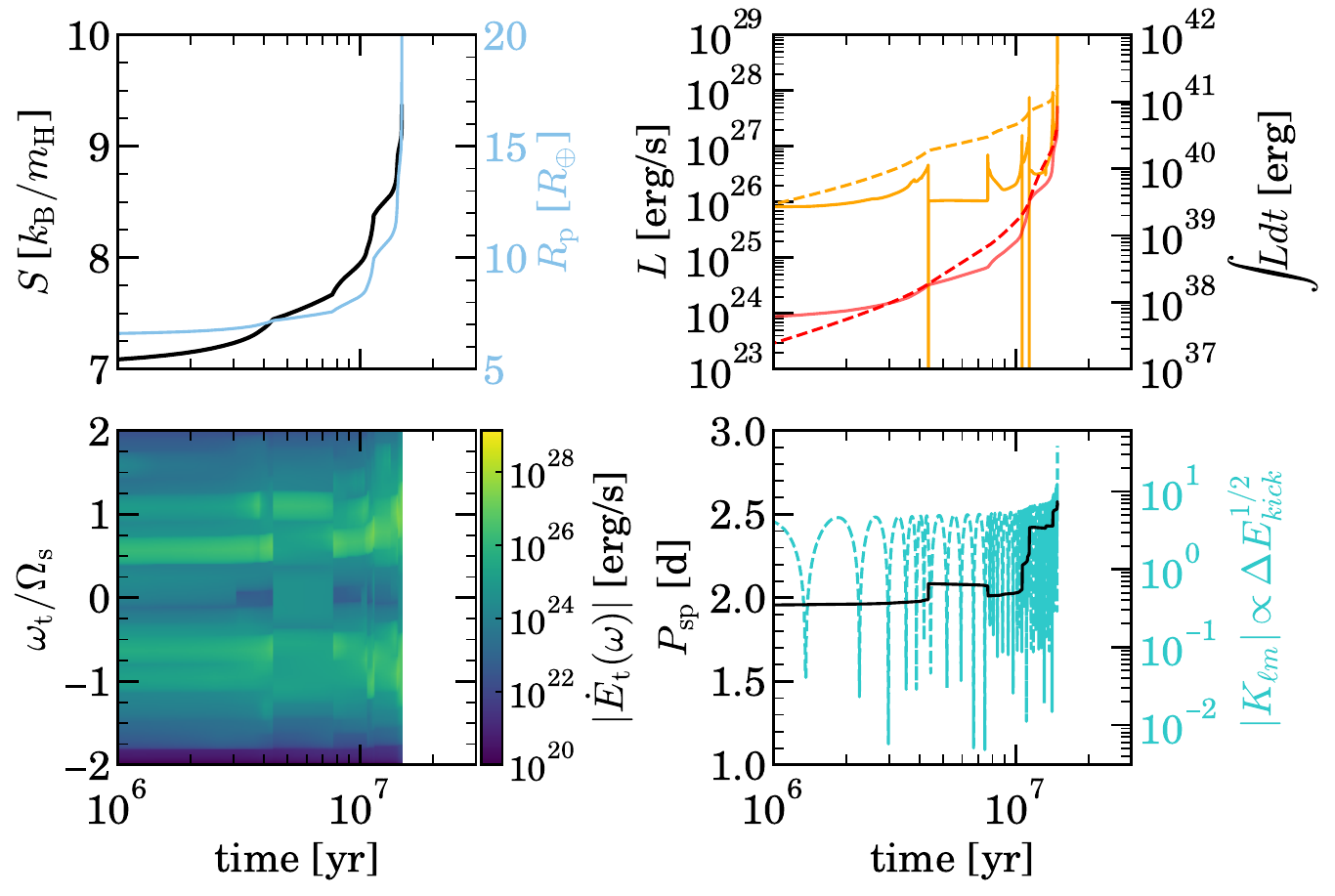}
\caption{Coupled thermal/structure/orbital evolution of the same planet from Figure \ref{figure:evo1}, except beginning with orbital semi-major axis and eccentricity $a{=}0.1$ au, $e{=}0.639$, respectively. Tidal evolution at this moderate eccentricity exhibits less fine structure (e.g.~in heating rate, and subsequently planet entropy) than at higher eccentricity, as the spectrum of tidal harmonics that excite inertial waves becomes sparser. Tidal radius inflation under inertial waves continues to run away.
\label{figure:evo3}}
\end{figure*}

\begin{figure*}[t]
\centering
\includegraphics[width=1\textwidth]{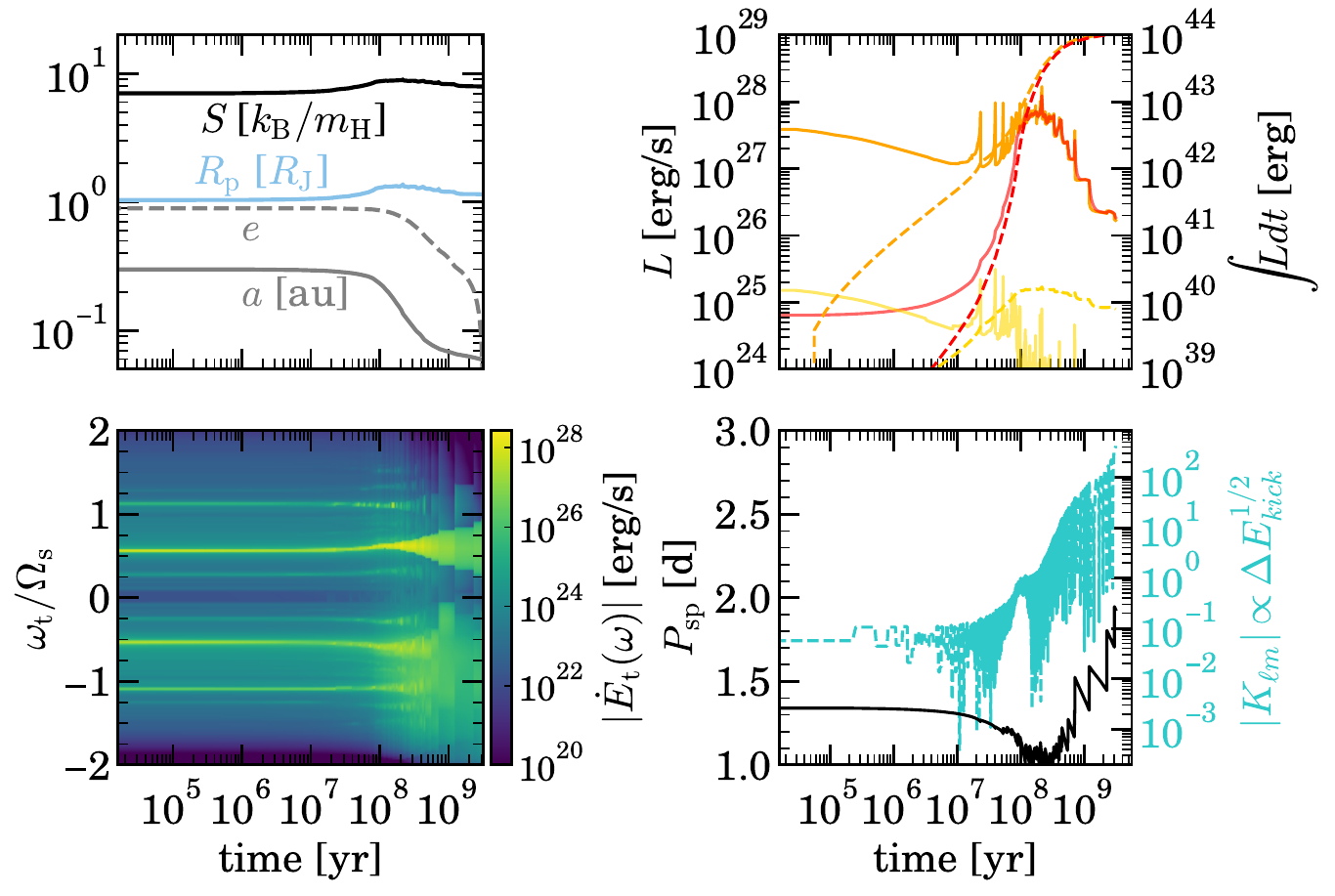}
\caption{Coupled thermal/structure/orbital evolution of a giant planet under inertial wave dissipation. This giant planet begins with initial semi-major axis $a{=}$0.3 au, eccentricity $e{=}0.896$, entropy $S{=}7 \ k_{\rm B}/m_{\rm H}$, and a spin period of ${\sim}$1.3 days, corresponding to $10\%$ of the dynamical period. The giant planet has a total mass of 300 $M_{\oplus}$, and a core mass of 10 $M_{\oplus}$. These quantities, along with planet radius, are plotted in the top left panel (in units depicted). Gas giants accommodate tidal heating, puffing up modestly before circularizing over ${\sim}$Gyr. 
\label{figure:evo4}}
\end{figure*}

\begin{figure*}[t]
\centering
\includegraphics[width=1\textwidth]{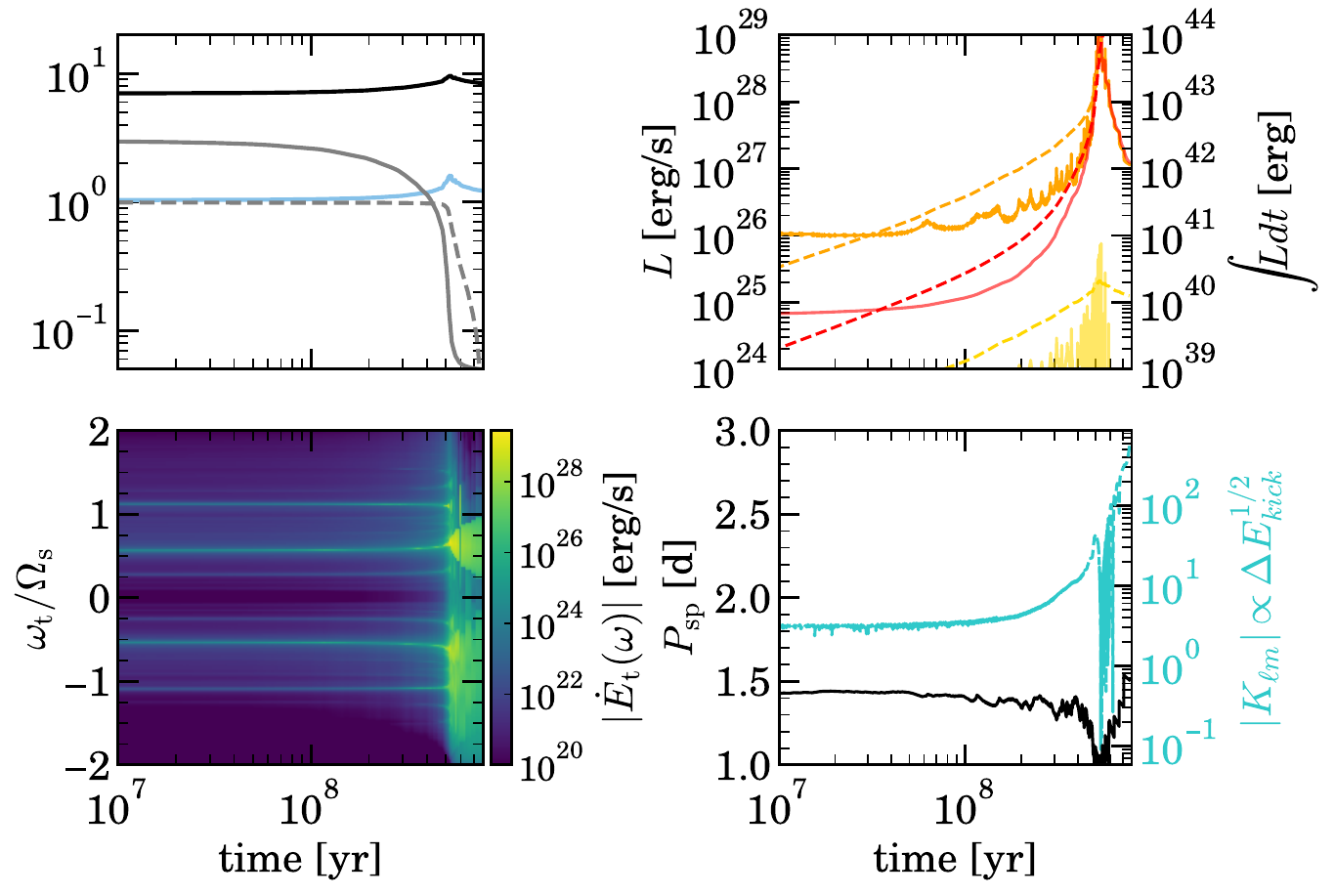}
\caption{Evolution for the same planet as used in Figure \ref{figure:evo4}, except beginning on a more distant ($a{=}3$ au), eccentric ($e{=}0.991$) orbit. Quantities in the top left panel are the same as labelled in Figure \ref{figure:evo4}. Highly eccentric gas giants accommodate tidal heating under inertial waves, circularizing over ${\sim}$1 Gyr (in this example).
\label{figure:evo5}}
\end{figure*}

\section{Coupled Orbit, Structure $\&$ Thermal Evolution: Results}\label{sec:evo}

\subsection{Evolution Under Inertial Waves $\&$ Runaway Inflation}\label{subsec:runaway}

We begin this section with a case study. We consider a Neptunian planet from Figure \ref{figure:om_dyn} on an eccentric orbit with $a{=}0.3$ au and $e{=}0.896$, with fiducial Ekman number $Ek{=}10^{-5}$. Circularization at constant angular momentum would then settle the planet at a distance $a_{\rm f}{=}a({1{-}e^{2}}){\sim}0.06$ au, or a period ${\sim}5$ days. This phase of circularization at constant angular momentum may arise, for example, after von Zeipel–Lidov–Kozai oscillations in eccentricity/inclination have been quenched by short-range forces \citep[e.g. general relativistic precession or tides;][]{fabtre07}. We begin the planet at an initial entropy $S{\sim}7 \ k_{\rm B}/m_{\rm H}$, which yields an initial cooling time $S/|\dot{S}|{\sim}10$ Gyr. This choice of initial cooling time is appropriate if the planet is kicked into an eccentric orbit after evolving for ${\sim}$Gyr, such that its cooling time is comparable to its age. 

Figure \ref{figure:evo1} displays the structural/tidal evolution for our fiducial Neptune. Tidal heating dwarfs radiative cooling for most of the evolution, producing runaway inflation (we turn off core thermal inertia in this example, to isolate its effects in Figure \ref{figure:evo2}; see below). Runaway heating occurs because the planet is deformed by tides more strongly as it puffs up (the non-wave displacement ${\propto}\epsilon_{\rm t}R_{\rm p}$ grows), which in turn seeds a larger inertial wave response (Equations \ref{equation:NS} and \ref{equation:f_IW}; the energy in IWs ${\propto}(\Omega_{\rm s}\epsilon_{\rm t}R_{\rm p})^{2}M_{\rm p}$). The planet's radiative cooling luminosity cannot keep pace with this feedback between the planet's weaker self-gravity and concomitant growth in tidal heating. Figure \ref{figure:evo1} confirms that runaway inflation not only occurs under frequency-dependent IW dissipation, but that runaway occurs rapidly relative to typical system ages or planets' pre-tidal cooling times (${\sim}$Gyr). The hypothesis stated in the introduction, namely that the ``bursty'' nature of IW dissipation may help planets avoid runaway inflation, is not borne out; we observe that resonant frequencies in $Q^{\prime}(\omega_{\rm t}/\Omega_{\rm s})$ are always excited to produce strong IW dissipation.

The bottom left panel of Figure \ref{figure:evo1} further illustrates the frequency bands that dominate the dissipation. As expected, these dominant frequencies correspond to resonant troughs in $Q^{\prime}$ where the planet's eigenmodes are efficiently excited. Meanwhile, tides sporadically torque the planet's spin period (initially at ${\sim}1.76$ days, or ${\sim}10\%$ of the dynamical frequency) into the nearest minimum-torque state ($\mathcal{T}_{\parallel}{\sim}0$); as the planet inflates, the spin readjusts quickly to a new torque minima, giving rise to the step function-like behavior exhibited by the spin period and tidal luminosity. The tidal luminosity dips upon each readjustment to the minimum-torque spin rate, before growing again as the spin departs from the instantaneous minimum-torque rate. While these equilibrium spin periods are numerically similar to the conventional ``pseudosynchronous'' spin rate of weak friction tidal theory \citep[e.g.][]{hut81}, spin evolution under IWs differs from this classical picture due to the existence of multiple spin equilibria \citep[][see also \citealt{stolai14}]{dew24}. Our calculations confirm that IWs produce a forest of equilibrium spin rates (see Figures \ref{figure:spin_equilibria} and \ref{figure:spin_equilibria2}), though during evolution only the nearest minimum-torque spin period is attained; the planet spins down until it encounters a period with a long evolution time $\tau_{\rm s}{=}\Omega_{\rm s}/|\dot{\Omega}_{\rm s}|{\sim}$Gyr, before evolving into the next nearest minimum-torque state. 

Figure \ref{figure:evo1} reveals that the orbit integral $K_{\ell m}$ also experiences runaway growth: coupling between the planet's $f$-mode and the tidal potential grows in lockstep with the radius. We end the integration once the orbit integral $|K_{lm}|{>}10^{2}$, as we are unable to calculate the non-linear evolution during this phase of $f$-mode excitation with the present linear approach.

We also conclude from Figure \ref{figure:evo1} that the amount of tidal heat that must be dissipated to excite the Neptunian planet's $f$-mode is a small fraction of the planet's binding energy. In other words, the energy requirement for $f$-mode excitation is modest. To confirm this, we emphasize in Figure \ref{figure:overlap} the steep dependence of $f$-mode excitation, quantified via the orbit integral $K_{\ell m}$, on planet entropy. Figure \ref{figure:overlap} indicates that an entropy $S{\gtrsim}8.75 \ k_{\rm B}/m_{\rm H}$ is required for the orbit integral to exceed ${\gtrsim}10$ (a rough threshold we choose to indicate strong tidal/$f$-mode coupling). The total energy injection needed to produce this degree of inflation, starting from an initial entropy $S_{0}{\sim}7 \ k_{\rm B}/m_{\rm H}$, is (expressed as a fraction of the planet's binding energy):
\begin{align}
\begin{split}\label{equation:E_inflate}
    \frac{\Delta E_{\rm inflate}}{|E_{\rm b}|}& \sim \frac{\Delta S \int Tdm}{G M^{2}_{\rm p}/R_{\rm p}},\\
    & \sim 0.1 \bigg(\frac{\Delta S}{2 \ k_{\rm B}/m_{\rm H}}\bigg)\bigg(\frac{\int Tdm}{4{\times}10^{32} \ \mathrm{K} \cdot g}\bigg)\\
    & \hspace{2cm} \times \bigg(\frac{R_{\rm p}}{6 \ R_{\oplus}}\bigg)\bigg(\frac{M_{\rm p}}{0.1 \ M_{\rm J}}\bigg)^{-2}.
\end{split}
\end{align}

\noindent Equation \ref{equation:E_inflate} is a lower limit on the amount of energy required, since a fraction will be removed by radiative cooling rather than directly increasing $S$. Comparing this to Figure \ref{figure:evo1}, however, indicates reasonable agreement for our Neptunian planets (${\Delta }E_{\rm inflate}{\sim}6{\times}10^{40} \ \rm erg$), as they do not efficiently radiate away tidal heat. The lesson from Equation \ref{equation:E_inflate} and Figure \ref{figure:overlap} is that a small amount of tidal heat relative to a Neptunian planet's binding energy can drive the planets' dynamical frequency into resonance with tidal forcing.

Inserting the parameters for a giant planet ($\int Tdm{\sim}2.5{\times}10^{34}$ K${\cdot} g$, $R_{\rm p}{=}12 \ R_{\oplus}$, $M_{\rm p}{=}M_{\rm J}$) into Equation \ref{equation:E_inflate} also yields the same ${\sim}10\%$ of planetary binding energy required to strongly couple the planet's $f$-mode with the tidal forcing. We will see, however, that giant planets bleed more of the tidal heat to space via radiative cooling, preventing significant radius expansion to a greater extent than for Neptune-mass planets. 

\subsection{Planetary Cores Do Not Help Avoid Runaway}\label{subsec:cores}

We have demonstrated that Neptunian planets undergo runaway tidal inflation, even at moderate orbital distances ($a{=}0.3$ au in Figure \ref{figure:evo1}). In Figure \ref{figure:evo2}, we test whether the efficiency of runaway inflation is influenced by a planetary core that may act as a heat sink. We find that core thermal inertia yields negligible braking of this runaway heating. Figure \ref{figure:evo2} also displays the evolutionary track for our Neptune with a smaller core; the puffier radii produced in this model exacerbates runaway inflation. We conclude from Figure \ref{figure:evo2} that planets' susceptibility to runaway inflation is not sensitive to the core's mass or its heat capacity.

\subsection{Tidal Stalling}\label{subsec:stalling}

The results of Sections \ref{subsec:runaway} and \ref{subsec:cores} show that if IWs are active, their dissipation runs away due to radius inflation. Here, we explore how a wider range of initial conditions in planet spin and orbital angular momentum affect how IW dissipation evolves. We will show that IW dissipation stalls if planets rotate sufficiently sluggishly.

Slow planet rotation may arise either before or after eccentricity excitation. In the first scenario, slow initial rotation may be expected if spin/orbit synchronization is attained prior to eccentricity excitation when the planet is on a wide, nearly circular orbit. In the second scenario, planet rotation can be sluggish due to a large periastron distance (large orbital angular momentum), which yields a slow pseudosynchronous spin rate \citep[similar to the weak friction model, our calculations produce pseudosynchronous spins ${\sim}$the periastron passage time;][]{hut81,tre23}. 

Tidal evolution peters out when planets spin slowly due to two effects. First, under IWs, the fraction of tidal energy dissipated per period $1/Q^{\prime}(\omega_{\rm t}/\Omega_{\rm s}){\propto}\Omega^{2}_{\rm s}R^{3}_{\rm p}$ (at fixed mass; Equation \ref{equation:eps_omega}). Second, slow planet spins decrease the number of tidal frequencies that may excite IWs (recall from Section \ref{sec:orbital_EOM} that the maximum harmonic $N_{+2}{\sim}4P_{\rm o}/P_{\rm s}$). Figure \ref{figure:pspin} showcases two examples in which slow planet rotation is achieved. In both panels, tidal evolution bifurcates: planets with slow enough spins (either as initial conditions, as in the top panel, or through a slow pseudosynchronous rate, in the bottom) hardly suffer tidal heating, while those rotating quickly enough undergo full-blown runaway inflation. The top panel tells us that initial rotation periods ${\gtrsim}14$ days preclude tidal spin-up to pseudosynchronism, as the spin-up timescale $\tau_{\rm s}{=}\Omega_{\rm s}/\dot{\Omega}_{\rm s}{\propto}Q^{\prime}$ grows prohibitively long when planets rotate more slowly. This tidal shutdown is reinforced as the radii shrink under radiative cooling, further increasing $Q^{\prime}$ and $\tau_{\rm s}$.

The bottom panel of Figure \ref{figure:pspin} shows an example wherein planets successfully attain pseudosynchronism, but still avoid runaway heating. Tidal heating stalls  due to the long pseudosynchronous rotation period, ${\sim}$4.5 days, which decreases the number of tidal harmonics that excite IWs from $N_{+2}{\sim}133{\rightarrow}50$. These fewer harmonics avoid sampling the resonant trough in $Q^{\prime}$ that drives runaway inflation in the higher eccentricity planet. This tidal stalling is exacerbated by the increase in $Q^{\prime}$ from rotation alone, which increases $Q^{\prime}$ by $Q^{\prime}{\propto}\Omega^{-2}_{\rm s}{\sim}(4.5/2)^{2}{=}5{\times}$. Altogether, the tidal heating rate drops by ${\sim}100{\times}$ relative to our fiducial planet.

The upshot of Figure \ref{figure:pspin} is that tidal evolution under IWs is not sensitive to the planet's initial spin, so long as it is fast enough for tidal heating to outpace planetary cooling. This means that IWs can only help circularize the orbits of Neptunes that park inside ${\lesssim}$0.1 au; wider orbits have larger orbital angular momentum, which precludes efficient IW heating, since such planets spin slower throughout their evolution.

\subsection{Runaway Also Occurs for Tighter Orbits}

We now extend our tidal calculations to tighter, less eccentric orbits. We seek to determine what, if anything, changes as the eccentricity declines, and ultimately whether runaway inflation still ensues.

As we have shown in Figure \ref{figure:evo3}, the tidal evolution of less eccentric orbits exhibits fewer step-like jumps and is dominated by a smaller number of harmonics (frequency/temporal). The relative lack of fine structure compared to Figure \ref{figure:evo1} follows from the fact that planets on short period orbits (or those with longer rotation periods) excite fewer tidal harmonics that are capable of driving IWs; the maximum temporal harmonic that yields IW excitation (tidal frequencies ${\in}[-2,2]$) is $N_{+2}{\sim}4P_{\rm o}/P_{\rm s}{\sim}136$ and ${\sim}26$ for our more and less eccentric cases in Figures \ref{figure:evo1} and \ref{figure:evo3}, respectively (using initial $P_{\rm s}{=}1.76$ days). 

Figure \ref{figure:evo3} confirms that IW heating also produces runaway inflation in less eccentric orbits. The orbit integral, $K_{\ell m}$, however, differs in two respects from the more eccentric case. First, it is larger in the pre-runaway phase as more of the orbit contributes to $K_{\ell m}$ than just the regions near periastron. Second, it rises less steeply as the planet undergoes inflation, as the planet's growing dynamical time takes longer to sweep into resonance with the longer periastron time; only radii ${\gtrsim}20\ R_{\oplus}$ produce such resonant dynamical times in our calculations. We conclude that, on the whole, tidal evolution under IWs proceeds similarly to that at higher eccentricity.

\subsection{IWs Circularize Gas Giant Orbits}

Having determined that Neptunian planets frequently undergo runaway tidal inflation, we now consider whether their gas giant counterparts suffer the same fate. 

We show in Figures \ref{figure:evo4} and \ref{figure:evo5} that gas giants at moderate and wide initial distances ($\{a,e\}{=}\{0.3\ \rm au,0.896\},\{3 \ \rm au, 0.991\}$ in Figures  \ref{figure:evo4} and \ref{figure:evo5} respectively) successfully circularize without undergoing runaway inflation. Giant planets' larger thermal inertia enables them to soak up the extra heat, allowing the radiative cooling luminosity to grow sufficiently to balance the tidal luminosity. The weak degree of radius inflation/lack of runaway inflation exhibited by these gas giants reflects their comparatively flat radius/entropy relationship, as discussed in Section \ref{subsec:total_mass}. 

Figures \ref{figure:evo4} and \ref{figure:evo5} inform us that gas giants undergoing migration driven by IWs dissipate more orbital energy than is contained in their structures, by radiating the excess heat to space. We further confirm that these gas giants are unlikely to be tidally disrupted, since their periastron distances $a_{\rm peri}{=}a(1{-}e)$ always exceed the tidal disruption limit $r_{\rm t}{=}\eta(M_{\star}/M_{\rm p})^{1/3}R_{\rm p}$, where $\eta{\sim}2.7$ \citep[i.e. the approximate periastron distance where direct tidal stripping of the atmosphere takes place;][see also \citealt{fanliu26} for updated estimates]{guiramlin11}. The survival of gas giants that migrate under IWs alleviates a long-perceived weakness of HEM theory, namely that gas giants may not survive the dissipation of more orbital energy than is contained in their structures \citep[e.g.][see also \citealt{tre23}]{fabtre07,wu18}. Our calculation suggests the energy budget issue for gas giants is not as dire as previously suggested. We note however that at the high eccentricity shown in Figure \ref{figure:evo5}, the orbit integral grows from $|K_{\ell m}|{\sim}3{\rightarrow}30$ as the planet puffs up, indicating the $f$-mode may dominate tidal circularization rather than IWs; nonetheless, our calculation demonstrates that dissipating via the $f$-mode is not the only way for gas giants to survive HEM.

This result extends previous work coupling gas giant orbital/structural evolution to determine whether they survive HEM. Previous efforts showed that hot Jupiters may survive if tidal heat is not deposited deep within the planet \cite[under the assumption that the $f$-mode breaks near the photosphere; see e.g.][who only include deep tidal heating when $e{<}0.2$]{rozglaper22,glarozper22}, or did not follow the tidally-heated, long-term phase of circularization after the first few initial periastron passages \cite[][]{welhannao26}. In contrast, we find that depositing tidal heat deep within a gas giant (thereby $\textit{maximizing}$ radius inflation) still does not produce runaway inflation/tidal destruction (under IWs). 

\subsection{Tidal Migration of Low-Viscosity Planets}\label{sec:evo_Ek}

We find that runaway inflation plays out in a similar manner to Figure \ref{figure:evo1} when using a viscosity $10^{5}{\times}$ smaller than our fiducial choice ($Ek{=}10^{-10}$; Figure \ref{figure:evo6}, placed in Appendix \ref{sec:low_Ek} for brevity). The same resonant harmonics drive tidal evolution in both the low and high viscosity cases, though runaway takes an additional ${\sim}10^{7}$ yr for the low viscosity case in this example. This slightly longer runaway time stems from the fact that resonant troughs in $Q^{\prime}$ are made narrower with lower viscosity (even though they may be deeper), so the system spends less time in them, while the non-resonant continuum between these peaks has smaller dissipation; the dissipation is more and more localized to the narrower and narrower peaks as $Ek$ is decreased. We conclude that our main results are not sensitive to the viscosity adopted to damp IWs in the planet. This is reassuring because the appropriate value of $\nu$ (and whether this is a microscopic or turbulent value) that should be applied to model damping of IWs is not well understood at present. 

\section{Discussion}\label{sec:discussion}

Our results suggest that tidal migration of Neptunian planets (to final circularized semi-major axes $a_{\rm f}{=}a(1{-}e^{2}){\lesssim}0.1$ au) under IWs leads to runaway inflation. In contrast, gas giants are able to accommodate this tidal heating and circularize without such drama. Here, we consider the implications of these results for high eccentricity migration theory (Section \ref{subsec:HEM}), and for population-level trends that observations may probe (Section \ref{subsec:populations}). We acknowledge some limitations of our calculations and avenues for future work in both Sections, as well as in Section \ref{subsec:fw}.

\subsection{Migration by Inflation}\label{subsec:HEM}

The starting point for our analysis was the last decade in orbital shrinkage required to place a hot Neptune/Jupiter in a short-period orbit (e.g. $a{\sim}0.5{\rightarrow}0.05$ au; Figure \ref{figure:schematic}). While tidal dissipation at very high eccentricities may be driven by planets' fundamental modes, their tidal excitation fails to be efficient once planetary eccentricities fall below $e{\lesssim}0.9$ \citep[e.g.][]{yuweiarr22}. Our work suggests that IWs pick up where $f$-modes leave: 
not only do IWs readily drain orbital energy (enough to circularize gas giants), but the runaway inflation they induce sweeps planets' fundamental modes back into resonance with tidal forcing. As $f$-modes shock at low pressures to bleed tidal heat to space under radiative diffusion \cite[e.g.][]{zanmacbry26}, this ``migration by inflation" offers a mechanism to remove more energy from Neptunes' orbits than is contained in their structures (Equation \ref{equation:Ebind}; our ``final parsec" problem). Below, we flesh out how this IW/$f$-mode coupling may operate.

The chief distinction between runaway inflation under IWs and dissipation via $f$-modes is their respective timescales: the tidal inflation time is ${\sim}10^{7}$ yr, whereas $f$-modes excite (and potentially break) orbit-to-orbit, over ${\sim}10^{2-5}$yr \citep[e.g.][]{mar95b,ivapap04}. We therefore expect that, once radius inflation sweeps fundamental modes into resonance with tidal forcing, the planet's entropy will not change substantially while $f$-modes excite and dissipate (if their dissipation is restricted to the lowest density surface layers). Recent hydrodynamic estimates \citep{zanmacbry26} further suggest that $f$-mode breaking dissipates a small fraction of the planet's binding and orbital energy, $\Delta E{\sim}10^{-3}E_{\rm b}$, translating to $\Delta E/E_{\rm orb}{\sim}10^{-4}(R_{\rm p}/15 \ R_{\oplus})^{-1}(a/0.3 \ \mathrm{au})(M_{\rm p}/30  \ M_{\oplus})$. Traversing the final decade in orbital shrinkage would therefore require ${\gtrsim}10^{4}$ episodes of wave breaking, translating to just ${\gtrsim}10^{3}$ yr (assuming each wave breaking event occurs once per orbit). Whether $f$-mode breaking can completely circularize the orbit after runaway inflation, or whether residual eccentricity remains, is uncertain. 

Future work should resolve this coupled IW/$f$-mode evolution, tying the orbit-to-orbit mapping required to follow $f$-modes with our secular orbit/structure method. Such a follow-up effort would also benefit from more accurately capturing the transition from IW to $f$-mode excitation 
\citep[e.g.~by extending the present inertial wave code to solve for the direct linear tidal response, similar to a compressible version of][]{ponbarhol24}, rather than using the orbit integral as a proxy. We plan to address these issues in a future paper.

\subsection{Population-Level Expectations}\label{subsec:populations}

For Neptunian planets, tidal dissipation via IWs occurs in a hurry; dissipation runs away in ${\sim}10^{7}$ yr. Gas giants, on the other hand, circularize over ${\sim}$Gyr, with comparatively little structural upheaval. Below, we outline how this distinction in tidal timescales may be imprinted in the population of warm, eccentric planets.

The predicted ${\sim}$Gyr circularization times for gas giants is consistent with the observation of warm ($a{\gtrsim}0.2$ au) Jupiters that exhibit orbital angular momenta on a high eccentricity migration track \citep[$a_{\rm f}{<}0.1$ au or $e{\gtrsim}0.8$; e.g.][]{dawmur13,donhuazho21,donhuadaw21,gupmil2024,livmilvic2026}. Under IW dissipation, Neptunian counterparts with such high eccentricities should be less common owing to their ${\sim}100{\times}$ shorter circularization times. In contrast, moderately eccentric warm Neptunes are predicted to be commonplace, as their slow equilibrium spin rates throttle IW dissipation (Figure \ref{figure:pspin}). Ongoing and near-future transit surveys such as $\textit{TESS}$ \citep[e.g.][]{ricwinvan14,fainabhua24,epsbhawin26}, $\textit{NGTS}$ \citep[e.g.][]{gunquedem17,kenulmgil26}, $\textit{PLATO}$ \citep[][]{rauaercab25}, and potentially $\textit{Roman}$ \citep[][]{wilbarpow23}, coupled with radial velocity follow-up when possible, may extend the planetary eccentricity census into the warm Neptune regime to test this prediction.\footnote{We confirm with the \textit{tepcat} exoplanet catalog \citep[][]{sou26} that only one warm Neptune/sub-Saturn with very high eccentricity ($e{>}0.8$) is currently validated -- Kepler 1656 b \citep[$a,e,M_{\rm p}\sin i{=}0.1974{\pm}0.0026\ \mathrm{au},0.838_{-0.029}^{+0.045},47.8_{-3.3}^{+6.2} \ M_{\oplus}$;][]{angnaopet22}. As argued by \cite{angnaopet22}, Kepler 1656 b does not appear to undergoing tidal migration, however, with $e$ oscillating over periods ${\sim}10^{5}$ yr (see their Figure 8).}

There is one caveat to this predicted dearth of very eccentric warm Neptunes: as explored in Section \ref{subsec:stalling} and Figure \ref{figure:pspin}, high eccentricity warm Neptunes may persist indefinitely if they possess rotation periods in excess of ${\gtrsim}$2 weeks. IW theory therefore predicts that any outlier warm Neptune with high eccentricity should possess negligible rotational oblateness, and a low entropy interior \citep[which may be probed by e.g.~$\textit{JWST}$ spectroscopy;][]{sinrustho24}. Specifically, we estimate the oblateness parameter $(R_{\rm eq}{-}R_{\rm pol})/R_{\rm eq}{=}3J_2/2{+}\Omega^{2}_{\rm s}R^{3}_{\rm eq}/2GM_{\rm p}{\lesssim}10^{-4}$, where $R_{\rm eq}$ and $R_{\rm p}$ are the equatorial and polar radii, and the quadrupole moment $J_{2}$ follows Equation 3 of \cite{ragwol09} (using a Love number from our structure grid). 

We next consider how our observational predictions for warm Neptunes may apply to the hot Neptune population (periods ${\sim}4{-}10$ days), many of which likely stem from warm Neptunes that undergo tidal migration. IW theory predicts that warm Neptunes undergoing HEM undergo rapid eccentricity damping due to runaway inflation. At first blush, the presence of a handful of hot Neptunes with elevated eccentricities \citep[$a{\lesssim}0.05$ au, $e{\lesssim}0.2$ but consistent with non-zero;][]{corboudel20} runs counter to the expectation that tidal migration occurs rapidly. Assuming the eccentricities are truly non-zero, it is unclear whether these few eccentric hot Neptunes constitute a genuine tension with theory. If runaway inflation does not completely circularize warm Neptune orbits, we may expect to observe such low-$e$ planets as they finalize circularization under IWs. To determine whether IW theory permits such low-$e$ Neptunes to exist, we construct a suite of circularization calculations varying initial rotation period, planet viscosity, and host stellar mass ($M_{\star}{=}\{0.5,0.8,1\} \ M_{\odot}$; several of these eccentric hot Neptunes orbit KM stars), for initial eccentricity of 0.2, orbital periods ${\sim}5$ days, and high entropy, as may be expected if they undergo tidal migration.\footnote{For these low-e runs, we employ numerically-computed Hansen coefficients, as the \cite{sulai22} approximations are not sufficiently accurate.} 

We find that initial rotation periods ${\gtrsim}$3 days produce indefinite tidal stalling at such low eccentricities. Whether such sluggish rotation can be achieved is an open question; planets are likely spun-up after runaway inflation as their $f$-modes absorb orbital angular momentum \citep{ivapap04,yuweiarr22}, but planetary spin-down could be affected by magnetized photoevaporative mass loss \citep[][]{webdav67,oweada14}. Alternatively, slow rotation may arise if these hot Neptunes scatter into their eccentric orbits from initial spin/orbit synchronism. Planet/planet scattering may be at odds with some hot Neptunes' polar obliquities \citep[which is a rare outcome of scattering; e.g.][]{beanes12,albdawwin22}, and lack of nearby companions \cite[][]{thoniecri26,almcasmar26}, however.

For fast-rotating Neptunes, we find that eccentricity damping times grow as the viscosity that damps IWs is lowered to $Ek{=}10^{-10}$, reaching ${\sim}0.5$ Gyr for eccentricities $e{\sim}0.2$ \citep[as exhibited by e.g.~HAT P 11 b;][]{bouattmall23,anlubra25}. As discussed in Section \ref{sec:evo_Ek}, eccentricity damping slows for low viscosity planets as resonant peaks in dissipation grow narrower, while the non-resonant background becomes less dissipative. This effect is exacerbated by the fact that, at such low eccentricities, there are fewer resonant tidal harmonics ($N_{+2}{\sim}10$), making it more unlikely for the planet to sample the narrow troughs in $Q^{\prime}$. For KM stars, we further find that eccentricities are not completely damped, but stall out at $e{\sim}0.05$. This stalling occurs as planetary spins are torqued toward synchronism at such low eccentricities, sharply limiting the energy dissipated by IWs. Such stalling at small but non-zero eccentricities ${\sim}$0.05 is formally consistent with a number of eccentric hot Neptunes within the measurement uncertainties. As an example, HAT P 26 b harbors $e{=}0.12^{+0.06}_{-0.06}$ \citep[at $a{=}0.047{\pm}0.0006$ au, $M_{\star}=0.796{\pm}0.03 \ M_{\odot}$;][]{stacolgau17,manespcov22,koktsiedw23}.

In summary, IW tidal theory may be compatible with the non-zero eccentricities of some hot Neptunes. Under IWs, low-viscosity hot Neptunes ($Ek{\sim}10^{-10}$) around KM stars may exhibit small, but non-zero $e{\sim}0.05$ indefinitely. Larger $e{\sim}0.2$ persist for ${\sim}0.5$ Gyr, a timescale that may permit observations to catch circularization in the act. Thermal atmospheric tides may help prolong this circularization time \citep[][]{corboudel20}. If planets are further assumed to rotate slowly, $e$-damping under IWs stalls out entirely.

\subsection{Future Work: non-convective interiors}\label{subsec:fw}

As discussed in Section \ref{sec:evo}, our calculations rely on the ``following the adiabats'' approach to follow planets' thermal/structural/orbital evolution. The main assumption is that Neptunian planets' atmospheres consist of a large convective zone overlaid by a thin radiative zone \citep[the conventional picture of gaseous planet structure; e.g.][]{arrbil06}. Future work should address whether non-convective layers would qualitatively change the runaway inflation mechanism we have described. We stress that our mechanism only requires a fraction of the planet to be convective: as long as a convective zone exists to host IWs, and the tidal luminosity exceeds the radiative cooling luminosity in the convective zone, we expect that layer (and those above it) to inflate, instigating the  tidal feedback loop. Interior models of Neptune that are in accordance with gravitational moment and magnetic field measurements indeed suggest the presence of sizable convection zones \citep[see e.g.][in particular their Figures 4 and 7]{morhel25}.

Furthermore, it is unclear whether radiative layers would remain 
non-convective in the presence of tidal dissipation. Radiative zones may be heated from below via core dissipation \citep[][]{der79,stolai14,petbha26}, and/or suffer heating from gravity/gravito-inertial modes that propagate within them \citep[][]{jertouogi17}. Future work generalizing our calculations to more complex interior structures is called for.

\section{Conclusion}\label{sec:conclusion}

Multiple observational lines of evidence suggest that some hot Neptunes and Jupiters are emplaced into their short-period orbits by high eccentricity migration. Motivated by these observations, this paper helps resolve what we term the ``Neptunian final parsec problem'': how do planets with modest masses drain ${\gtrsim}10{\times}$ more orbital energy than is contained in their structures? By coupling Neptunes' and gas giants' thermal/structural evolution with dissipation under inertial waves, we determined how the energy budget of a tidally migrating Neptune/gas giant evolves and, by extension, how they may survive in the face of tidal dissipation. Our main conclusions are summarized below.

\begin{itemize}
\item Evolution under IWs is dominated by resonances between tidal forcing and free inertial eigenmodes in the planet. The amount of tidal power injected at each resonant frequency is correlated with
the spatial overlap (i.e.~degree of spatial alignment) between tidal forcing and the mode's velocity field, normalized by the mode damping rate.
\item Dissipation under IWs becomes more efficient for puffier planets (those with higher entropy and/or smaller cores), and for larger viscosities that damp IWs. 
\item IWs produce deep atmospheric tidal heating (assuming atmospheres are convective), powering radius inflation.
\item Neptunes experience runaway radius inflation under IW heating; their structures are deformed by tides more strongly as they inflate, seeding a larger IW response. Runaway inflation robustly occurs across a wide range in viscosities (Ekman numbers $Ek{=}\{10^{-5},10^{-10}\}$).
\item As Neptunes undergo runaway tidal inflation, their fundamental modes sweep into resonance with tidal forcing. A small amount of tidal heat relative to a planet’s binding energy can sweep the planet’s dynamical frequency into resonance with tidal forcing. If the $f$-mode bleeds tidal energy to space rather than the planet's interior, this ``migration by inflation'' may permit Neptunes to dissipate orbital energy without catastrophically harming the planet.
\item Highly eccentric ($e{\sim}0.9$) gas giants circularize over ${\sim}$Gyr under IWs, with modest structural upheaval; giant planets successfully radiate more orbital energy than is contained in their structures. 
The survival of gas giants that migrate under IWs alleviates a long-perceived weakness of HEM theory, namely that gas giants may not survive the dissipation of more orbital energy than is contained in their structures. Such inflated gas giants avoid the tidal disruption limit en route to becoming a hot Jupiter.
\item Tidally-migrating planets may be spun into a forest of ``equilibrium''/minimum torque spin rates. There is no single ``equilibrium'' (pseudosynchronous) spin rate, in contrast to the simplistic predictions of weak friction tidal theory \citep[such as in the constant time-lag tidal model that is commonly adopted, e.g.][thereby highlighting one of its limitations]{hut81}.
\item IWs can only help circularize the orbits of Neptunes that park inside ${\lesssim}0.1$ au; a larger orbital angular momentum throttles inertial wave heating,  
as the longer periastron passage time spins planets down. Such slowly rotating planets sample fewer tidal harmonics and exhibit larger tidal quality factor $Q^{\prime}$.
\end{itemize}

Our calculations make the following observational predictions:

\begin{itemize}
\item Warm Neptunes with orbital angular momenta consistent with HEM (i.e.~with post-circularization distances $a_{\rm f}{\lesssim}$0.1 au) should be rare, as runaway inflation under IWs happens quickly (${\sim}10^{7}$ yr). In contrast, moderately eccentric warm Neptunes should be commonplace, as IW dissipation stalls 
for such slowly rotating planets.
\item If they exist, highly eccentric warm Neptunes should possess negligible rotational oblateness (for rotation periods ${\gtrsim}$2 weeks, or oblateness parameters ${\lesssim}10^{-4}$), and a low-entropy interior.
\item Assuming $f$-modes do not completely circularize Neptunes' orbits after runaway inflation, hot Neptunes orbiting KM stars may possess small but non-zero eccentricities ($e{\sim}0.05$) indefinitely in the absence of other efficient tidal mechanisms (e.g.~gravity waves in the planet's envelope or stellar tides), as IW dissipation falls drastically as planets spin down toward synchronism. 
\end{itemize}

Observational confirmation of our predictions would not only establish that HEM governs the hot Neptune population, but would also constrain the interior structures of Neptunian planets via tidal theory. 

\appendix{}\label{sec:appendix}

\section{Equations of Motion for the Tidal Flow}\label{sec:EOM}

Following the discussion in Section \ref{subsec:tidal_problem}, in this appendix we provide the full equations describing the linearized tidal response. Our treatment builds on that of \cite{ogi09,ogi13} and Vanon et al. (under review).

The hydrodynamic equations governing the tidal flow are (assuming spatially constant kinematic viscosity, rigid rotation, and dropping the non-linear advective term):
\begin{equation}\label{equation:full_fluid_eom}
\begin{split}
    \partial_{t}\boldsymbol{u}_{\rm t}+2\boldsymbol{\Omega}_{\rm s}\times \boldsymbol{u}_{\rm t}&=-\boldsymbol{\nabla}W_{\rm t}+\frac{1}{\rho_{\rm t}}\boldsymbol{\nabla}\cdot\sigma_{\rm t},\\\boldsymbol{\nabla}\cdot(\rho_{\rm t}\boldsymbol{u}_{\rm t})&=-\partial_{t}\rho_{\rm t},\\
    W_{\rm t}&=h_{\rm t}+\Phi_{\rm t}+\Psi,\\
    \sigma_{\rm t}&=\nu\rho_{\rm t}\bigg[\boldsymbol{\nabla}\boldsymbol{u}_{\rm t}+(\boldsymbol{\nabla}\boldsymbol{u}_{\rm t})^{T}-\frac{2}{3}(\boldsymbol{\nabla}\cdot\boldsymbol{u}_{\rm t})\rm I\bigg],\\
    \nabla^{2}\Phi_{\rm t}&=4\pi G\rho_{\rm t},
\end{split}
\end{equation}

\noindent where the total velocity $\boldsymbol{u}_{\rm t}{=}\boldsymbol{u}_{\rm nw}{+}\boldsymbol{u}_{\rm w}$ is composed of the non-wavelike and wavelike contributions, respectively, and both the total density $\rho_{\rm t}{=}\bar{\rho}(r){+}\rho_{\rm nw}$ and the gravitational potential $\Phi_{\rm t}{=}\bar{\Phi}(r){+}\Phi_{\rm nw}$ contain contributions from the background and non-wavelike flow, respectively. The wavelike component in the low-frequency approximation satisfies the Cowling approximation, i.e.~we neglect the wavelike gravitational potential perturbation, $\Phi_{\rm w}$ \citep{cow41}.\footnote{See also the lecture notes by J. Christensen-Dalsgaard \href{https://users-phys.au.dk/jcd/oscilnotes/Lecture_Notes_on_Stellar_Oscillations.pdf}{here} for an overview of the Cowling approximation.}  The total enthalpy is $h_{\rm t}{=}\bar{h}{+}h$ (note $dh{=}dP/\rho$ for our homentropic fluid), where $h{=}h_{\rm nw}{+}h_{\rm w}$ is the Eulerian tidal enthalpy perturbation, decomposed into its non-wavelike and wavelike parts. We similarly define $W_{\rm t}{=}\bar{h}{+}\bar{\Phi}{+}W$, where
$W{=}h{+}\Phi_{\rm nw}{+}\Psi$.

Specializing to the low-frequency limit (tidal frequency ${\ll}$ the planet's dynamical frequency), 
Equations \ref{equation:full_fluid_eom} simplify to: 
\begin{align}\label{equation:lowf_eom}
    \partial_{t}\boldsymbol{u}_{\rm t}+2\boldsymbol{\Omega}_{\rm s}\times \boldsymbol{u}_{\rm t}&=-\boldsymbol{\nabla}W+\frac{1}{\bar{\rho}}\boldsymbol{\nabla}\cdot\sigma,\\\label{equation:anelastic_lowf}\boldsymbol{\nabla}\cdot(\bar{\rho}\boldsymbol{u}_{\rm t})&=-\partial_{t}\rho_{\rm nw},\\\label{equation:HSE_pert}
    0&=h_{\rm nw}+\Phi_{\rm nw}+\Psi,\\
    \sigma&=\nu\bar{\rho}\bigg[\boldsymbol{\nabla}\boldsymbol{u}_{\rm t}+(\boldsymbol{\nabla}\boldsymbol{u}_{\rm t})^{T}-\frac{2}{3}(\boldsymbol{\nabla}\cdot\boldsymbol{u}_{\rm t})\rm I\bigg],\\
    \nabla^{2}\Phi_{\rm nw}&=4\pi G\rho_{\rm nw},
\end{align}

\noindent where Equation \ref{equation:HSE_pert} arises from the linearized hydrostatic equation in the low-frequency limit \citep[e.g.][]{ogi14}. We have used $\boldsymbol{\nabla}(\bar{h}{+}\bar{\Phi}){=}0$ under hydrostatic balance of the background state, and $\nabla^{2}\bar{\Phi}{=}4\pi G\bar{\rho}$. 

Below we describe our method of solution of Equations A2-A6, by separating the flow into wavelike and non-wavelike responses.

\subsection{Equations of Motion for Non-Wavelike Tides}

The linearized equations describing the non-wavelike response are:
\begin{align}
&\partial_{t}\boldsymbol{u}_{\rm nw}=-\boldsymbol{\nabla}W_{\rm nw},&\\
&\boldsymbol{\nabla}\cdot (\bar{\rho}\boldsymbol{u}_{\rm nw})=-\partial_{t}\rho_{\rm nw},&\\
&0=h_{\rm nw}+\Phi_{\rm nw}+\Psi,&\\
&\nabla^{2}\Phi_{\rm nw}=4\pi G \rho_{\rm nw},&
\end{align}

\noindent where $W_{\rm nw}$ is the potential associated with the acceleration of the non-wavelike flow, and we have neglected viscous damping of this non-wavelike tide (i.e.~dropped the $(1/\bar{\rho})\boldsymbol{\nabla}{\cdot}\sigma$ term where $\sigma$ contains gradients of $\boldsymbol{u}_{\rm nw}$).

Defining $\boldsymbol{u}_{\rm nw}{=}{-}\boldsymbol{\nabla}\dot{X}$ with $X$ a scalar potential field, and expanding $\dot{X}$ in spherical harmonics via $\dot{X}{=}\mathrm{Re}[\dot{X}_{\ell}(r)Y_{\ell}^{m}(\theta,\phi)e^{-\mathrm{i}\omega t}]$, these equations yield the following governing expression for $\dot{X}_{\ell}$ \citep[e.g.][]{ogi14,bar20}:
\begin{equation}
    \frac{1}{r^{2}}\frac{d}{dr}(r^{2}\bar{\rho}\frac{d\dot{X}_{\ell}}{dr})-\frac{\ell(\ell+1)}{r^{2}}\bar{\rho}\dot{X}_{\ell}=-\mathrm{i}\omega\frac{1}{g}\frac{d\bar{\rho}}{dr}\mathcal{A}_{\ell}r^{\ell}.
\end{equation}
Here we have neglected the non-wavelike potential $\Phi_{\rm nw}$ when determining the non-wavelike deformation, and we have used the enthalpy perturbation $h_{\rm nw}{=}c^{2}_{\rm s}\rho_{\rm nw}/\bar{\rho}$. The quantity $\dot{X}_{\ell}$ is subject to the boundary conditions:
\begin{align}
    \frac{d\dot{X}_{\ell}}{dr}\bigg|_{R_{c}}&=0,\\
    \frac{d\dot{X}_{\ell}}{dr}\bigg|_{\rm RCB}&=-\mathrm{i}\omega\frac{\mathcal{A}_{\ell}R_{\rm RCB}^{\ell}}{g(R_{\rm RCB})},
\end{align}

\noindent i.e., the radial component of the non-wavelike flow matches that of the conventional equilibrium tide at the outer boundary (the RCB), again neglecting the non-wavelike potential perturbation $\Phi_{\rm nw}$.

\subsection{Equations of Motion for Wavelike Tides}

The momentum equation for the wavelike flow is given by Equation \ref{equation:NS}. We decompose the velocity field $\boldsymbol{u}_{\rm w}$ into vector spherical harmonics \citep[e.g.][]{barestgir85,tre23}. Using the convention of \cite{ogi13} to decompose the velocity field into spheroidal/toroidal contributions \citep[see also][]{rieval97}, we write $\boldsymbol{u}_{\rm w}(r,\theta,\phi,t){=}\sum_{n}[a_{n}(r)Y_{n}^{m}(\theta,\phi)\boldsymbol{\hat{r}}+r^{2}b_{n}(r)\boldsymbol{\nabla}Y^{m}_{n}(\theta,\phi)+r^{2}c_{n}(r)(\boldsymbol{\hat{r}}\times\boldsymbol{\nabla}Y^{m}_{n}(\theta,\phi))]\mathrm{e}^{-i\omega t}$. The velocity field $\boldsymbol{u}_\mathrm{w}$ contains the following spherical polar components, and we also expand the pressure \citep[where we drop the subscript $t$;][]{ogi09}:
\begin{equation}\label{equation:expansions}
\begin{split}
    u_{r}&=\sum_{n} a_{n}(r)Y^{m}_{n}(\theta,\phi)e^{-\mathrm{i}\omega t},\\
    u_{\theta}&=r\sum_{n}\bigg[b_{n}(r)\partial_{\theta}+\frac{c_{n}(r)}{\sin\theta}\partial_{\phi}\bigg]Y^{m}_{n}(\theta,\phi)e^{-\mathrm{i}\omega t},\\
    u_{\phi}&=r\sum_{n}\bigg[\frac{b_{n}(r)}{\sin \theta}\partial_{\phi}-c_{n}(r)\partial_{\theta}\bigg]Y^{m}_{n}(\theta,\phi)e^{-\mathrm{i}\omega t},\\
    W_{\mathrm{w}}&=\sum_{n} W_{n}(r)Y^{m}_{n}(\theta,\phi)e^{-\mathrm{i}\omega t}.
\end{split}
\end{equation}

\noindent With this projection of the velocity field and $W$, the anelastic approximation (Equation \ref{equation:anelastic_lowf}) reads (for a single $n$ component):
\begin{equation}\label{equation:anelastic_comp}
    \frac{1}{r^{2}}\partial_{r}(r^{2}a_{n}(r))-n(n+1)b_n(r)+(\partial_{r}\ln\bar{\rho})a_{n}=0,
\end{equation}

\noindent where we have used $\nabla^{2}_{\perp}Y^{m}_{n}{=}{-}n(n{+}1)r^{-2}Y^{m}_{n}$, with $\nabla^{2}_{\perp}{=}(r^{2}\sin^{2}\theta)^{-1}[\sin\theta\partial_{\theta}(\sin\theta\partial_{\theta})+\partial^{2}_{\phi}]$ the horizontal component of the Laplacian in spherical coordinates \citep[e.g.][]{unn79}.

Inserting expansions \ref{equation:expansions} into Equations \ref{equation:NS} and \ref{equation:NS2}, and applying Equation \ref{equation:anelastic_comp}, the radial and horizontal components of the momentum equation are \citep[generalizing Equations 19-24 of][]{ogi09}:
\begin{flalign*}
    & -\mathrm{i}\omega a_{n}-2\mathrm{i}\Omega_{\rm s}mr{b_{n}}+2\Omega_{\rm s}[(n-1)q_{n}rc_{n-1}-(n+2)q_{n+1}rc_{n+1}]
    =&\\
    &\hspace{1cm} -\frac{dW_{n}}{dr}-\frac{1}{3}\frac{\nu}{r}\bigg[\bigg(\frac{8+3n(n+1)}{r}+4\partial_{r} \ln \bar{\rho}\bigg)a_{n}-2n(n+1)(3+r\partial_{r}\ln\bar{\rho} )b_{n}\bigg]+\frac{4}{3}\nu\bigg[\bigg(\frac{2}{r}+\partial_{r} \ln\bar{\rho}\bigg)\frac{da_{n}}{dr}-\frac{1}{4}n(n+1)\frac{d b_{n}}{dr}\bigg]&\\
    &\hspace{1cm}+\frac{4}{3}\nu\partial_{r}^{2}a_{n}-2\Omega_{\rm s}\omega m \frac{X}{r}\delta_{n2}.&\\
&-\mathrm{i}\omega r^{2}b_{n}
-\frac{2\mathrm{i}\Omega_{\rm s}mr}{n(n+1)}(a_{n}+rb_{n})
+2\Omega_{\rm s}r^{2}
\left[(n-1)\frac{q_{n}}{n}c_{n-1}
+q_{n+1}\frac{(n+2)}{n+1}c_{n+1}\right]
=&\\
&\hspace{1cm}
-W_{n}
+\frac{8}{3}\frac{\nu}{r}a_{n}
+\nu\left[
(\partial_{r}\ln\bar{\rho})a_{n}
+\frac{1}{3}\partial_{r}a_{n}
+\left(2-\frac{4}{3}n(n+1)\right)b_{n}
+r(4+r\partial_{r}\ln\bar{\rho})\partial_{r}b_{n}
+r^{2}\partial^{2}_{r}b_{n}
\right]&\\
&\hspace{1cm}
-2\Omega_{\rm s}\omega
\frac{m}{n(n+1)}
\left(r\frac{dX}{dr}+X\right)\delta_{n2}.&\\
&-\mathrm{i}\omega r^{2}c_{n}
-\frac{2\mathrm{i}\Omega_{\rm s}mr^{2}c_{n}}{n(n+1)}
+2\Omega_{\rm s}r
\left[
\frac{q_{n}}{n}\left(a_{n-1}-(n-1)rb_{n-1}\right)
-\frac{q_{n+1}}{n+1}\left(a_{n+1}-(n+2)rb_{n+1}\right)
\right]
=&\\
&\hspace{1cm}
(2-n-n^{2})\nu c_{n}
+r\nu\left(4+r\partial_{r}\ln\bar{\rho}\right)\partial_{r}c_{n}
+r^{2}\nu\partial^{2}_{r}c_{n}
-\mathrm{i}\Omega_{\rm s}\omega m
\frac{q_{n}}{n}
\left(r\frac{dX}{dr}-2X\right)\delta_{n3}.&
\end{flalign*}

\noindent where we have defined
\begin{flalign*}
q_{n}=\sqrt{\frac{n^{2}-m^{2}}{4n^{2}-1}},\quad\text{and}\quad 
q_{n+1}=\sqrt{\frac{(n+1)^{2}-m^{2}}{4(n+1)^{2}-1}}.
\end{flalign*}

\noindent We assemble this system of equations into matrix form for use in our linear algebra solver \citep[e.g.][]{rieval97}.

Our impenetrable and stress-free top and bottom boundary conditions correspond to $a_{n}|_{\rm boundary}{=}0$, $db_{n}/dr|_{\rm boundary}{=}dc_{n}/dr|_{\rm boundary}{=}0$.

\section{Tidal Evolution at Low Viscosity}\label{sec:low_Ek}

\begin{figure*}[t]
\centering
\includegraphics[width=1\textwidth]{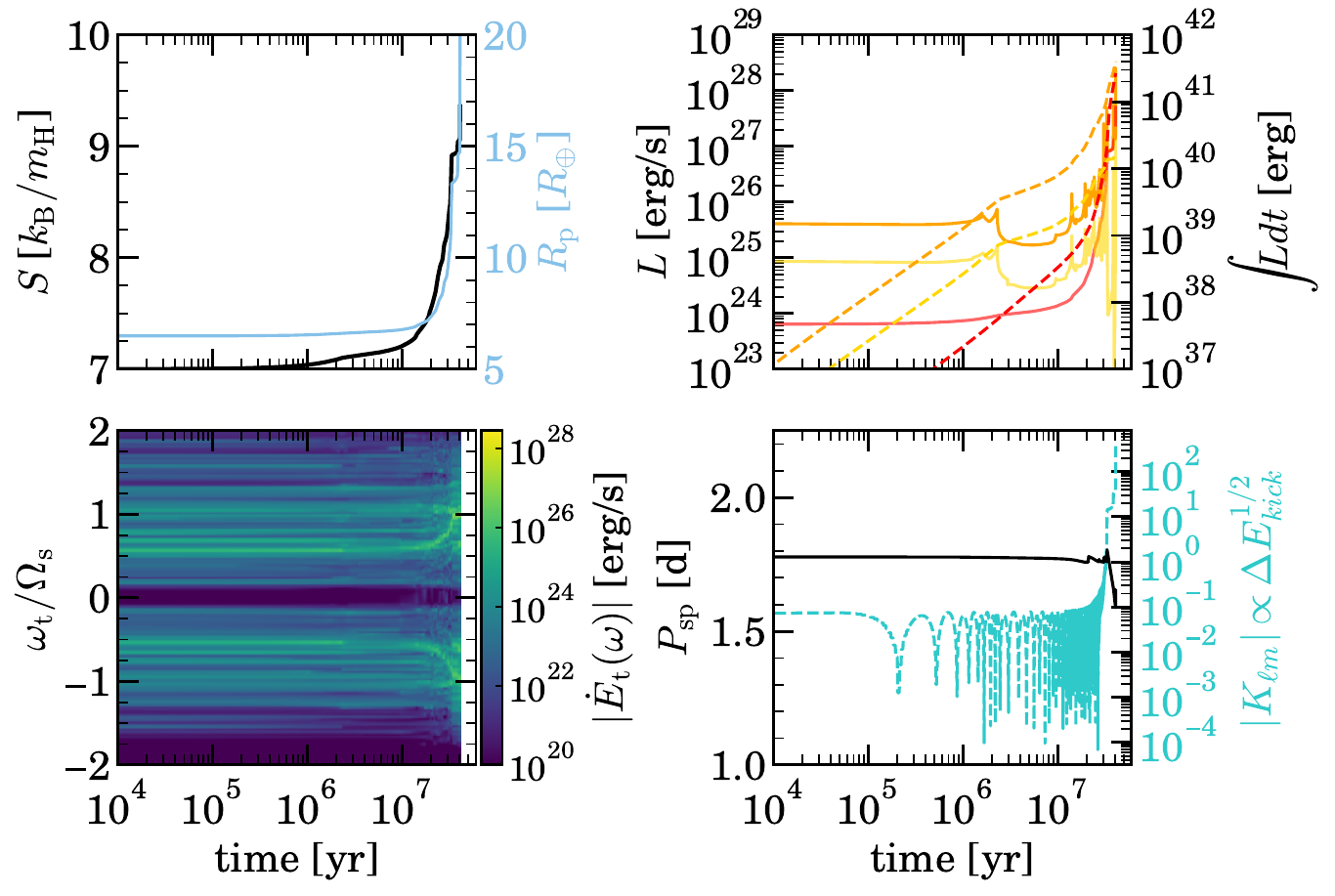}
\caption{The same as Figure \ref{figure:evo1}, except using a low planet viscosity (Ekman number $Ek{=}10^{-10}$). Runaway inflation ensues, despite the much finer structure in the low viscosity planet's inertial wave dissipation spectrum.
\label{figure:evo6}}
\end{figure*}

In Figure \ref{figure:evo6} we display the evolution of our fiducial Neptunian planet, but with our lower choice of viscosity ($Ek{=}10^{-10}$). Features remain largely the same as in Figure \ref{figure:evo1} with $Ek{=}10^{-5}$, except for finer structure evident in the tidal dissipation spectrum.

\section{A Forest of Spin Equilibria}

\begin{figure*}[t]
\centering
\includegraphics[width=1\textwidth]{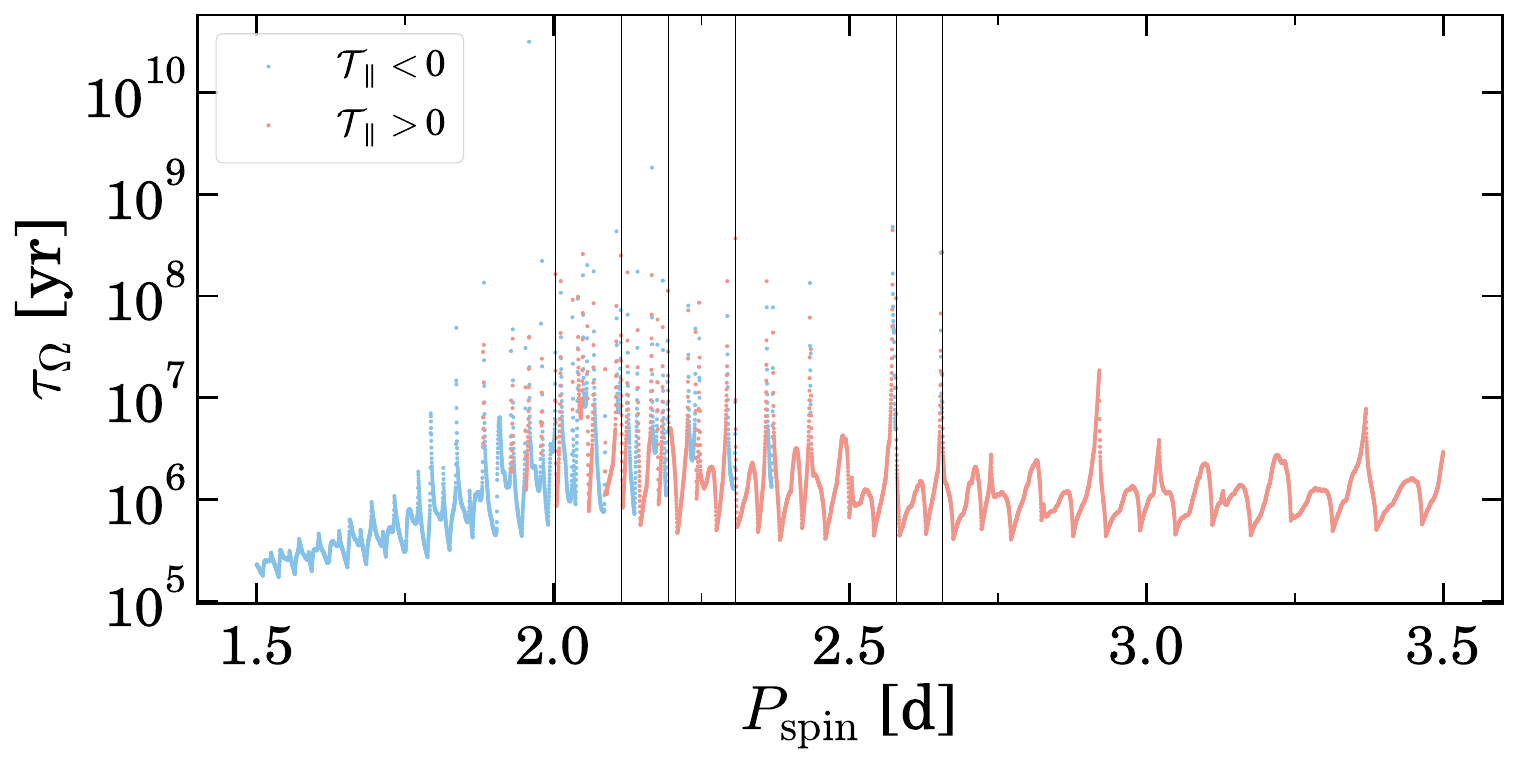}
\caption{The tidal torque timescale $\tau_{\Omega}{=}\Omega_{\rm spin}/|\dot{\Omega}_{\rm spin}|$, where $\dot{\Omega}_{\rm spin}{=}\mathcal{T}_{\parallel}/I_{\rm p}$, as a function of the planet's spin period $P_{\rm spin}$, for our fiducial initial conditions ($a{=}0.3$ au, $e{=}0.896, S{=}7 \ k_{\rm B}/m_{\rm H}$, $Ek{=}10^{-5}$). Spin periods that yield a negative torque ($\mathcal{T}_{\parallel}{<}0$; spin-down to longer periods) are shown in blue, whereas those with positive torque (spin-up) are in red. Vertical black lines denote the location of stable spin equilibria ($\mathcal{T}_{\parallel}{\sim}0$, $d\mathcal{T}_{\parallel}/dP_{\rm spin}{>}0$) with a tidal torque timescale that exceeds 10$^{8}$ yr; i.e., the deepest equilibria. Inertial waves produce a forest of local spin equilibria.
\label{figure:spin_equilibria}}
\end{figure*}

\begin{figure*}[t]
\centering
\includegraphics[width=1\textwidth]{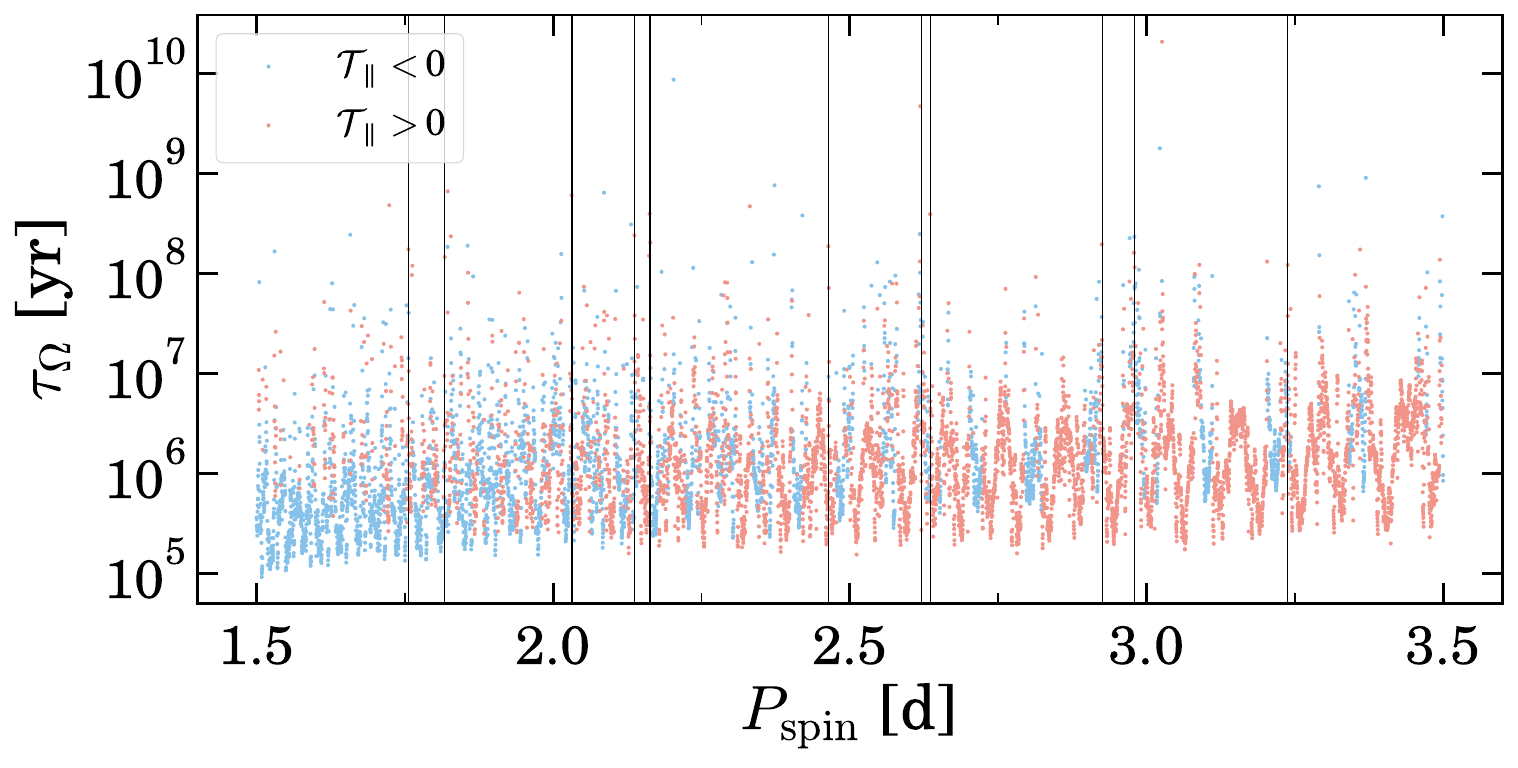}
\caption{The same as Figure \ref{figure:spin_equilibria}, except using a low viscosity ($Ek{=}10^{-10}$). Fine structure in the tidal response produces more spin equilibria distributed over a wider range in spin period.
\label{figure:spin_equilibria2}}
\end{figure*}

In Figures \ref{figure:spin_equilibria} and \ref{figure:spin_equilibria2}, we illustrate the tidal torque timescale $\tau_{\Omega}{=}\Omega_{\rm spin}/|\dot{\Omega}_{\rm spin}|$ as a function of planet spin period, for our low and high choices of viscosity ($Ek{=}10^{-5},10^{-10}$ in Figures \ref{figure:spin_equilibria} and \ref{figure:spin_equilibria2}, respectively). A forest of local spin equilibria is produced in both cases of low and high viscosity, in contrast to the single ``equilibrium'' spin period predicted in weak friction tidal theory \citep[][]{hut81}. The planet's spin rate stops evolving whenever the torque timescale becomes long.

\acknowledgments

T.H. thanks Vincent Bourrier, Adam Burrows, Andrew Cumming, Jeremy Goodman, Ravit Helled, Doug Lin, Morgan MacLeod, David Nesvorný, James Owen, Mor Rozner, Scott Tremaine, Michelle Vick, Shreyas Vissapragada, and J.J. Zanazzi for helpful discussions. T.H. acknowledges the use of Claude Code for translating code to python, and for aid with $\texttt{matplotlib}$. T.H. futher thanks the organizers of ``The formation, life and fate of exo-Neptunian systems'' workshop in Leiden, N.L. The authors acknowledge the MIT Office of Research Computing and Data for providing high performance computing resources that have contributed to the research results reported in this paper. AJB was supported by STFC grants ST/W000873/1 and UKRI1179. This material is based upon work supported by the National Science Foundation under grant No. 2306391. S.C.M. acknowledges support from an Alfred P. Sloan Research Fellowship. 

\software{Numerical integrations were carried out with \texttt{scipy} \citep[][]{virgomoli20} and \texttt{numpy} \citep[][]{har20}, while plots were created using \texttt{matplotlib} \citep[][]{hun07}.}

\bibliography{hallatt}{}

@ARTICLE{Nils2023,
       author = {{de Vries}, Nils B. and {Barker}, Adrian J. and {Hollerbach}, Rainer},
        title = "{Tidal dissipation due to the elliptical instability and turbulent viscosity in convection zones in rotating giant planets and stars}",
      journal = {\mnras},
         year = 2023,
        month = sep,
       volume = {524},
       number = {2},
        pages = {2661-2683},
          doi = {10.1093/mnras/stad1990},
archivePrefix = {arXiv},
       eprint = {2306.17622},
 primaryClass = {astro-ph.EP},
       adsurl = {https://ui.adsabs.harvard.edu/abs/2023MNRAS.524.2661D}
}

@ARTICLE{gunquedem17,
       author = {{G{\"u}nther}, Maximilian N. and {Queloz}, Didier and {Demory}, Brice-Olivier and {Bouchy}, Francois},
        title = "{A new yield simulator for transiting planets and false positives: application to the Next Generation Transit Survey}",
      journal = {\mnras},
         year = 2017,
        month = mar,
       volume = {465},
       number = {3},
        pages = {3379-3389},
          doi = {10.1093/mnras/stw2908},
archivePrefix = {arXiv},
       eprint = {1611.02526},
 primaryClass = {astro-ph.EP},
       adsurl = {https://ui.adsabs.harvard.edu/abs/2017MNRAS.465.3379G}
}

@ARTICLE{sou26,
       author = {{Southworth}, John},
        title = "{TEPCat: The Transiting Extrasolar Planet Catalogue}",
      journal = {Universe},
         year = 2026,
        month = feb,
       volume = {12},
       number = {3},
          eid = {62},
        pages = {62},
          doi = {10.3390/universe12030062},
archivePrefix = {arXiv},
       eprint = {2602.23102},
 primaryClass = {astro-ph.EP},
       adsurl = {https://ui.adsabs.harvard.edu/abs/2026Univ...12...62S}
}

@ARTICLE{petbha26,
       author = {{Petrovich}, Cristobal and {Bhaskar}, Hareesh Gautham},
        title = "{Thermally Regulated Viscoelastic Tidal Migration of Eccentric Planets}",
      journal = {\apjl},
         year = 2026,
        month = aug,
       volume = {1007},
       number = {2},
          eid = {L38},
        pages = {L38},
          doi = {10.3847/2041-8213/ae909d},
archivePrefix = {arXiv},
       eprint = {2606.14983},
 primaryClass = {astro-ph.EP},
       adsurl = {https://ui.adsabs.harvard.edu/abs/2026ApJ..1007L..38P}
}

@ARTICLE{ivapap04,
       author = {{Ivanov}, P.~B. and {Papaloizou}, J.~C.~B.},
        title = "{On the tidal interaction of massive extrasolar planets on highly eccentric orbits}",
      journal = {\mnras},
         year = 2004,
        month = jan,
       volume = {347},
       number = {2},
        pages = {437-453},
          doi = {10.1111/j.1365-2966.2004.07238.x},
archivePrefix = {arXiv},
       eprint = {astro-ph/0303669},
 primaryClass = {astro-ph},
       adsurl = {https://ui.adsabs.harvard.edu/abs/2004MNRAS.347..437I}
}

@INPROCEEDINGS{milmer03,
       author = {{Milosavljevi{\'c}}, Milo{\v{s}} and {Merritt}, David},
        title = "{The Final Parsec Problem}",
    booktitle = {The Astrophysics of Gravitational Wave Sources},
         year = 2003,
       editor = {{Centrella}, Joan M.},
       series = {American Institute of Physics Conference Series},
       volume = {686},
        month = oct,
    publisher = {AIP},
        pages = {201-210},
          doi = {10.1063/1.1629432},
archivePrefix = {arXiv},
       eprint = {astro-ph/0212270},
 primaryClass = {astro-ph},
       adsurl = {https://ui.adsabs.harvard.edu/abs/2003AIPC..686..201M}
}

@ARTICLE{espwinbra26,
       author = {{Espinoza-Retamal}, Juan I. and {Winn}, Joshua N. and {Brahm}, Rafael and {Petrovich}, Cristobal and {Stef{\'a}nsson}, Gu{\dj}mundur and {Bhaskar}, Hareesh and {Koo}, Elise and {Jord{\'a}n}, Andr{\'e}s and {Tala Pinto}, Marcelo and {Hobson}, Melissa J. and {Veldhuis}, Hugo and {Rojas}, Felipe I. and {Teske}, Johanna K. and {Butler}, R. Paul and {Crane}, Jeffrey D. and {Shectman}, Stephen and {Vissapragada}, Shreyas and {Boyle}, Gavin and {Leiva}, Rodrigo and {Suc}, Vincent},
        title = "{POSEIDON I: The Dynamical Origins of Transiting Neptunes}",
      journal = {\aj},
         year = 2026,
        month = jul,
       volume = {172},
       number = {1},
          eid = {57},
        pages = {57},
          doi = {10.3847/1538-3881/ae75da},
archivePrefix = {arXiv},
       eprint = {2602.18553},
 primaryClass = {astro-ph.EP},
       adsurl = {https://ui.adsabs.harvard.edu/abs/2026AJ....172...57E}
}

@ARTICLE{verful19,
       author = {{Veras}, Dimitri and {Fuller}, Jim},
        title = "{Tidal circularization of gaseous planets orbiting white dwarfs}",
      journal = {\mnras},
         year = 2019,
        month = oct,
       volume = {489},
       number = {2},
        pages = {2941-2953},
          doi = {10.1093/mnras/stz2339},
archivePrefix = {arXiv},
       eprint = {1908.08052},
 primaryClass = {astro-ph.EP},
       adsurl = {https://ui.adsabs.harvard.edu/abs/2019MNRAS.489.2941V}
}

@ARTICLE{teylaivic19,
       author = {{Teyssandier}, Jean and {Lai}, Dong and {Vick}, Michelle},
        title = "{Formation of hot Jupiters through secular chaos and dynamical tides}",
      journal = {\mnras},
         year = 2019,
        month = jun,
       volume = {486},
       number = {2},
        pages = {2265-2280},
          doi = {10.1093/mnras/stz1011},
archivePrefix = {arXiv},
       eprint = {1901.05006},
 primaryClass = {astro-ph.EP},
       adsurl = {https://ui.adsabs.harvard.edu/abs/2019MNRAS.486.2265T}
}

@ARTICLE{dew24,
       author = {{Dewberry}, Janosz W.},
        title = "{Tidal Synchronization Trapping in Stars and Planets with Convective Envelopes}",
      journal = {\apj},
         year = 2024,
        month = may,
       volume = {966},
       number = {2},
          eid = {180},
        pages = {180},
          doi = {10.3847/1538-4357/ad344d},
archivePrefix = {arXiv},
       eprint = {2403.06979},
 primaryClass = {astro-ph.SR},
       adsurl = {https://ui.adsabs.harvard.edu/abs/2024ApJ...966..180D}
}

@ARTICLE{anlubra25,
       author = {{An}, Qier and {Lu}, Tiger and {Brandt}, G. Mirek and {Brandt}, Timothy D. and {Li}, Gongjie},
        title = "{Significant Mutual Inclinations Between the Stellar Spin and the Orbits of Both Planets in the HAT-P-11 System}",
      journal = {\aj},
         year = 2025,
        month = jan,
       volume = {169},
       number = {1},
          eid = {22},
        pages = {22},
          doi = {10.3847/1538-3881/ad90b4},
archivePrefix = {arXiv},
       eprint = {2405.19510},
 primaryClass = {astro-ph.EP},
       adsurl = {https://ui.adsabs.harvard.edu/abs/2025AJ....169...22A}
}

@ARTICLE{thoniecri26,
       author = {{Thomas}, Luis and {Nielsen}, Louise D. and {Cridland}, Alex J. and {Schlecker}, Martin and {Vach}, Sydney},
        title = "{Companion Architectures of Sub-Saturns: Distinct Migration Pathways Across the Neptunian Landscape}",
      journal = {arXiv e-prints},
         year = 2026,
        month = jul,
          eid = {arXiv:2607.09451},
        pages = {arXiv:2607.09451},
          doi = {10.48550/arXiv.2607.09451},
archivePrefix = {arXiv},
       eprint = {2607.09451},
 primaryClass = {astro-ph.EP},
       adsurl = {https://ui.adsabs.harvard.edu/abs/2026arXiv260709451T}
}

@ARTICLE{albdawwin22,
       author = {{Albrecht}, Simon H. and {Dawson}, Rebekah I. and {Winn}, Joshua N.},
        title = "{Stellar Obliquities in Exoplanetary Systems}",
      journal = {\pasp},
         year = 2022,
        month = aug,
       volume = {134},
       number = {1038},
          eid = {082001},
        pages = {082001},
          doi = {10.1088/1538-3873/ac6c09},
archivePrefix = {arXiv},
       eprint = {2203.05460},
 primaryClass = {astro-ph.EP},
       adsurl = {https://ui.adsabs.harvard.edu/abs/2022PASP..134h2001A}
}

@ARTICLE{fanliu26,
       author = {{Fan}, Qianli and {Liu}, Shang-Fei},
        title = "{Where Do Hot Jupiters Come from? Revisiting Tidal Disruption and Ejection in High-eccentricity Migration}",
      journal = {\apjl},
         year = 2026,
        month = may,
       volume = {1003},
       number = {1},
          eid = {L7},
        pages = {L7},
          doi = {10.3847/2041-8213/ae6659},
archivePrefix = {arXiv},
       eprint = {2605.14433},
 primaryClass = {astro-ph.EP},
       adsurl = {https://ui.adsabs.harvard.edu/abs/2026ApJ..1003L...7F}
}

@ARTICLE{welhannao26,
       author = {{Weldon}, Grant C. and {Hansen}, Bradley M.~S. and {Naoz}, Smadar},
        title = "{Saving Doomed Planets: Mass Loss and Angular Momentum Return Boost Hot Jupiter Survival Rates}",
      journal = {\apj},
         year = 2026,
        month = apr,
       volume = {1001},
       number = {2},
          eid = {139},
        pages = {139},
          doi = {10.3847/1538-4357/ae548c},
archivePrefix = {arXiv},
       eprint = {2510.26882},
 primaryClass = {astro-ph.EP},
       adsurl = {https://ui.adsabs.harvard.edu/abs/2026ApJ..1001..139W}
}

@ARTICLE{surtejsu25,
       author = {{Sur}, Ankan and {Tejada Arevalo}, Roberto and {Su}, Yubo and {Burrows}, Adam},
        title = "{Simultaneous Evolutionary Fits for Jupiter and Saturn Incorporating Fuzzy Cores}",
      journal = {\apjl},
         year = 2025,
        month = feb,
       volume = {980},
       number = {1},
          eid = {L5},
        pages = {L5},
          doi = {10.3847/2041-8213/adad62},
archivePrefix = {arXiv},
       eprint = {2412.17127},
 primaryClass = {astro-ph.EP},
       adsurl = {https://ui.adsabs.harvard.edu/abs/2025ApJ...980L...5S}
}

@ARTICLE{morhel25,
       author = {{Morf}, Luca and {Helled}, Ravit},
        title = "{Icy or rocky? Convective or stable?: New interior models of Uranus and Neptune}",
      journal = {\aap},
         year = 2025,
        month = dec,
       volume = {704},
          eid = {A183},
        pages = {A183},
          doi = {10.1051/0004-6361/202556911},
archivePrefix = {arXiv},
       eprint = {2510.00175},
 primaryClass = {astro-ph.EP},
       adsurl = {https://ui.adsabs.harvard.edu/abs/2025A&A...704A.183M}
}

@ARTICLE{chr08,
       author = {{Christensen-Dalsgaard}, J{\o}rgen},
        title = "{ADIPLS{\textemdash}the Aarhus adiabatic oscillation package}",
      journal = {\apss},
         year = 2008,
        month = aug,
       volume = {316},
       number = {1-4},
        pages = {113-120},
          doi = {10.1007/s10509-007-9689-z},
archivePrefix = {arXiv},
       eprint = {0710.3106},
 primaryClass = {astro-ph},
       adsurl = {https://ui.adsabs.harvard.edu/abs/2008Ap&SS.316..113C}
}

@ARTICLE{webdav67,
       author = {{Weber}, Edmund J. and {Davis}, Jr., Leverett},
        title = "{The Angular Momentum of the Solar Wind}",
      journal = {\apj},
         year = 1967,
        month = apr,
       volume = {148},
        pages = {217-227},
          doi = {10.1086/149138},
       adsurl = {https://ui.adsabs.harvard.edu/abs/1967ApJ...148..217W}
}

@ARTICLE{halmil26b,
       author = {{Hallatt}, Tim and {Millholland}, Sarah},
        title = "{Shedding Light on Desert Dwellers}",
      journal = {\apj},
         year = 2026,
        month = feb,
       volume = {997},
       number = {2},
          eid = {139},
        pages = {139},
          doi = {10.3847/1538-4357/adfb75},
archivePrefix = {arXiv},
       eprint = {2509.22893},
 primaryClass = {astro-ph.EP},
       adsurl = {https://ui.adsabs.harvard.edu/abs/2026ApJ...997..139H}
}

@Article{hun07,
  Author    = {Hunter, J. D.},
  Title     = {Matplotlib: A 2D graphics environment},
  Journal   = {Computing in Science \& Engineering},
  Volume    = {9},
  Number    = {3},
  Pages     = {90--95},
  publisher = {IEEE COMPUTER SOC},
  doi       = {10.1109/MCSE.2007.55},
  year      = 2007
}

@Article{har20,
 title         = {Array programming with {NumPy}},
 author        = {Charles R. Harris and K. Jarrod Millman and St{\'{e}}fan J.
                 van der Walt and Ralf Gommers and Pauli Virtanen and David
                 Cournapeau and Eric Wieser and Julian Taylor and Sebastian
                 Berg and Nathaniel J. Smith and Robert Kern and Matti Picus
                 and Stephan Hoyer and Marten H. van Kerkwijk and Matthew
                 Brett and Allan Haldane and Jaime Fern{\'{a}}ndez del
                 R{\'{i}}o and Mark Wiebe and Pearu Peterson and Pierre
                 G{\'{e}}rard-Marchant and Kevin Sheppard and Tyler Reddy and
                 Warren Weckesser and Hameer Abbasi and Christoph Gohlke and
                 Travis E. Oliphant},
 year          = {2020},
 month         = sep,
 journal       = {Nature},
 volume        = {585},
 number        = {7825},
 pages         = {357--362},
 doi           = {10.1038/s41586-020-2649-2},
 publisher     = {Springer Science and Business Media {LLC}},
 url           = {https://doi.org/10.1038/s41586-020-2649-2}
}

@ARTICLE{riegeoval01,
       author = {{Rieutord}, M. and {Georgeot}, B. and {Valdettaro}, L.},
        title = "{Inertial waves in a rotating spherical shell: attractors and asymptotic spectrum}",
      journal = {Journal of Fluid Mechanics},
         year = 2001,
        month = may,
       volume = {435},
       number = {1},
        pages = {103-144},
          doi = {10.1017/S0022112001003718},
archivePrefix = {arXiv},
       eprint = {physics/0007007},
 primaryClass = {physics.flu-dyn},
       adsurl = {https://ui.adsabs.harvard.edu/abs/2001JFM...435..103R}
}

@ARTICLE{almcasmar26,
       author = {{Almenara}, J.~M. and {Castro-Gonz{\'a}lez}, A. and {Mardling}, R. and {Guillot}, T. and {Bonfils}, X. and {Thomas}, L. and {Gauvrit}, L. and {Abe}, L. and {Agabi}, A. and {Aubert}, J. and {Bouchy}, F. and {Cadieux}, C. and {Deline}, A. and {D{\'\i}az}, R.~F. and {Ehrenreich}, D. and {Forveille}, T. and {Heitzmann}, A. and {Hirling}, P. and {Hoogenboom}, R.~M. and {Leleu}, A. and {Lendl}, M. and {M{\'e}karnia}, D. and {Pelletier}, S. and {Psaridi}, A. and {Rea}, E. and {Song}, X. and {Su{\'a}rez}, O. and {Triaud}, A.~H.~M.~J. and {Udry}, S.},
        title = "{HD 148797: A bright F-type star with two moderate-period low-density sub-Jovian planets. Compact multi-planet architectures are common in the Neptunian savanna}",
      journal = {arXiv e-prints},
         year = 2026,
        month = jul,
          eid = {arXiv:2607.09656},
        pages = {arXiv:2607.09656},
          doi = {10.48550/arXiv.2607.09656},
archivePrefix = {arXiv},
       eprint = {2607.09656},
 primaryClass = {astro-ph.EP},
       adsurl = {https://ui.adsabs.harvard.edu/abs/2026arXiv260709656A}
}

@ARTICLE{manespcov22,
       author = {{Mancini}, L. and {Esposito}, M. and {Covino}, E. and {Southworth}, J. and {Poretti}, E. and {Andreuzzi}, G. and {Barbato}, D. and {Biazzo}, K. and {Borsato}, L. and {Bruni}, I. and {Damasso}, M. and {Di Fabrizio}, L. and {Evans}, D.~F. and {Granata}, V. and {Lanza}, A.~F. and {Naponiello}, L. and {Nascimbeni}, V. and {Pinamonti}, M. and {Sozzetti}, A. and {Tregloan-Reed}, J. and {Basilicata}, M. and {Bignamini}, A. and {Bonomo}, A.~S. and {Claudi}, R. and {Cosentino}, R. and {Desidera}, S. and {Fiorenzano}, A.~F.~M. and {Giacobbe}, P. and {Harutyunyan}, A. and {Henning}, Th. and {Knapic}, C. and {Maggio}, A. and {Micela}, G. and {Molinari}, E. and {Pagano}, I. and {Pedani}, M. and {Piotto}, G.},
        title = "{The GAPS Programme at TNG. XXXVI. Measurement of the Rossiter-McLaughlin effect and revising the physical and orbital parameters of the HAT-P-15, HAT-P-17, HAT-P-21, HAT-P-26, HAT-P-29 eccentric planetary systems}",
      journal = {\aap},
         year = 2022,
        month = aug,
       volume = {664},
          eid = {A162},
        pages = {A162},
          doi = {10.1051/0004-6361/202243742},
archivePrefix = {arXiv},
       eprint = {2205.10549},
 primaryClass = {astro-ph.EP},
       adsurl = {https://ui.adsabs.harvard.edu/abs/2022A&A...664A.162M}
}

@ARTICLE{stacolgau17,
       author = {{Stassun}, Keivan G. and {Collins}, Karen A. and {Gaudi}, B. Scott},
        title = "{Accurate Empirical Radii and Masses of Planets and Their Host Stars with Gaia Parallaxes}",
      journal = {\aj},
         year = 2017,
        month = mar,
       volume = {153},
       number = {3},
          eid = {136},
        pages = {136},
          doi = {10.3847/1538-3881/aa5df3},
archivePrefix = {arXiv},
       eprint = {1609.04389},
 primaryClass = {astro-ph.EP},
       adsurl = {https://ui.adsabs.harvard.edu/abs/2017AJ....153..136S}
}

@ARTICLE{ivapap07,
       author = {{Ivanov}, P.~B. and {Papaloizou}, J.~C.~B.},
        title = "{Dynamic tides in rotating objects: orbital circularization of extrasolar planets for realistic planet models}",
      journal = {\mnras},
         year = 2007,
        month = apr,
       volume = {376},
       number = {2},
        pages = {682-704},
          doi = {10.1111/j.1365-2966.2007.11463.x},
archivePrefix = {arXiv},
       eprint = {astro-ph/0512150},
 primaryClass = {astro-ph},
       adsurl = {https://ui.adsabs.harvard.edu/abs/2007MNRAS.376..682I}
}

@ARTICLE{angnaopet22,
       author = {{Angelo}, Isabel and {Naoz}, Smadar and {Petigura}, Erik and {MacDougall}, Mason and {Stephan}, Alexander P. and {Isaacson}, Howard and {Howard}, Andrew W.},
        title = "{Kepler-1656b's Extreme Eccentricity: Signature of a Gentle Giant}",
      journal = {\aj},
         year = 2022,
        month = may,
       volume = {163},
       number = {5},
          eid = {227},
        pages = {227},
          doi = {10.3847/1538-3881/ac6094},
archivePrefix = {arXiv},
       eprint = {2204.00019},
 primaryClass = {astro-ph.EP},
       adsurl = {https://ui.adsabs.harvard.edu/abs/2022AJ....163..227A}
}

@ARTICLE{sinrustho24,
       author = {{Sing}, David K. and {Rustamkulov}, Zafar and {Thorngren}, Daniel P. and {Barstow}, Joanna K. and {Tremblin}, Pascal and {Alves de Oliveira}, Catarina and {Beck}, Tracy L. and {Birkmann}, Stephan M. and {Challener}, Ryan C. and {Crouzet}, Nicolas and {Espinoza}, N{\'e}stor and {Ferruit}, Pierre and {Giardino}, Giovanna and {Gressier}, Am{\'e}lie and {Lee}, Elspeth K.~H. and {Lewis}, Nikole K. and {Maiolino}, Roberto and {Manjavacas}, Elena and {Rauscher}, Bernard J. and {Sirianni}, Marco and {Valenti}, Jeff A.},
        title = "{A warm Neptune's methane reveals core mass and vigorous atmospheric mixing}",
      journal = {\nat},
         year = 2024,
        month = jun,
       volume = {630},
       number = {8018},
        pages = {831-835},
          doi = {10.1038/s41586-024-07395-z},
archivePrefix = {arXiv},
       eprint = {2405.11027},
 primaryClass = {astro-ph.EP},
       adsurl = {https://ui.adsabs.harvard.edu/abs/2024Natur.630..831S}
}

@ARTICLE{corboudel20,
       author = {{Correia}, A.~C.~M. and {Bourrier}, V. and {Delisle}, J.-B.},
        title = "{Why do warm Neptunes present nonzero eccentricity?}",
      journal = {\aap},
         year = 2020,
        month = mar,
       volume = {635},
          eid = {A37},
        pages = {A37},
          doi = {10.1051/0004-6361/201936967},
archivePrefix = {arXiv},
       eprint = {2005.02700},
 primaryClass = {astro-ph.EP},
       adsurl = {https://ui.adsabs.harvard.edu/abs/2020A&A...635A..37C}
}

@ARTICLE{ragwol09,
       author = {{Ragozzine}, Darin and {Wolf}, Aaron S.},
        title = "{Probing the Interiors of very Hot Jupiters Using Transit Light Curves}",
      journal = {\apj},
         year = 2009,
        month = jun,
       volume = {698},
       number = {2},
        pages = {1778-1794},
          doi = {10.1088/0004-637X/698/2/1778},
archivePrefix = {arXiv},
       eprint = {0807.2856},
 primaryClass = {astro-ph},
       adsurl = {https://ui.adsabs.harvard.edu/abs/2009ApJ...698.1778R}
}

@ARTICLE{mar95b,
       author = {{Mardling}, Rosemary A.},
        title = "{The Role of Chaos in the Circularization of Tidal Capture Binaries. II. Long-Time Evolution}",
      journal = {\apj},
         year = 1995,
        month = sep,
       volume = {450},
        pages = {732},
          doi = {10.1086/176179},
       adsurl = {https://ui.adsabs.harvard.edu/abs/1995ApJ...450..732M}
}

@ARTICLE{mar95,
       author = {{Mardling}, Rosemary A.},
        title = "{The Role of Chaos in the Circularization of Tidal Capture Binaries. I. The Chaos Boundary}",
      journal = {\apj},
         year = 1995,
        month = sep,
       volume = {450},
        pages = {722},
          doi = {10.1086/176178},
       adsurl = {https://ui.adsabs.harvard.edu/abs/1995ApJ...450..722M}
}

@ARTICLE{fainabhua24,
       author = {{Fairnington}, Tyler R. and {Nabbie}, Emma and {Huang}, Chelsea X. and {Zhou}, George and {Foo}, Orion and {Millholland}, Sarah and {Wright}, Duncan and {Belinski}, Alexandre A. and {Bieryla}, Allyson and {Ciardi}, David R. and {Collins}, Karen A. and {Collins}, Kevin I. and {Everett}, Mark and {Howell}, Steve B. and {Lissauer}, Jack J. and {Lund}, Michael B. and {Murgas}, Felipe and {Palle}, Enric and {Quinn}, Samuel N. and {Relles}, Howard M. and {Safonov}, Boris S. and {Schwarz}, Richard P. and {Scott}, Nicholas J. and {Srdoc}, Gregor and {Ricker}, George and {Vanderspek}, Roland and {Seager}, Sara and {Latham}, David W. and {Winn}, Joshua W. and {Jenkins}, Jon M. and {Bouma}, Luke G. and {Shporer}, Avi and {Ting}, Eric B. and {Dragomir}, Diana and {Kunimoto}, Michelle and {Eisner}, Nora L.},
        title = "{TOI-5126: a hot super-Neptune and warm Neptune pair discovered by TESS and CHEOPS}",
      journal = {\mnras},
         year = 2024,
        month = jan,
       volume = {527},
       number = {3},
        pages = {8768-8783},
          doi = {10.1093/mnras/stad3036},
archivePrefix = {arXiv},
       eprint = {2310.08890},
 primaryClass = {astro-ph.EP},
       adsurl = {https://ui.adsabs.harvard.edu/abs/2024MNRAS.527.8768F}
}

@ARTICLE{epsbhawin26,
       author = {{Espinoza-Retamal}, Juan I. and {Bhaskar}, Hareesh and {Winn}, Joshua N. and {Petrovich}, Cristobal and {Brahm}, Rafael and {Lammers}, Caleb and {Stef{\'a}nsson}, Gu{\dj}mundur and {Koo}, Elise and {Jord{\'a}n}, Andr{\'e}s and {Rojas}, Felipe I.},
        title = "{POSEIDON. II. The Antialigned Orbit of the Warm Neptune TOI-1710 A b}",
      journal = {\apjl},
         year = 2026,
        month = jul,
       volume = {1005},
       number = {1},
          eid = {L15},
        pages = {L15},
          doi = {10.3847/2041-8213/ae7711},
archivePrefix = {arXiv},
       eprint = {2604.03364},
 primaryClass = {astro-ph.EP},
       adsurl = {https://ui.adsabs.harvard.edu/abs/2026ApJ..1005L..15E}
}

@ARTICLE{kenulmgil26,
       author = {{Kendall}, Alicia and {Ulmer-Moll}, Sol{\`e}ne and {Gill}, Samuel and {Burleigh}, Matthew R. and {Goad}, Michael R. and {Anderson}, David R. and {Bryant}, Edward M. and {Lavie}, Baptiste and {Bugatti}, Maddalena and {Acevedo Barroso}, Javier A. and {Steiner}, Michal and {Dragomir}, Diana and {Villanueva}, Jr., Steven and {Stevens}, Daniel J. and {Gupta}, Arvind F. and {Gaudi}, Scott and {Sun}, Guoyou and {Claringbold}, Alastair and {Doyle}, Lauren and {Guillot}, Tristan and {Suarez}, Olga and {M{\'e}karnia}, Djamel and {Triaud}, Amaury H.~M.~J. and {Bendjoya}, Philippe and {Ziegler}, Carl and {Mann}, Andrew W. and {Howell}, Steve B. and {Fajardo-Acosta}, Sergio B. and {Littlefield}, Colin and {Caldwell}, Douglas A. and {Kunimoto}, Michelle and {Rowden}, Pamela and {Kostov}, Veselin and {Villase{\~n}or}, Jesus Noel and {Alves}, Douglas and {Apergis}, Ioannis and {Armstrong}, David J. and {Battley}, Matthew P. and {Bayliss}, Daniel and {Bouchy}, Fran{\c{c}}ois and {Casewell}, Sarah L. and {G{\"u}nther}, Maximilian N. and {Harvey}, George T. and {Hawthorn}, Faith and {Jenkins}, James S. and {Lendl}, Monika and {McCormac}, James and {Moyano}, Maximilano and {Nielsen}, Louise D. and {Osborn}, Ares and {Rodel}, Toby and {Saha}, Suman and {Udry}, Stephane and {Vines}, Jose I. and {Wheatley}, Peter J. and {Zivave}, Tafadzwa},
        title = "{A 43 d transiting Neptune and two 25 d Saturns from TESS, NGTS, and ASTEP}",
      journal = {\mnras},
         year = 2026,
        month = apr,
       volume = {547},
       number = {2},
          eid = {staf2189},
        pages = {staf2189},
          doi = {10.1093/mnras/staf2189},
archivePrefix = {arXiv},
       eprint = {2512.07716},
 primaryClass = {astro-ph.EP},
       adsurl = {https://ui.adsabs.harvard.edu/abs/2026MNRAS.547f2189K}
}

@ARTICLE{donhuadaw21,
       author = {{Dong}, Jiayin and {Huang}, Chelsea X. and {Dawson}, Rebekah I. and {Foreman-Mackey}, Daniel and {Collins}, Karen A. and {Quinn}, Samuel N. and {Lissauer}, Jack J. and {Beatty}, Thomas and {Quarles}, Billy and {Sha}, Lizhou and {Shporer}, Avi and {Guo}, Zhao and {Kane}, Stephen R. and {Abe}, Lyu and {Barkaoui}, Khalid and {Benkhaldoun}, Zouhair and {Brahm}, Rafael and {Bouchy}, Fran{\c{c}}ois and {Carmichael}, Theron W. and {Collins}, Kevin I. and {Conti}, Dennis M. and {Crouzet}, Nicolas and {Dransfield}, Georgina and {Evans}, Phil and {Gan}, Tianjun and {Ghachoui}, Mourad and {Gillon}, Micha{\"e}l and {Grieves}, Nolan and {Guillot}, Tristan and {Hellier}, Coel and {Jehin}, Emmanu{\"e}l and {Jensen}, Eric L.~N. and {Jord{\'a}n}, Andres and {Kamler}, Jacob and {Kielkopf}, John F. and {M{\'e}karnia}, Djamel and {Nielsen}, Louise D. and {Pozuelos}, Francisco J. and {Radford}, Don J. and {Schmider}, Fran{\c{c}}ois-Xavier and {Schwarz}, Richard P. and {Stockdale}, Chris and {Tan}, Thiam-Guan and {Timmermans}, Mathilde and {Triaud}, Amaury H.~M.~J. and {Wang}, Gavin and {Ricker}, George and {Vanderspek}, Roland and {Latham}, David W. and {Seager}, Sara and {Winn}, Joshua N. and {Jenkins}, Jon M. and {Mireles}, Ismael and {Yahalomi}, Daniel A. and {Morgan}, Edward H. and {Vezie}, Michael and {Quintana}, Elisa V. and {Rose}, Mark E. and {Smith}, Jeffrey C. and {Shiao}, Bernie},
        title = "{Warm Jupiters in TESS Full-frame Images: A Catalog and Observed Eccentricity Distribution for Year 1}",
      journal = {\apjs},
         year = 2021,
        month = jul,
       volume = {255},
       number = {1},
          eid = {6},
        pages = {6},
          doi = {10.3847/1538-4365/abf73c},
archivePrefix = {arXiv},
       eprint = {2104.01970},
 primaryClass = {astro-ph.EP},
       adsurl = {https://ui.adsabs.harvard.edu/abs/2021ApJS..255....6D}
}

@ARTICLE{dawmur13,
       author = {{Dawson}, Rebekah I. and {Murray-Clay}, Ruth A.},
        title = "{Giant Planets Orbiting Metal-rich Stars Show Signatures of Planet-Planet Interactions}",
      journal = {\apjl},
         year = 2013,
        month = apr,
       volume = {767},
       number = {2},
          eid = {L24},
        pages = {L24},
          doi = {10.1088/2041-8205/767/2/L24},
archivePrefix = {arXiv},
       eprint = {1302.6244},
 primaryClass = {astro-ph.EP},
       adsurl = {https://ui.adsabs.harvard.edu/abs/2013ApJ...767L..24D}
}

@ARTICLE{donhuazho21,
       author = {{Dong}, Jiayin and {Huang}, Chelsea X. and {Zhou}, George and {Dawson}, Rebekah I. and {Rodriguez}, Joseph E. and {Eastman}, Jason D. and {Collins}, Karen A. and {Quinn}, Samuel N. and {Shporer}, Avi and {Triaud}, Amaury H.~M.~J. and {Wang}, Songhu and {Beatty}, Thomas and {Jackson}, Jonathon M. and {Collins}, Kevin I. and {Abe}, Lyu and {Suarez}, Olga and {Crouzet}, Nicolas and {M{\'e}karnia}, Djamel and {Dransfield}, Georgina and {Jensen}, Eric L.~N. and {Stockdale}, Chris and {Barkaoui}, Khalid and {Heitzmann}, Alexis and {Wright}, Duncan J. and {Addison}, Brett C. and {Wittenmyer}, Robert A. and {Okumura}, Jack and {Bowler}, Brendan P. and {Horner}, Jonathan and {Kane}, Stephen R. and {Kielkopf}, John and {Liu}, Huigen and {Plavchan}, Peter and {Mengel}, Matthew W. and {Ricker}, George R. and {Vanderspek}, Roland and {Latham}, David W. and {Seager}, S. and {Winn}, Joshua N. and {Jenkins}, Jon M. and {Christiansen}, Jessie L. and {Paegert}, Martin},
        title = "{TOI-3362b: A Proto Hot Jupiter Undergoing High-eccentricity Tidal Migration}",
      journal = {\apjl},
         year = 2021,
        month = oct,
       volume = {920},
       number = {1},
          eid = {L16},
        pages = {L16},
          doi = {10.3847/2041-8213/ac2600},
archivePrefix = {arXiv},
       eprint = {2109.03771},
 primaryClass = {astro-ph.EP},
       adsurl = {https://ui.adsabs.harvard.edu/abs/2021ApJ...920L..16D}
}

@ARTICLE{bouefr19,
       author = {{Bou{\'e}}, Gwena{\"e}l and {Efroimsky}, Michael},
        title = "{Tidal evolution of the Keplerian elements}",
      journal = {Celestial Mechanics and Dynamical Astronomy},
         year = 2019,
        month = jul,
       volume = {131},
       number = {7},
          eid = {30},
        pages = {30},
          doi = {10.1007/s10569-019-9908-2},
archivePrefix = {arXiv},
       eprint = {1904.02253},
 primaryClass = {astro-ph.EP},
       adsurl = {https://ui.adsabs.harvard.edu/abs/2019CeMDA.131...30B}
}

@ARTICLE{gol63,
       author = {{Goldreich}, P.},
        title = "{On the eccentricity of satellite orbits in the solar system}",
      journal = {\mnras},
         year = 1963,
        month = jan,
       volume = {126},
        pages = {257},
          doi = {10.1093/mnras/126.3.257},
       adsurl = {https://ui.adsabs.harvard.edu/abs/1963MNRAS.126..257G}
}

@ARTICLE{polsme90,
       author = {{Polfliet}, R. and {Smeyers}, P.},
        title = "{Dynamic tides in stars as forced isentropic oscillations and their effects on free oscillations.}",
      journal = {\aap},
         year = 1990,
        month = oct,
       volume = {237},
        pages = {110-124},
       adsurl = {https://ui.adsabs.harvard.edu/abs/1990A&A...237..110P}
}

@ARTICLE{zholai26,
       author = {{Zhou}, Hongzhe and {Lai}, Dong},
        title = "{Efficiency of Tidal Dissipation in Convective Flow Under Rapid Tidal Forcing}",
      journal = {arXiv e-prints},
         year = 2026,
        month = jul,
          eid = {arXiv:2607.14637},
        pages = {arXiv:2607.14637},
          doi = {10.48550/arXiv.2607.14637},
archivePrefix = {arXiv},
       eprint = {2607.14637},
 primaryClass = {astro-ph.SR},
       adsurl = {https://ui.adsabs.harvard.edu/abs/2026arXiv260714637Z}
}

@ARTICLE{hug81,
       author = {{Hughes}, S.},
        title = "{The Computation of Tables of Hansen Coefficients}",
      journal = {Celestial Mechanics},
         year = 1981,
        month = sep,
       volume = {25},
       number = {1},
        pages = {101-107},
          doi = {10.1007/BF01301812},
       adsurl = {https://ui.adsabs.harvard.edu/abs/1981CeMec..25..101H}
}

@BOOK{tre23,
       author = {{Tremaine}, Scott},
        title = "{Dynamics of Planetary Systems}",
         year = 2023,
         publisher = "{Princeton University Press}",
       adsurl = {https://ui.adsabs.harvard.edu/abs/2023dyps.book.....T}
}

@ARTICLE{kau61,
       author = {{Kaula}, W.~M.},
        title = "{Analysis of Gravitational and Geometric Aspects of Geodetic Utilization of Satellites}",
      journal = {Geophysical Journal},
         year = 1961,
        month = jul,
       volume = {5},
       number = {2},
        pages = {104-133},
          doi = {10.1111/j.1365-246X.1961.tb00417.x},
       adsurl = {https://ui.adsabs.harvard.edu/abs/1961GeoJ....5..104K}
}

@ARTICLE{gilsti26,
       author = {{Gilmore}, Travis and {Stixrude}, Lars},
        title = "{Core─envelope miscibility in sub-Neptunes and super-Earths}",
      journal = {\nat},
         year = 2026,
        month = feb,
       volume = {650},
       number = {8100},
        pages = {60-64},
          doi = {10.1038/s41586-025-09970-4},
       adsurl = {https://ui.adsabs.harvard.edu/abs/2026Natur.650...60G}
}

@ARTICLE{golkee77,
       author = {{Goldreich}, P. and {Keeley}, D.~A.},
        title = "{Solar seismology. I. The stability of the solar p-modes.}",
      journal = {\apj},
         year = 1977,
        month = feb,
       volume = {211},
        pages = {934-942},
          doi = {10.1086/155005},
       adsurl = {https://ui.adsabs.harvard.edu/abs/1977ApJ...211..934G}
}

@ARTICLE{zah89,
       author = {{Zahn}, J.-P.},
        title = "{Tidal evolution of close binary stars. I - Revisiting the theory of the equilibrium tide}",
      journal = {\aap},
         year = 1989,
        month = aug,
       volume = {220},
       number = {1-2},
        pages = {112-116},
       adsurl = {https://ui.adsabs.harvard.edu/abs/1989A&A...220..112Z}
}

@INPROCEEDINGS{zah08,
       author = {{Zahn}, J.-P.},
        title = "{Tidal dissipation in binary systems}",
    booktitle = {EAS Publications Series},
         year = 2008,
       editor = {{Goupil}, M.-J. and {Zahn}, J.-P.},
       series = {EAS Publications Series},
       volume = {29},
        month = jan,
    publisher = {EDP},
        pages = {67-90},
          doi = {10.1051/eas:0829002},
archivePrefix = {arXiv},
       eprint = {0807.4870},
 primaryClass = {astro-ph},
       adsurl = {https://ui.adsabs.harvard.edu/abs/2008EAS....29...67Z}
}

@ARTICLE{zah66,
       author = {{Zahn}, J.~P.},
        title = "{Les mar{\'e}es dans une {\'e}toile double serr{\'e}e}",
      journal = {\anap},
         year = 1966,
        month = feb,
       volume = {29},
        pages = {313},
       adsurl = {https://ui.adsabs.harvard.edu/abs/1966AnAp...29..313Z}
}

@BOOK{bat00,
       author = {{Batchelor}, G.~K.},
        title = "{An Introduction to Fluid Dynamics}",
         year = 2000,
         publisher = {Cambridge University Press},
       adsurl = {https://ui.adsabs.harvard.edu/abs/2000ifd..book.....B}
}

@ARTICLE{terpapnel98,
       author = {{Terquem}, C. and {Papaloizou}, J.~C.~B. and {Nelson}, R.~P. and {Lin}, D.~N.~C.},
        title = "{On the Tidal Interaction of a Solar-Type Star with an Orbiting Companion: Excitation of g-Mode Oscillation and Orbital Evolution}",
      journal = {\apj},
         year = 1998,
        month = aug,
       volume = {502},
       number = {2},
        pages = {788-801},
          doi = {10.1086/305927},
archivePrefix = {arXiv},
       eprint = {astro-ph/9801280},
 primaryClass = {astro-ph},
       adsurl = {https://ui.adsabs.harvard.edu/abs/1998ApJ...502..788T}
}

@ARTICLE{goodic98,
       author = {{Goodman}, Jeremy and {Dickson}, Eric S.},
        title = "{Dynamical Tide in Solar-Type Binaries}",
      journal = {\apj},
         year = 1998,
        month = nov,
       volume = {507},
       number = {2},
        pages = {938-944},
          doi = {10.1086/306348},
archivePrefix = {arXiv},
       eprint = {astro-ph/9801289},
 primaryClass = {astro-ph},
       adsurl = {https://ui.adsabs.harvard.edu/abs/1998ApJ...507..938G}
}

@ARTICLE{bolbra13,
       author = {{Bolmatov}, Dima and {Brazhkin}, V.~V. and {Trachenko}, K.},
        title = "{Thermodynamic behaviour of supercritical matter}",
      journal = {Nature Communications},
         year = 2013,
        month = aug,
       volume = {4},
          eid = {2331},
        pages = {2331},
          doi = {10.1038/ncomms3331},
archivePrefix = {arXiv},
       eprint = {1303.3153},
 primaryClass = {physics.chem-ph},
       adsurl = {https://ui.adsabs.harvard.edu/abs/2013NatCo...4.2331B}
}

@ARTICLE{halmil26,
       author = {{Hallatt}, Tim and {Millholland}, Sarah},
        title = "{Coupled Planetary Interior and Tidal Evolution}",
      journal = {\apj},
         year = 2026,
        month = feb,
       volume = {997},
       number = {2},
          eid = {138},
        pages = {138},
          doi = {10.3847/1538-4357/ae129d},
archivePrefix = {arXiv},
       eprint = {2509.22923},
 primaryClass = {astro-ph.EP},
       adsurl = {https://ui.adsabs.harvard.edu/abs/2026ApJ...997..138H}
}

@BOOK{preflateu86,
       author = {{Press}, William H. and {Flannery}, Brian P. and {Teukolsky}, Saul A. and {Vetterling}, William T.},
        title = "{Numerical recipes: The art of scientific computing}",
         year = 1986,
         publisher = "{Cambridge University Press}",
       adsurl = {https://ui.adsabs.harvard.edu/abs/1986nras.book.....P}
}

@ARTICLE{dugbarjon20,
       author = {{Duguid}, Craig D. and {Barker}, Adrian J. and {Jones}, C.~A.},
        title = "{Convective turbulent viscosity acting on equilibrium tidal flows: new frequency scaling of the effective viscosity}",
      journal = {\mnras},
         year = 2020,
        month = sep,
       volume = {497},
       number = {3},
        pages = {3400-3417},
          doi = {10.1093/mnras/staa2216},
archivePrefix = {arXiv},
       eprint = {2007.12624},
 primaryClass = {astro-ph.EP},
       adsurl = {https://ui.adsabs.harvard.edu/abs/2020MNRAS.497.3400D}
}

@ARTICLE{vidbar20,
       author = {{Vidal}, J{\'e}r{\'e}mie and {Barker}, Adrian J.},
        title = "{Turbulent Viscosity Acting on the Equilibrium Tidal Flow in Convective Stars}",
      journal = {\apjl},
         year = 2020,
        month = jan,
       volume = {888},
       number = {2},
          eid = {L31},
        pages = {L31},
          doi = {10.3847/2041-8213/ab6219},
archivePrefix = {arXiv},
       eprint = {1912.07910},
 primaryClass = {astro-ph.SR},
       adsurl = {https://ui.adsabs.harvard.edu/abs/2020ApJ...888L..31V}
}

@ARTICLE{bar20,
       author = {{Barker}, A.~J.},
        title = "{Tidal dissipation in evolving low-mass and solar-type stars with predictions for planetary orbital decay}",
      journal = {\mnras},
         year = 2020,
        month = oct,
       volume = {498},
       number = {2},
        pages = {2270-2294},
          doi = {10.1093/mnras/staa2405},
archivePrefix = {arXiv},
       eprint = {2008.03262},
 primaryClass = {astro-ph.EP},
       adsurl = {https://ui.adsabs.harvard.edu/abs/2020MNRAS.498.2270B}
}

@ARTICLE{ogi13,
       author = {{Ogilvie}, Gordon I.},
        title = "{Tides in rotating barotropic fluid bodies: the contribution of inertial waves and the role of internal structure}",
      journal = {\mnras},
         year = 2013,
        month = feb,
       volume = {429},
       number = {1},
        pages = {613-632},
          doi = {10.1093/mnras/sts362},
archivePrefix = {arXiv},
       eprint = {1211.0837},
 primaryClass = {astro-ph.EP},
       adsurl = {https://ui.adsabs.harvard.edu/abs/2013MNRAS.429..613O}
}

@ARTICLE{Dewberry2023,
       author = {{Dewberry}, Janosz W.},
        title = "{Dynamical tides in Jupiter and other rotationally flattened planets and stars with stable stratification}",
      journal = {\mnras},
         year = 2023,
        month = jun,
       volume = {521},
       number = {4},
        pages = {5991-6004},
          doi = {10.1093/mnras/stad546},
archivePrefix = {arXiv},
       eprint = {2301.07097},
 primaryClass = {astro-ph.EP},
       adsurl = {https://ui.adsabs.harvard.edu/abs/2023MNRAS.521.5991D}
}

@ARTICLE{Lin2023,
       author = {{Lin}, Yufeng},
        title = "{Dynamical tides in Jupiter and the role of interior structure}",
      journal = {\aap},
         year = 2023,
        month = mar,
       volume = {671},
          eid = {A37},
        pages = {A37},
          doi = {10.1051/0004-6361/202245112},
archivePrefix = {arXiv},
       eprint = {2301.02418},
 primaryClass = {astro-ph.EP},
       adsurl = {https://ui.adsabs.harvard.edu/abs/2023A&A...671A..37L}
}

@ARTICLE{2024Dhouib,
       author = {{Dhouib}, H. and {Baruteau}, C. and {Mathis}, S. and {Debras}, F. and {Astoul}, A. and {Rieutord}, M.},
        title = "{Hydrodynamic modelling of dynamical tide dissipation in Jupiter's interior as revealed by Juno}",
      journal = {\aap},
         year = 2024,
        month = feb,
       volume = {682},
          eid = {A85},
        pages = {A85},
          doi = {10.1051/0004-6361/202347703},
archivePrefix = {arXiv},
       eprint = {2311.03288},
 primaryClass = {astro-ph.EP},
       adsurl = {https://ui.adsabs.harvard.edu/abs/2024A&A...682A..85D}
}

@ARTICLE{ogi09,
       author = {{Ogilvie}, Gordon I.},
        title = "{Tidal dissipation in rotating fluid bodies: a simplified model}",
      journal = {\mnras},
         year = 2009,
        month = jun,
       volume = {396},
       number = {2},
        pages = {794-806},
          doi = {10.1111/j.1365-2966.2009.14814.x},
archivePrefix = {arXiv},
       eprint = {0903.4103},
 primaryClass = {astro-ph.EP},
       adsurl = {https://ui.adsabs.harvard.edu/abs/2009MNRAS.396..794O}
}

@ARTICLE{BDL2014,
       author = {{Barker}, Adrian J. and {Dempsey}, Adam M. and {Lithwick}, Yoram},
        title = "{Theory and Simulations of Rotating Convection}",
      journal = {\apj},
         year = 2014,
        month = aug,
       volume = {791},
       number = {1},
          eid = {13},
        pages = {13},
          doi = {10.1088/0004-637X/791/1/13},
archivePrefix = {arXiv},
       eprint = {1403.7207},
 primaryClass = {astro-ph.SR},
       adsurl = {https://ui.adsabs.harvard.edu/abs/2014ApJ...791...13B}
}

@ARTICLE{cow41,
       author = {{Cowling}, T.~G.},
        title = "{The non-radial oscillations of polytropic stars}",
      journal = {\mnras},
         year = 1941,
        month = jan,
       volume = {101},
        pages = {367},
          doi = {10.1093/mnras/101.8.367},
       adsurl = {https://ui.adsabs.harvard.edu/abs/1941MNRAS.101..367C}
}

@ARTICLE{ponbarhol23,
       author = {{Pontin}, Christina M. and {Barker}, Adrian J. and {Hollerbach}, Rainer},
        title = "{Tidal Dissipation in Stratified and Semi-convective Regions of Giant Planets}",
      journal = {\apj},
         year = 2023,
        month = jun,
       volume = {950},
       number = {2},
          eid = {176},
        pages = {176},
          doi = {10.3847/1538-4357/accd67},
archivePrefix = {arXiv},
       eprint = {2304.11898},
 primaryClass = {astro-ph.EP},
       adsurl = {https://ui.adsabs.harvard.edu/abs/2023ApJ...950..176P}
}

@ARTICLE{ponbarhol24,
       author = {{Pontin}, Christina M. and {Barker}, Adrian J. and {Hollerbach}, Rainer},
        title = "{Tidal Dissipation in Stably Stratified and Semiconvective Regions of Rotating Giant Planets: Incorporating Coriolis Forces}",
      journal = {\apj},
         year = 2024,
        month = jan,
       volume = {960},
       number = {1},
          eid = {32},
        pages = {32},
          doi = {10.3847/1538-4357/ad0a90},
archivePrefix = {arXiv},
       eprint = {2311.03273},
 primaryClass = {astro-ph.EP},
       adsurl = {https://ui.adsabs.harvard.edu/abs/2024ApJ...960...32P}
}

@ARTICLE{sulai22,
       author = {{Su}, Yubo and {Lai}, Dong},
        title = "{Dynamical tides in eccentric binaries containing massive main-sequence stars: analytical expressions}",
      journal = {\mnras},
         year = 2022,
        month = mar,
       volume = {510},
       number = {4},
        pages = {4943-4951},
          doi = {10.1093/mnras/stab3698},
archivePrefix = {arXiv},
       eprint = {2110.12030},
 primaryClass = {astro-ph.SR},
       adsurl = {https://ui.adsabs.harvard.edu/abs/2022MNRAS.510.4943S}
}

@BOOK{unn79,
       author = {{Unno}, W. and {Osaki}, Y. and {Ando}, H. and {Shibahashi}, H.},
        title = "{Nonradial oscillations of stars}",
         year = 1979,
         publisher = "{University of Tokyo Press}",
       adsurl = {https://ui.adsabs.harvard.edu/abs/1979nos..book.....U}
}

@BOOK{gre68,
       author = {{Greenspan}, H. P.},
        title = "{The Theory of Rotating Fluids}",
         year = 1968,
         publisher = "{Cambridge University Press}",
       adsurl = {}
}

@ARTICLE{lazbarvri24,
       author = {{Lazovik}, Yaroslav A. and {Barker}, Adrian J. and {de Vries}, Nils B. and {Astoul}, Aur{\'e}lie},
        title = "{Tidal dissipation in rotating and evolving giant planets with application to exoplanet systems}",
      journal = {\mnras},
         year = 2024,
        month = jan,
       volume = {527},
       number = {3},
        pages = {8245-8256},
          doi = {10.1093/mnras/stad3689},
archivePrefix = {arXiv},
       eprint = {2311.15815},
 primaryClass = {astro-ph.EP},
       adsurl = {https://ui.adsabs.harvard.edu/abs/2024MNRAS.527.8245L}
}

@ARTICLE{viclai19,
       author = {{Vick}, Michelle and {Lai}, Dong and {Anderson}, Kassandra R.},
        title = "{Chaotic tides in migrating gas giants: forming hot and transient warm Jupiters via Lidov-Kozai migration}",
      journal = {\mnras},
         year = 2019,
        month = apr,
       volume = {484},
       number = {4},
        pages = {5645-5668},
          doi = {10.1093/mnras/stz354},
archivePrefix = {arXiv},
       eprint = {1812.05618},
 primaryClass = {astro-ph.EP},
       adsurl = {https://ui.adsabs.harvard.edu/abs/2019MNRAS.484.5645V}
}

@ARTICLE{yuweiarr22,
       author = {{Yu}, Hang and {Weinberg}, Nevin N. and {Arras}, Phil},
        title = "{Tidal Evolution and Diffusive Growth During High-eccentricity Planet Migration: Revisiting the Eccentricity Distribution of Hot Jupiters}",
      journal = {\apj},
         year = 2022,
        month = apr,
       volume = {928},
       number = {2},
          eid = {140},
        pages = {140},
          doi = {10.3847/1538-4357/ac5627},
archivePrefix = {arXiv},
       eprint = {2111.04649},
 primaryClass = {astro-ph.EP},
       adsurl = {https://ui.adsabs.harvard.edu/abs/2022ApJ...928..140Y}
}

@ARTICLE{quaferkas16,
       author = {{Quataert}, Eliot and {Fern{\'a}ndez}, Rodrigo and {Kasen}, Daniel and {Klion}, Hannah and {Paxton}, Bill},
        title = "{Super-Eddington stellar winds driven by near-surface energy deposition}",
      journal = {\mnras},
         year = 2016,
        month = may,
       volume = {458},
       number = {2},
        pages = {1214-1233},
          doi = {10.1093/mnras/stw365},
archivePrefix = {arXiv},
       eprint = {1509.06370},
 primaryClass = {astro-ph.SR},
       adsurl = {https://ui.adsabs.harvard.edu/abs/2016MNRAS.458.1214Q}
}

@ARTICLE{lopformil12,
       author = {{Lopez}, Eric D. and {Fortney}, Jonathan J. and {Miller}, Neil},
        title = "{How Thermal Evolution and Mass-loss Sculpt Populations of Super-Earths and Sub-Neptunes: Application to the Kepler-11 System and Beyond}",
      journal = {\apj},
         year = 2012,
        month = dec,
       volume = {761},
       number = {1},
          eid = {59},
        pages = {59},
          doi = {10.1088/0004-637X/761/1/59},
archivePrefix = {arXiv},
       eprint = {1205.0010},
 primaryClass = {astro-ph.EP},
       adsurl = {https://ui.adsabs.harvard.edu/abs/2012ApJ...761...59L}
}

@ARTICLE{fabtre07,
       author = {{Fabrycky}, Daniel and {Tremaine}, Scott},
        title = "{Shrinking Binary and Planetary Orbits by Kozai Cycles with Tidal Friction}",
      journal = {\apj},
         year = 2007,
        month = nov,
       volume = {669},
       number = {2},
        pages = {1298-1315},
          doi = {10.1086/521702},
archivePrefix = {arXiv},
       eprint = {0705.4285},
 primaryClass = {astro-ph},
       adsurl = {https://ui.adsabs.harvard.edu/abs/2007ApJ...669.1298F}
}

@ARTICLE{beanes12,
       author = {{Beaug{\'e}}, C. and {Nesvorn{\'y}}, D.},
        title = "{Multiple-planet Scattering and the Origin of Hot Jupiters}",
      journal = {\apj},
         year = 2012,
        month = jun,
       volume = {751},
       number = {2},
          eid = {119},
        pages = {119},
          doi = {10.1088/0004-637X/751/2/119},
archivePrefix = {arXiv},
       eprint = {1110.4392},
 primaryClass = {astro-ph.EP},
       adsurl = {https://ui.adsabs.harvard.edu/abs/2012ApJ...751..119B}
}

@ARTICLE{stolai14,
       author = {{Storch}, Natalia I. and {Lai}, Dong},
        title = "{Viscoelastic tidal dissipation in giant planets and formation of hot Jupiters through high-eccentricity migration}",
      journal = {\mnras},
         year = 2014,
        month = feb,
       volume = {438},
       number = {2},
        pages = {1526-1534},
          doi = {10.1093/mnras/stt2292},
archivePrefix = {arXiv},
       eprint = {1308.4968},
 primaryClass = {astro-ph.EP},
       adsurl = {https://ui.adsabs.harvard.edu/abs/2014MNRAS.438.1526S}
}

@ARTICLE{paxbildot11,
       author = {{Paxton}, Bill and {Bildsten}, Lars and {Dotter}, Aaron and
         {Herwig}, Falk and {Lesaffre}, Pierre and {Timmes}, Frank},
        title = "{Modules for Experiments in Stellar Astrophysics (MESA)}",
      journal = {\apjs},
         year = 2011,
        month = jan,
       volume = {192},
       number = {1},
          eid = {3},
        pages = {3},
          doi = {10.1088/0067-0049/192/1/3},
archivePrefix = {arXiv},
       eprint = {1009.1622},
 primaryClass = {astro-ph.SR},
       adsurl = {https://ui.adsabs.harvard.edu/abs/2011ApJS..192....3P}
}

@ARTICLE{paxcanarr13,
       author = {{Paxton}, Bill and {Cantiello}, Matteo and {Arras}, Phil and
         {Bildsten}, Lars and {Brown}, Edward F. and {Dotter}, Aaron and
         {Mankovich}, Christopher and {Montgomery}, M.~H. and {Stello}, Dennis and
         {Timmes}, F.~X. and {Townsend}, Richard},
        title = "{Modules for Experiments in Stellar Astrophysics (MESA): Planets, Oscillations, Rotation, and Massive Stars}",
      journal = {\apjs},
         year = 2013,
        month = sep,
       volume = {208},
       number = {1},
          eid = {4},
        pages = {4},
          doi = {10.1088/0067-0049/208/1/4},
archivePrefix = {arXiv},
       eprint = {1301.0319},
 primaryClass = {astro-ph.SR},
       adsurl = {https://ui.adsabs.harvard.edu/abs/2013ApJS..208....4P}
}

@ARTICLE{paxmarsch15,
       author = {{Paxton}, Bill and {Marchant}, Pablo and {Schwab}, Josiah and
         {Bauer}, Evan B. and {Bildsten}, Lars and {Cantiello}, Matteo and
         {Dessart}, Luc and {Farmer}, R. and {Hu}, H. and {Langer}, N. and
         {Townsend}, R.~H.~D. and {Townsley}, Dean M. and {Timmes}, F.~X.},
        title = "{Modules for Experiments in Stellar Astrophysics (MESA): Binaries, Pulsations, and Explosions}",
      journal = {\apjs},
         year = 2015,
        month = sep,
       volume = {220},
       number = {1},
          eid = {15},
        pages = {15},
          doi = {10.1088/0067-0049/220/1/15},
archivePrefix = {arXiv},
       eprint = {1506.03146},
 primaryClass = {astro-ph.SR},
       adsurl = {https://ui.adsabs.harvard.edu/abs/2015ApJS..220...15P}
}

@ARTICLE{paxschbau18,
       author = {{Paxton}, Bill and {Schwab}, Josiah and {Bauer}, Evan B. and
         {Bildsten}, Lars and {Blinnikov}, Sergei and {Duffell}, Paul and
         {Farmer}, R. and {Goldberg}, Jared A. and {Marchant}, Pablo and
         {Sorokina}, Elena and {Thoul}, Anne and {Townsend}, Richard H.~D. and
         {Timmes}, F.~X.},
        title = "{Modules for Experiments in Stellar Astrophysics (MESA): Convective Boundaries, Element Diffusion, and Massive Star Explosions}",
      journal = {\apjs},
         year = 2018,
        month = feb,
       volume = {234},
       number = {2},
          eid = {34},
        pages = {34},
          doi = {10.3847/1538-4365/aaa5a8},
archivePrefix = {arXiv},
       eprint = {1710.08424},
 primaryClass = {astro-ph.SR},
       adsurl = {https://ui.adsabs.harvard.edu/abs/2018ApJS..234...34P}
}

@ARTICLE{paxsmosch19,
       author = {{Paxton}, Bill and {Smolec}, R. and {Schwab}, Josiah and {Gautschy}, A. and
         {Bildsten}, Lars and {Cantiello}, Matteo and {Dotter}, Aaron and
         {Farmer}, R. and {Goldberg}, Jared A. and {Jermyn}, Adam S. and
         {Kanbur}, S.~M. and {Marchant}, Pablo and {Thoul}, Anne and
         {Townsend}, Richard H.~D. and {Wolf}, William M. and {Zhang}, Michael and
         {Timmes}, F.~X.},
        title = "{Modules for Experiments in Stellar Astrophysics (MESA): Pulsating Variable Stars, Rotation, Convective Boundaries, and Energy Conservation}",
      journal = {\apjs},
         year = 2019,
        month = jul,
       volume = {243},
       number = {1},
          eid = {10},
        pages = {10},
          doi = {10.3847/1538-4365/ab2241},
archivePrefix = {arXiv},
       eprint = {1903.01426},
 primaryClass = {astro-ph.SR},
       adsurl = {https://ui.adsabs.harvard.edu/abs/2019ApJS..243...10P}
}

@ARTICLE{guiramlin11,
       author = {{Guillochon}, James and {Ramirez-Ruiz}, Enrico and {Lin}, Douglas},
        title = "{Consequences of the Ejection and Disruption of Giant Planets}",
      journal = {\apj},
         year = 2011,
        month = may,
       volume = {732},
       number = {2},
          eid = {74},
        pages = {74},
          doi = {10.1088/0004-637X/732/2/74},
archivePrefix = {arXiv},
       eprint = {1012.2382},
 primaryClass = {astro-ph.EP},
       adsurl = {https://ui.adsabs.harvard.edu/abs/2011ApJ...732...74G}
}

@ARTICLE{oweada14,
       author = {{Owen}, James E. and {Adams}, Fred C.},
        title = "{Magnetically controlled mass-loss from extrasolar planets in close orbits}",
      journal = {\mnras},
         year = 2014,
        month = nov,
       volume = {444},
       number = {4},
        pages = {3761-3779},
          doi = {10.1093/mnras/stu1684},
archivePrefix = {arXiv},
       eprint = {1408.3636},
 primaryClass = {astro-ph.EP},
       adsurl = {https://ui.adsabs.harvard.edu/abs/2014MNRAS.444.3761O}
}

@ARTICLE{rieval97,
       author = {{Rieutord}, M. and {Valdettaro}, L.},
        title = "{Inertial waves in a rotating spherical shell}",
      journal = {Journal of Fluid Mechanics},
         year = 1997,
        month = jun,
       volume = {341},
       number = {1},
        pages = {77-99},
          doi = {10.1017/S0022112097005491},
       adsurl = {https://ui.adsabs.harvard.edu/abs/1997JFM...341...77R}
}

@ARTICLE{barestgir85,
       author = {{Barrera}, R.~G. and {Estevez}, G.~A. and {Giraldo}, J.},
        title = "{Vector spherical harmonics and their application to magnetostatics}",
      journal = {European Journal of Physics},
         year = 1985,
        month = oct,
       volume = {6},
       number = {4},
        pages = {287-294},
          doi = {10.1088/0143-0807/6/4/014},
       adsurl = {https://ui.adsabs.harvard.edu/abs/1985EJPh....6..287B}
}

@ARTICLE{zanmacbry26,
       author = {{Zanazzi}, J.~J. and {MacLeod}, Morgan and {Bryan}, Marta L. and {Mahadevan}, Suvrath},
        title = "{Dynamical Tides during High-Eccentricity Migration produces the Hot Jupiter Pile-up, Neptune Ridge, and Neptune Desert}",
      journal = {arXiv e-prints},
         year = 2026,
        month = jun,
          eid = {arXiv:2606.20789},
        pages = {arXiv:2606.20789},
          doi = {10.48550/arXiv.2606.20789},
archivePrefix = {arXiv},
       eprint = {2606.20789},
 primaryClass = {astro-ph.EP},
       adsurl = {https://ui.adsabs.harvard.edu/abs/2026arXiv260620789Z}
}

@ARTICLE{jertouogi17,
       author = {{Jermyn}, Adam S. and {Tout}, Christopher A. and {Ogilvie}, Gordon I.},
        title = "{Tidal heating and stellar irradiation of hot Jupiters}",
      journal = {\mnras},
         year = 2017,
        month = aug,
       volume = {469},
       number = {2},
        pages = {1768-1782},
          doi = {10.1093/mnras/stx831},
archivePrefix = {arXiv},
       eprint = {1704.01126},
 primaryClass = {astro-ph.EP},
       adsurl = {https://ui.adsabs.harvard.edu/abs/2017MNRAS.469.1768J}
}

@ARTICLE{wilbarpow23,
       author = {{Wilson}, Robert F. and {Barclay}, Thomas and {Powell}, Brian P. and {Schlieder}, Joshua and {Hedges}, Christina and {Montet}, Benjamin T. and {Quintana}, Elisa and {Mcdonald}, Iain and {Penny}, Matthew T. and {Espinoza}, N{\'e}stor and {Kerins}, Eamonn},
        title = "{Transiting Exoplanet Yields for the Roman Galactic Bulge Time Domain Survey Predicted from Pixel-level Simulations}",
      journal = {\apjs},
         year = 2023,
        month = nov,
       volume = {269},
       number = {1},
          eid = {5},
        pages = {5},
          doi = {10.3847/1538-4365/acf3df},
archivePrefix = {arXiv},
       eprint = {2305.16204},
 primaryClass = {astro-ph.EP},
       adsurl = {https://ui.adsabs.harvard.edu/abs/2023ApJS..269....5W}
}

@ARTICLE{preteu77,
       author = {{Press}, W.~H. and {Teukolsky}, S.~A.},
        title = "{On formation of close binaries by two-body tidal capture.}",
      journal = {\apj},
         year = 1977,
        month = apr,
       volume = {213},
        pages = {183-192},
          doi = {10.1086/155143},
       adsurl = {https://ui.adsabs.harvard.edu/abs/1977ApJ...213..183P}
}

@INPROCEEDINGS{ricwinvan14,
       author = {{Ricker}, George R. and {Winn}, Joshua N. and {Vanderspek}, Roland and {Latham}, David W. and {Bakos}, G{\'a}sp{\'a}r. {\'A}. and {Bean}, Jacob L. and {Berta-Thompson}, Zachory K. and {Brown}, Timothy M. and {Buchhave}, Lars and {Butler}, Nathaniel R. and {Butler}, R. Paul and {Chaplin}, William J. and {Charbonneau}, David and {Christensen-Dalsgaard}, J{\o}rgen and {Clampin}, Mark and {Deming}, Drake and {Doty}, John and {De Lee}, Nathan and {Dressing}, Courtney and {Dunham}, E.~W. and {Endl}, Michael and {Fressin}, Francois and {Ge}, Jian and {Henning}, Thomas and {Holman}, Matthew J. and {Howard}, Andrew W. and {Ida}, Shigeru and {Jenkins}, Jon and {Jernigan}, Garrett and {Johnson}, John A. and {Kaltenegger}, Lisa and {Kawai}, Nobuyuki and {Kjeldsen}, Hans and {Laughlin}, Gregory and {Levine}, Alan M. and {Lin}, Douglas and {Lissauer}, Jack J. and {MacQueen}, Phillip and {Marcy}, Geoffrey and {McCullough}, P.~R. and {Morton}, Timothy D. and {Narita}, Norio and {Paegert}, Martin and {Palle}, Enric and {Pepe}, Francesco and {Pepper}, Joshua and {Quirrenbach}, Andreas and {Rinehart}, S.~A. and {Sasselov}, Dimitar and {Sato}, Bun'ei and {Seager}, Sara and {Sozzetti}, Alessandro and {Stassun}, Keivan G. and {Sullivan}, Peter and {Szentgyorgyi}, Andrew and {Torres}, Guillermo and {Udry}, Stephane and {Villasenor}, Joel},
        title = "{Transiting Exoplanet Survey Satellite (TESS)}",
    booktitle = {Space Telescopes and Instrumentation 2014: Optical, Infrared, and Millimeter Wave},
         year = 2014,
       editor = {{Oschmann}, Jacobus M., Jr. and {Clampin}, Mark and {Fazio}, Giovanni G. and {MacEwen}, Howard A.},
       series = {Society of Photo-Optical Instrumentation Engineers (SPIE) Conference Series},
       volume = {9143},
        month = aug,
          eid = {914320},
        pages = {914320},
          doi = {10.1117/12.2063489},
archivePrefix = {arXiv},
       eprint = {1406.0151},
 primaryClass = {astro-ph.EP},
       adsurl = {https://ui.adsabs.harvard.edu/abs/2014SPIE.9143E..20R}
}

@ARTICLE{hub77,
       author = {{Hubbard}, W.~B.},
        title = "{The Jovian Surface Condition and Cooling Rate}",
      journal = {\icarus},
         year = 1977,
        month = feb,
       volume = {30},
       number = {2},
        pages = {305-310},
          doi = {10.1016/0019-1035(77)90164-6},
       adsurl = {https://ui.adsabs.harvard.edu/abs/1977Icar...30..305H}
}

@ARTICLE{arrbil06,
       author = {{Arras}, Phil and {Bildsten}, Lars},
        title = "{Thermal Structure and Radius Evolution of Irradiated Gas Giant Planets}",
      journal = {\apj},
         year = 2006,
        month = oct,
       volume = {650},
       number = {1},
        pages = {394-407},
          doi = {10.1086/506011},
archivePrefix = {arXiv},
       eprint = {astro-ph/0601317},
 primaryClass = {astro-ph},
       adsurl = {https://ui.adsabs.harvard.edu/abs/2006ApJ...650..394A}
}

@ARTICLE{formarbar07,
       author = {{Fortney}, J.~J. and {Marley}, M.~S. and {Barnes}, J.~W.},
        title = "{Planetary Radii across Five Orders of Magnitude in Mass and Stellar Insolation: Application to Transits}",
      journal = {\apj},
         year = 2007,
        month = apr,
       volume = {659},
       number = {2},
        pages = {1661-1672},
          doi = {10.1086/512120},
archivePrefix = {arXiv},
       eprint = {astro-ph/0612671},
 primaryClass = {astro-ph},
       adsurl = {https://ui.adsabs.harvard.edu/abs/2007ApJ...659.1661F}
}

@ARTICLE{komyou17,
       author = {{Komacek}, Thaddeus D. and {Youdin}, Andrew N.},
        title = "{Structure and Evolution of Internally Heated Hot Jupiters}",
      journal = {\apj},
         year = 2017,
        month = aug,
       volume = {844},
       number = {2},
          eid = {94},
        pages = {94},
          doi = {10.3847/1538-4357/aa7b75},
archivePrefix = {arXiv},
       eprint = {1706.07605},
 primaryClass = {astro-ph.EP},
       adsurl = {https://ui.adsabs.harvard.edu/abs/2017ApJ...844...94K}
}

@ARTICLE{huacum12,
       author = {{Huang}, Xu and {Cumming}, Andrew},
        title = "{Ohmic Dissipation in the Interiors of Hot Jupiters}",
      journal = {\apj},
         year = 2012,
        month = sep,
       volume = {757},
       number = {1},
          eid = {47},
        pages = {47},
          doi = {10.1088/0004-637X/757/1/47},
archivePrefix = {arXiv},
       eprint = {1207.3278},
 primaryClass = {astro-ph.EP},
       adsurl = {https://ui.adsabs.harvard.edu/abs/2012ApJ...757...47H}
}

@ARTICLE{hut81,
       author = {{Hut}, P.},
        title = "{Tidal evolution in close binary systems.}",
      journal = {\aap},
         year = 1981,
        month = jun,
       volume = {99},
        pages = {126-140},
       adsurl = {https://ui.adsabs.harvard.edu/abs/1981A&A....99..126H}
}

@ARTICLE{pacsie72,
       author = {{Paczy{\'n}ski}, B. and {Sienkiewicz}, R.},
        title = "{Evolution of Close Binaries VIII. Mass Exchange on the Dynamical Time Scale}",
      journal = {\actaa},
         year = 1972,
        month = jan,
       volume = {22},
        pages = {73-91},
       adsurl = {https://ui.adsabs.harvard.edu/abs/1972AcA....22...73P}
}

@ARTICLE{barogi09,
       author = {{Barker}, Adrian J. and {Ogilvie}, Gordon I.},
        title = "{On the tidal evolution of Hot Jupiters on inclined orbits}",
      journal = {\mnras},
         year = 2009,
        month = jun,
       volume = {395},
       number = {4},
        pages = {2268-2287},
          doi = {10.1111/j.1365-2966.2009.14694.x},
archivePrefix = {arXiv},
       eprint = {0902.4563},
 primaryClass = {astro-ph.EP},
       adsurl = {https://ui.adsabs.harvard.edu/abs/2009MNRAS.395.2268B}
}

@ARTICLE{visbeh25,
       author = {{Vissapragada}, Shreyas and {Behmard}, Aida},
        title = "{The Hottest Neptunes Orbit Metal-rich Stars}",
      journal = {\aj},
         year = 2025,
        month = feb,
       volume = {169},
       number = {2},
          eid = {117},
        pages = {117},
          doi = {10.3847/1538-3881/ada143},
archivePrefix = {arXiv},
       eprint = {2412.13245},
 primaryClass = {astro-ph.EP},
       adsurl = {https://ui.adsabs.harvard.edu/abs/2025AJ....169..117V}
}

@ARTICLE{casboulil24,
       author = {{Castro-Gonz{\'a}lez}, A. and {Bourrier}, V. and {Lillo-Box}, J. and {Delisle}, J. -B. and {Armstrong}, D.~J. and {Barrado}, D. and {Correia}, A.~C.~M.},
        title = "{Mapping the exo-Neptunian landscape: A ridge between the desert and savanna}",
      journal = {\aap},
         year = 2024,
        month = sep,
       volume = {689},
          eid = {A250},
        pages = {A250},
          doi = {10.1051/0004-6361/202450957},
archivePrefix = {arXiv},
       eprint = {2409.10517},
 primaryClass = {astro-ph.EP},
       adsurl = {https://ui.adsabs.harvard.edu/abs/2024A&A...689A.250C}
}

@ARTICLE{bouattmall23,
       author = {{Bourrier}, V. and {Attia}, M. and {Mallonn}, M. and {Marret}, A. and {Lendl}, M. and {Konig}, P. -C. and {Krenn}, A. and {Cretignier}, M. and {Allart}, R. and {Henry}, G. and {Bryant}, E. and {Leleu}, A. and {Nielsen}, L. and {Hebrard}, G. and {Hara}, N. and {Ehrenreich}, D. and {Seidel}, J. and {dos Santos}, L. and {Lovis}, C. and {Bayliss}, D. and {Cegla}, H.~M. and {Dumusque}, X. and {Boisse}, I. and {Boucher}, A. and {Bouchy}, F. and {Pepe}, F. and {Lavie}, B. and {Rey Cerda}, J. and {S{\'e}gransan}, D. and {Udry}, S. and {Vrignaud}, T.},
        title = "{DREAM: I. Orbital architecture orrery}",
      journal = {\aap},
         year = 2023,
        month = jan,
       volume = {669},
          eid = {A63},
        pages = {A63},
          doi = {10.1051/0004-6361/202245004},
archivePrefix = {arXiv},
       eprint = {2301.07727},
 primaryClass = {astro-ph.EP},
       adsurl = {https://ui.adsabs.harvard.edu/abs/2023A&A...669A..63B}
}

@ARTICLE{doyarmacu25,
       author = {{Doyle}, Lauren and {Armstrong}, David J. and {Acu{\~n}a}, Lorena and {Osborn}, Ares and {Sousa}, S{\'e}rgio A.~G. and {Castro-Gonz{\'a}lez}, Amadeo and {Bourrier}, Vincent and {Alves}, Douglas and {Barrado}, David and {Barros}, Susana C.~C. and {Bayliss}, Daniel and {Cui}, Kaiming and {Demangeon}, Olivier and {D{\'\i}az}, Rodrigo F. and {Dumusque}, Xavier and {Eeles-Nolle}, Fintan and {Gill}, Samuel and {Hacker}, Alejandro and {Jenkins}, James S. and {Keniger}, Marcelo Aron Fetzner and {Lafarga}, Marina and {Lillo-Box}, Jorge and {Lockley}, Isobel and {Nielsen}, Louise D. and {Parc}, L{\'e}na and {Rodrigues}, Jos{\'e} and {Santerne}, Alexandre and {Santos}, Nuno C. and {Wheatley}, Peter J.},
        title = "{Exploring the Neptunian Desert: Insights from a Homogeneous Planetary Sample}",
      journal = {\mnras},
         year = 2025,
        month = apr,
          doi = {10.1093/mnras/staf670},
archivePrefix = {arXiv},
       eprint = {2504.16164},
 primaryClass = {astro-ph.EP},
       adsurl = {https://ui.adsabs.harvard.edu/abs/2025MNRAS.tmp..623D}
}

@ARTICLE{koktsiedw23,
       author = {{Kokori}, A. and {Tsiaras}, A. and {Edwards}, B. and {Jones}, A. and {Pantelidou}, G. and {Tinetti}, G. and {Bewersdorff}, L. and {Iliadou}, A. and {Jongen}, Y. and {Lekkas}, G. and {Nastasi}, A. and {Poultourtzidis}, E. and {Sidiropoulos}, C. and {Walter}, F. and {W{\"u}nsche}, A. and {Abraham}, R. and {Agnihotri}, V.~K. and {Albanesi}, R. and {Arce-Mansego}, E. and {Arnot}, D. and {Audejean}, M. and {Aumasson}, C. and {Bachschmidt}, M. and {Baj}, G. and {Barroy}, P.~R. and {Belinski}, A.~A. and {Bennett}, D. and {Benni}, P. and {Bernacki}, K. and {Betti}, L. and {Biagini}, A. and {Bosch}, P. and {Brandebourg}, P. and {Br{\'a}t}, L. and {Bretton}, M. and {Brincat}, S.~M. and {Brouillard}, S. and {Bruzas}, A. and {Bruzzone}, A. and {Buckland}, R.~A. and {Cal{\'o}}, M. and {Campos}, F. and {Carre{\~n}o}, A. and {Carrion Rodrigo}, J.~A. and {Casali}, R. and {Casalnuovo}, G. and {Cataneo}, M. and {Chang}, C. -M. and {Changeat}, L. and {Chowdhury}, V. and {Ciantini}, R. and {Cilluffo}, M. and {Coliac}, J. -F. and {Conzo}, G. and {Correa}, M. and {Coulon}, G. and {Crouzet}, N. and {Crow}, M.~V. and {Curtis}, I.~A. and {Daniel}, D. and {Dauchet}, B. and {Dawes}, S. and {Deldem}, M. and {Deligeorgopoulos}, D. and {Dransfield}, G. and {Dymock}, R. and {Eenm{\"a}e}, T. and {Esseiva}, N. and {Evans}, P. and {Falco}, C. and {Farf{\'a}n}, R.~G. and {Fern{\'a}ndez-Laj{\'u}s}, E. and {Ferratfiat}, S. and {Ferreira}, S.~L. and {Ferretti}, A. and {Fio{\l}ka}, J. and {Fowler}, M. and {Futcher}, S.~R. and {Gabellini}, D. and {Gainey}, T. and {Gaitan}, J. and {Gajdo{\v{s}}}, P. and {Garc{\'\i}a-S{\'a}nchez}, A. and {Garlitz}, J. and {Gillier}, C. and {Gison}, C. and {Gonzales}, J. and {Gorshanov}, D. and {Grau Horta}, F. and {Grivas}, G. and {Guerra}, P. and {Guillot}, T. and {Haswell}, C.~A. and {Haymes}, T. and {Hentunen}, V. -P. and {Hills}, K. and {Hose}, K. and {Humbert}, T. and {Hurter}, F. and {Hynek}, T. and {Irzyk}, M. and {Jacobsen}, J. and {Jannetta}, A.~L. and {Johnson}, K. and {J{\'o}{\'z}wik-Wabik}, P. and {Kaeouach}, A.~E. and {Kang}, W. and {Kiiskinen}, H. and {Kim}, T. and {Kivila}, {\"U}. and {Koch}, B. and {Kolb}, U. and {Ku{\v{c}}{\'a}kov{\'a}}, H. and {Lai}, S. -P. and {Laloum}, D. and {Lasota}, S. and {Lewis}, L.~A. and {Liakos}, G. -I. and {Libotte}, F. and {Lomoz}, F. and {Lopresti}, C. and {Majewski}, R. and {Malcher}, A. and {Mallonn}, M. and {Mannucci}, M. and {Marchini}, A. and {Mari}, J. -M. and {Marino}, A. and {Marino}, G. and {Mario}, J. -C. and {Marquette}, J. -B. and {Mart{\'\i}nez-Bravo}, F.~A. and {Ma{\v{s}}ek}, M. and {Matassa}, P. and {Michel}, P. and {Michelet}, J. and {Miller}, M. and {Miny}, E. and {Molina}, D. and {Mollier}, T. and {Monteleone}, B. and {Montigiani}, N. and {Morales-Aimar}, M. and {Mortari}, F. and {Morvan}, M. and {Mugnai}, L.~V. and {Murawski}, G. and {Naponiello}, L. and {Naudin}, J. -L. and {Naves}, R. and {N{\'e}el}, D. and {Neito}, R. and {Neveu}, S. and {Noschese}, A. and {{\"O}{\u{g}}men}, Y. and {Ohshima}, O. and {Orbanic}, Z. and {Pace}, E.~P. and {Pantacchini}, C. and {Paschalis}, N.~I. and {Pereira}, C. and {Peretto}, I. and {Perroud}, V. and {Phillips}, M. and {Pintr}, P. and {Pioppa}, J. -B. and {Plazas}, J. and {Poelarends}, A.~J. and {Popowicz}, A. and {Purcell}, J. and {Quinn}, N. and {Raetz}, M. and {Rees}, D. and {Regembal}, F. and {Rocchetto}, M. and {Rocci}, P. -F. and {Rockenbauer}, M. and {Roth}, R. and {Rousselot}, L. and {Rubia}, X. and {Ruocco}, N. and {Russo}, E. and {Salisbury}, M. and {Salvaggio}, F. and {Santos}, A. and {Savage}, J. and {Scaggiante}, F. and {Sedita}, D. and {Shadick}, S. and {Silva}, A.~F. and {Sioulas}, N. and {{\v{S}}koln{\'\i}k}, V. and {Smith}, M. and {Smolka}, M. and {Solmaz}, A. and {Stanbury}, N. and {Stouraitis}, D. and {Tan}, T. -G. and {Theusner}, M. and {Thurston}, G.},
        title = "{ExoClock Project. III. 450 New Exoplanet Ephemerides from Ground and Space Observations}",
      journal = {\apjs},
         year = 2023,
        month = mar,
       volume = {265},
       number = {1},
          eid = {4},
        pages = {4},
          doi = {10.3847/1538-4365/ac9da4},
archivePrefix = {arXiv},
       eprint = {2209.09673},
 primaryClass = {astro-ph.EP},
       adsurl = {https://ui.adsabs.harvard.edu/abs/2023ApJS..265....4K}
}

@ARTICLE{der79,
       author = {{Dermott}, S.~F.},
        title = "{Tidal Dissipation in the Solid Cores of the Major Planets}",
      journal = {\icarus},
         year = 1979,
        month = jan,
       volume = {37},
       number = {1},
        pages = {310-321},
          doi = {10.1016/0019-1035(79)90137-4},
       adsurl = {https://ui.adsabs.harvard.edu/abs/1979Icar...37..310D}
}

@ARTICLE{towtei13,
       author = {{Townsend}, R.~H.~D. and {Teitler}, S.~A.},
        title = "{GYRE: an open-source stellar oscillation code based on a new Magnus Multiple Shooting scheme}",
      journal = {\mnras},
         year = 2013,
        month = nov,
       volume = {435},
       number = {4},
        pages = {3406-3418},
          doi = {10.1093/mnras/stt1533},
archivePrefix = {arXiv},
       eprint = {1308.2965},
 primaryClass = {astro-ph.SR},
       adsurl = {https://ui.adsabs.harvard.edu/abs/2013MNRAS.435.3406T}
}

@ARTICLE{farchi26,
       author = {{Farhat}, Mohammad and {Chiang}, Eugene},
        title = "{Magma Ocean Waves and Thermal Variability on Lava Worlds}",
      journal = {\apj},
         year = 2026,
        month = jun,
       volume = {1003},
       number = {2},
          eid = {208},
        pages = {208},
          doi = {10.3847/1538-4357/ae6504},
archivePrefix = {arXiv},
       eprint = {2601.07080},
 primaryClass = {astro-ph.EP},
       adsurl = {https://ui.adsabs.harvard.edu/abs/2026ApJ..1003..208F}
}

@ARTICLE{ogilin04,
       author = {{Ogilvie}, G.~I. and {Lin}, D.~N.~C.},
        title = "{Tidal Dissipation in Rotating Giant Planets}",
      journal = {\apj},
         year = 2004,
        month = jul,
       volume = {610},
       number = {1},
        pages = {477-509},
          doi = {10.1086/421454},
archivePrefix = {arXiv},
       eprint = {astro-ph/0310218},
 primaryClass = {astro-ph},
       adsurl = {https://ui.adsabs.harvard.edu/abs/2004ApJ...610..477O}
}

@ARTICLE{corboulas14,
       author = {{Correia}, Alexandre C.~M. and {Bou{\'e}}, Gwena{\"e}l and {Laskar}, Jacques and {Rodr{\'\i}guez}, Adri{\'a}n},
        title = "{Deformation and tidal evolution of close-in planets and satellites using a Maxwell viscoelastic rheology}",
      journal = {\aap},
         year = 2014,
        month = nov,
       volume = {571},
          eid = {A50},
        pages = {A50},
          doi = {10.1051/0004-6361/201424211},
archivePrefix = {arXiv},
       eprint = {1411.1860},
 primaryClass = {astro-ph.EP},
       adsurl = {https://ui.adsabs.harvard.edu/abs/2014A&A...571A..50C}
}

@ARTICLE{ivapap04b,
       author = {{Ivanov}, P.~B. and {Papaloizou}, J.~C.~B.},
        title = "{On equilibrium tides in fully convective planets and stars}",
      journal = {\mnras},
         year = 2004,
        month = oct,
       volume = {353},
       number = {4},
        pages = {1161-1175},
          doi = {10.1111/j.1365-2966.2004.08136.x},
archivePrefix = {arXiv},
       eprint = {astro-ph/0312593},
 primaryClass = {astro-ph},
       adsurl = {https://ui.adsabs.harvard.edu/abs/2004MNRAS.353.1161I}
}

@ARTICLE{viclai18,
       author = {{Vick}, Michelle and {Lai}, Dong},
        title = "{Dynamical tides in highly eccentric binaries: chaos, dissipation, and quasi-steady state}",
      journal = {\mnras},
         year = 2018,
        month = may,
       volume = {476},
       number = {1},
        pages = {482-495},
          doi = {10.1093/mnras/sty225},
archivePrefix = {arXiv},
       eprint = {1708.09392},
 primaryClass = {astro-ph.SR},
       adsurl = {https://ui.adsabs.harvard.edu/abs/2018MNRAS.476..482V}
}

@ARTICLE{wu18,
       author = {{Wu}, Yanqin},
        title = "{Diffusive Tidal Evolution for Migrating Hot Jupiters}",
      journal = {\aj},
         year = 2018,
        month = mar,
       volume = {155},
       number = {3},
          eid = {118},
        pages = {118},
          doi = {10.3847/1538-3881/aaa970},
archivePrefix = {arXiv},
       eprint = {1710.02542},
 primaryClass = {astro-ph.EP},
       adsurl = {https://ui.adsabs.harvard.edu/abs/2018AJ....155..118W}
}

@BOOK{murder99,
       author = {{Murray}, Carl D. and {Dermott}, Stanley F.},
        title = "{Solar System Dynamics}",
         year = 1999,
          doi = {10.1017/CBO9781139174817},
       adsurl = {https://ui.adsabs.harvard.edu/abs/1999ssd..book.....M}
}

@ARTICLE{bar25,
       author = {{Barker}, Adrian J.},
        title = "{Tidal interactions in stellar and planetary systems}",
      journal = {arXiv e-prints},
         year = 2025,
        month = apr,
          eid = {arXiv:2504.10941},
        pages = {arXiv:2504.10941},
          doi = {10.48550/arXiv.2504.10941},
archivePrefix = {arXiv},
       eprint = {2504.10941},
 primaryClass = {astro-ph.EP},
       adsurl = {https://ui.adsabs.harvard.edu/abs/2025arXiv250410941B}
}

@ARTICLE{mataucgue16,
       author = {{Mathis}, S. and {Auclair-Desrotour}, P. and {Guenel}, M. and {Gallet}, F. and {Le Poncin-Lafitte}, C.},
        title = "{The impact of rotation on turbulent tidal friction in stellar and planetary convective regions}",
      journal = {\aap},
         year = 2016,
        month = jul,
       volume = {592},
          eid = {A33},
        pages = {A33},
          doi = {10.1051/0004-6361/201527545},
archivePrefix = {arXiv},
       eprint = {1604.08570},
 primaryClass = {astro-ph.SR},
       adsurl = {https://ui.adsabs.harvard.edu/abs/2016A&A...592A..33M}
}

@ARTICLE{golnic77,
       author = {{Goldreich}, P. and {Nicholson}, P.~D.},
        title = "{Turbulent Viscosity and Jupiter's Tidal Q}",
      journal = {\icarus},
         year = 1977,
        month = feb,
       volume = {30},
       number = {2},
        pages = {301-304},
          doi = {10.1016/0019-1035(77)90163-4},
       adsurl = {https://ui.adsabs.harvard.edu/abs/1977Icar...30..301G}
}

@ARTICLE{golnic89,
       author = {{Goldreich}, Peter and {Nicholson}, Philip D.},
        title = "{Tidal Friction in Early-Type Stars}",
      journal = {\apj},
         year = 1989,
        month = jul,
       volume = {342},
        pages = {1079},
          doi = {10.1086/167665},
       adsurl = {https://ui.adsabs.harvard.edu/abs/1989ApJ...342.1079G}
}

@ARTICLE{lubtouliv97,
       author = {{Lubow}, S.~H. and {Tout}, C.~A. and {Livio}, M.},
        title = "{Resonant Tides in Close Orbiting Planets}",
      journal = {\apj},
         year = 1997,
        month = jul,
       volume = {484},
       number = {2},
        pages = {866-870},
          doi = {10.1086/304369},
       adsurl = {https://ui.adsabs.harvard.edu/abs/1997ApJ...484..866L}
}

@ARTICLE{ogi14,
       author = {{Ogilvie}, Gordon I.},
        title = "{Tidal Dissipation in Stars and Giant Planets}",
      journal = {\araa},
         year = 2014,
        month = aug,
       volume = {52},
        pages = {171-210},
          doi = {10.1146/annurev-astro-081913-035941},
archivePrefix = {arXiv},
       eprint = {1406.2207},
 primaryClass = {astro-ph.SR},
       adsurl = {https://ui.adsabs.harvard.edu/abs/2014ARA&A..52..171O}
}

@ARTICLE{virgomoli20,
       author = {{Virtanen}, Pauli and {Gommers}, Ralf and {Oliphant}, Travis E. and {Haberland}, Matt and {Reddy}, Tyler and {Cournapeau}, David and {Burovski}, Evgeni and {Peterson}, Pearu and {Weckesser}, Warren and {Bright}, Jonathan and {van der Walt}, St{\'e}fan J. and {Brett}, Matthew and {Wilson}, Joshua and {Millman}, K. Jarrod and {Mayorov}, Nikolay and {Nelson}, Andrew R.~J. and {Jones}, Eric and {Kern}, Robert and {Larson}, Eric and {Carey}, C.~J. and {Polat}, {\.I}lhan and {Feng}, Yu and {Moore}, Eric W. and {VanderPlas}, Jake and {Laxalde}, Denis and {Perktold}, Josef and {Cimrman}, Robert and {Henriksen}, Ian and {Quintero}, E.~A. and {Harris}, Charles R. and {Archibald}, Anne M. and {Ribeiro}, Ant{\^o}nio H. and {Pedregosa}, Fabian and {van Mulbregt}, Paul and {SciPy 1. 0 Contributors}},
        title = "{SciPy 1.0: fundamental algorithms for scientific computing in Python}",
      journal = {Nature Methods},
         year = 2020,
        month = feb,
       volume = {17},
        pages = {261-272},
          doi = {10.1038/s41592-019-0686-2},
archivePrefix = {arXiv},
       eprint = {1907.10121},
 primaryClass = {cs.MS},
       adsurl = {https://ui.adsabs.harvard.edu/abs/2020NatMe..17..261V}
}

@ARTICLE{youstirog24,
       author = {{Young}, Edward D. and {Stixrude}, Lars and {Rogers}, James G. and {Schlichting}, Hilke E. and {Marcum}, Sarah P.},
        title = "{Phase Equilibria of Sub-Neptunes and Super-Earths}",
      journal = {Planetary Science Journal},
         year = 2024,
        month = dec,
       volume = {5},
       number = {12},
          eid = {268},
        pages = {268},
          doi = {10.3847/PSJ/ad8c40},
archivePrefix = {arXiv},
       eprint = {2408.11321},
 primaryClass = {astro-ph.EP},
       adsurl = {https://ui.adsabs.harvard.edu/abs/2024PSJ.....5..268Y}
}

@ARTICLE{marguiste22,
       author = {{Markham}, S. and {Guillot}, T. and {Stevenson}, D.},
        title = "{Convective inhibition with an ocean. I. Supercritical cores on sub-Neptunes/super-Earths}",
      journal = {\aap},
         year = 2022,
        month = sep,
       volume = {665},
          eid = {A12},
        pages = {A12},
          doi = {10.1051/0004-6361/202243359},
archivePrefix = {arXiv},
       eprint = {2207.04708},
 primaryClass = {astro-ph.EP},
       adsurl = {https://ui.adsabs.harvard.edu/abs/2022A&A...665A..12M}
}

@ARTICLE{glarozper22,
       author = {{Glanz}, Hila and {Rozner}, Mor and {Perets}, Hagai B. and {Grishin}, Evgeni},
        title = "{Inflated Eccentric Migration of Evolving Gas Giants II - Numerical Methodology and Basic Concepts}",
      journal = {\apj},
         year = 2022,
        month = may,
       volume = {931},
       number = {1},
          eid = {11},
        pages = {11},
          doi = {10.3847/1538-4357/ac6807},
archivePrefix = {arXiv},
       eprint = {2111.12714},
 primaryClass = {astro-ph.EP},
       adsurl = {https://ui.adsabs.harvard.edu/abs/2022ApJ...931...11G}
}

@ARTICLE{rozglaper22,
       author = {{Rozner}, Mor and {Glanz}, Hila and {Perets}, Hagai B. and {Grishin}, Evgeni},
        title = "{Inflated Eccentric Migration of Evolving Gas Giants I - Accelerated Formation and Destruction of Hot and Warm Jupiters}",
      journal = {\apj},
         year = 2022,
        month = may,
       volume = {931},
       number = {1},
          eid = {10},
        pages = {10},
          doi = {10.3847/1538-4357/ac6808},
archivePrefix = {arXiv},
       eprint = {2111.12718},
 primaryClass = {astro-ph.EP},
       adsurl = {https://ui.adsabs.harvard.edu/abs/2022ApJ...931...10R}
}

@ARTICLE{sursutej24,
       author = {{Sur}, Ankan and {Su}, Yubo and {Tejada Arevalo}, Roberto and {Chen}, Yi-Xian and {Burrows}, Adam},
        title = "{APPLE: An Evolution Code for Modeling Giant Planets}",
      journal = {\apj},
         year = 2024,
        month = aug,
       volume = {971},
       number = {1},
          eid = {104},
        pages = {104},
          doi = {10.3847/1538-4357/ad57c3},
archivePrefix = {arXiv},
       eprint = {2404.14483},
 primaryClass = {astro-ph.EP},
       adsurl = {https://ui.adsabs.harvard.edu/abs/2024ApJ...971..104S}
}

@ARTICLE{rauaercab25,
       author = {{Rauer}, Heike and {Aerts}, Conny and {Cabrera}, Juan and {Deleuil}, Magali and {Erikson}, Anders and {Gizon}, Laurent and {Goupil}, Mariejo and {Heras}, Ana and {Walloschek}, Thomas and {Lorenzo-Alvarez}, Jose and {Marliani}, Filippo and {Martin-Garcia}, C{\'e}sar and {Mas-Hesse}, J. Miguel and {O'Rourke}, Laurence and {Osborn}, Hugh and {Pagano}, Isabella and {Piotto}, Giampaolo and {Pollacco}, Don and {Ragazzoni}, Roberto and {Ramsay}, Gavin and {Udry}, St{\'e}phane and {Appourchaux}, Thierry and {Benz}, Willy and {Brandeker}, Alexis and {G{\"u}del}, Manuel and {Janot-Pacheco}, Eduardo and {Kabath}, Petr and {Kjeldsen}, Hans and {Min}, Michiel and {Santos}, Nuno and {Smith}, Alan and {Suarez}, Juan-Carlos and {Werner}, Stephanie C. and {Aboudan}, Alessio and {Abreu}, Manuel and {Acu{\~n}a}, Lorena and {Adams}, Moritz and {Adibekyan}, Vardan and {Affer}, Laura and {Agneray}, Fran{\c{c}}ois and {Agnor}, Craig and {Aguirre B{\o}rsen-Koch}, Victor and {Ahmed}, Saad and {Aigrain}, Suzanne and {Al-Bahlawan}, Ashraf and {Alcacera Gil}, Ma de los Angeles and {Alei}, Eleonora and {Alencar}, Silvia and {Alexander}, Richard and {Alfonso-Garz{\'o}n}, Julia and {Alibert}, Yann and {Allende Prieto}, Carlos and {Almeida}, Leonardo and {Alonso Sobrino}, Roi and {Altavilla}, Giuseppe and {Althaus}, Christian and {Alvarez Trujillo}, Luis Alonso and {Amarsi}, Anish and {Ammler-von Eiff}, Matthias and {Am{\^o}res}, Eduardo and {Andrade}, Laerte and {Antoniadis-Karnavas}, Alexandros and {Ant{\'o}nio}, Carlos and {Aparicio del Moral}, Beatriz and {Appolloni}, Matteo and {Arena}, Claudio and {Armstrong}, David and {Aroca Aliaga}, Jose and {Asplund}, Martin and {Audenaert}, Jeroen and {Auricchio}, Natalia and {Avelino}, Pedro and {Baeke}, Ann and {Bailli{\'e}}, Kevin and {Balado}, Ana and {Ballber Balaguer{\'o}}, Pau and {Balestra}, Andrea and {Ball}, Warrick and {Ballans}, Herve and {Ballot}, Jerome and {Barban}, Caroline and {Barbary}, Ga{\"e}le and {Barbieri}, Mauro and {Barcel{\'o} Forteza}, Sebasti{\`a} and {Barker}, Adrian and {Barklem}, Paul and {Barnes}, Sydney and {Barrado Navascues}, David and {Barragan}, Oscar and {Baruteau}, Cl{\'e}ment and {Basu}, Sarbani and {Baudin}, Frederic and {Baumeister}, Philipp and {Bayliss}, Daniel and {Bazot}, Michael and {Beck}, Paul G. and {Belkacem}, Kevin and {Bellinger}, Earl and {Benatti}, Serena and {Benomar}, Othman and {B{\'e}rard}, Diane and {Bergemann}, Maria and {Bergomi}, Maria and {Bernardo}, Pierre and {Biazzo}, Katia and {Bignamini}, Andrea and {Bigot}, Lionel and {Billot}, Nicolas and {Binet}, Martin and {Biondi}, David and {Biondi}, Federico and {Birch}, Aaron C. and {Bitsch}, Bertram and {Bluhm Ceballos}, Paz Victoria and {B{\'o}di}, Attila and {Bogn{\'a}r}, Zs{\'o}fia and {Boisse}, Isabelle and {Bolmont}, Emeline and {Bonanno}, Alfio and {Bonavita}, Mariangela and {Bonfanti}, Andrea and {Bonfils}, Xavier and {Bonito}, Rosaria and {Bonomo}, Aldo Stefano and {B{\"o}rner}, Anko and {Boro Saikia}, Sudeshna and {Borreguero Mart{\'\i}n}, Elisa and {Borsa}, Francesco and {Borsato}, Luca and {Bossini}, Diego and {Bouchy}, Francois and {Bou{\'e}}, Gwena{\"e}l and {Boufleur}, Rodrigo and {Boumier}, Patrick and {Bourrier}, Vincent and {Bowman}, Dominic M. and {Bozzo}, Enrico and {Bradley}, Louisa and {Bray}, John and {Bressan}, Alessandro and {Breton}, Sylvain and {Brienza}, Daniele and {Brito}, Ana and {Brogi}, Matteo and {Brown}, Beverly and {Brown}, David J.~A. and {Brun}, Allan Sacha and {Bruno}, Giovanni and {Bruns}, Michael and {Buchhave}, Lars A. and {Bugnet}, Lisa and {Buldgen}, Ga{\"e}l and {Burgess}, Patrick and {Busatta}, Andrea and {Busso}, Giorgia and {Buzasi}, Derek and {Caballero}, Jos{\'e} A. and {Cabral}, Alexandre and {Cabrero Gomez}, Juan-Francisco and {Calderone}, Flavia and {Cameron}, Robert and {Cameron}, Andrew and {Campante}, Tiago and {Campos Gestal}, N{\'e}stor and {Canto Martins}, Bruno Leonardo and {Cara}, Christophe and {Carone}, Ludmila and {Carrasco}, Josep Manel and {Casagrande}, Luca and {Casewell}, Sarah L. and {Cassisi}, Santi and {Castellani}, Marco and {Castro}, Matthieu and {Catala}, Claude and {Catal{\'a}n Fern{\'a}ndez}, Irene and {Catelan}, M{\'a}rcio and {Cegla}, Heather and {Cerruti}, Chiara and {Cessa}, Virginie and {Chadid}, Merieme and {Chaplin}, William and {Charpinet}, Stephane and {Chiappini}, Cristina and {Chiarucci}, Simone and {Chiavassa}, Andrea and {Chinellato}, Simonetta and {Chirulli}, Giovanni and {Christensen-Dalsgaard}, J{\o}rgen and {Church}, Ross and {Claret}, Antonio and {Clarke}, Cathie and {Claudi}, Riccardo and {Clermont}, Lionel and {Coelho}, Hugo and {Coelho}, Joao and {Cogato}, Fabrizio and {Colom{\'e}}, Josep and {Condamin}, Mathieu and {Conde Garc{\'\i}a}, Fernando and {Conseil}, Simon},
        title = "{The PLATO mission}",
      journal = {Experimental Astronomy},
         year = 2025,
        month = jun,
       volume = {59},
       number = {3},
          eid = {26},
        pages = {26},
          doi = {10.1007/s10686-025-09985-9},
archivePrefix = {arXiv},
       eprint = {2406.05447},
 primaryClass = {astro-ph.IM},
       adsurl = {https://ui.adsabs.harvard.edu/abs/2025ExA....59...26R}
}

@ARTICLE{gupmil2024,
       author = {{Gupta}, Arvind F. and {Millholland}, Sarah C. and {Im}, Haedam and {Dong}, Jiayin and {Jackson}, Jonathan M. and {Carleo}, Ilaria and {Libby-Roberts}, Jessica and {Delamer}, Megan and {Giovinazzi}, Mark R. and {Lin}, Andrea S.~J. and {Kanodia}, Shubham and {Wang}, Xian-Yu and {Stassun}, Keivan and {Masseron}, Thomas and {Dragomir}, Diana and {Mahadevan}, Suvrath and {Wright}, Jason and {Alvarado-Montes}, Jaime A. and {Bender}, Chad and {Blake}, Cullen H. and {Caldwell}, Douglas and {Ca{\~n}as}, Caleb I. and {Cochran}, William D. and {Dalba}, Paul and {Everett}, Mark E. and {Fernandez}, Pipa and {Golub}, Eli and {Guillet}, Bruno and {Halverson}, Samuel and {Hebb}, Leslie and {Higuera}, Jesus and {Huang}, Chelsea X. and {Klusmeyer}, Jessica and {Knight}, Rachel and {Leroux}, Liouba and {Logsdon}, Sarah E. and {Loose}, Margaret and {McElwain}, Michael W. and {Monson}, Andrew and {Ninan}, Joe P. and {Nowak}, Grzegorz and {Palle}, Enric and {Patel}, Yatrik and {Pepper}, Joshua and {Primm}, Michael and {Rajagopal}, Jayadev and {Robertson}, Paul and {Roy}, Arpita and {Schneider}, Donald P. and {Schwab}, Christian and {Schweiker}, Heidi and {Sgro}, Lauren and {Shimizu}, Masao and {Simard}, Georges and {Stef{\'a}nsson}, Gudmundur and {Stevens}, Daniel J. and {Villanueva}, Steven and {Wisniewski}, John and {Will}, Stefan and {Ziegler}, Carl},
        title = "{A hot-Jupiter progenitor on a super-eccentric retrograde orbit}",
      journal = {\nat},
         year = 2024,
        month = aug,
       volume = {632},
       number = {8023},
        pages = {50-54},
          doi = {10.1038/s41586-024-07688-3},
       adsurl = {https://ui.adsabs.harvard.edu/abs/2024Natur.632...50G}
}

@ARTICLE{livmilvic2026,
       author = {{Liveoak}, Donald and {Millholland}, Sarah C. and {Vick}, Michelle},
        title = "{Case Study of a Super-eccentric Warm Jupiter Migrating via Equilibrium and Dynamical Tides}",
      journal = {\apj},
         year = 2026,
        month = aug,
       volume = {1007},
       number = {2},
          eid = {152},
        pages = {152},
          doi = {10.3847/1538-4357/ae8b99},
archivePrefix = {arXiv},
       eprint = {2607.16406},
 primaryClass = {astro-ph.EP},
       adsurl = {https://ui.adsabs.harvard.edu/abs/2026ApJ..1007..152L}
}
\bibliographystyle{aasjournal}

\end{document}